\documentclass[english,aip,letter,superscriptaddress,floatfix,jcp,11pt]{revtex4}
\usepackage[T1]{fontenc}
\usepackage[latin9]{inputenc}
\usepackage{color}
\usepackage{verbatim}
\usepackage{amsmath}
\usepackage{amssymb}
\usepackage{graphicx}
\usepackage{esint}

\makeatletter

\providecommand{\tabularnewline}{\\}

\@ifundefined{textcolor}{}
{%
 \definecolor{BLACK}{gray}{0}
 \definecolor{WHITE}{gray}{1}
 \definecolor{RED}{rgb}{1,0,0}
 \definecolor{GREEN}{rgb}{0,1,0}
 \definecolor{BLUE}{rgb}{0,0,1}
 \definecolor{CYAN}{cmyk}{1,0,0,0}
 \definecolor{MAGENTA}{cmyk}{0,1,0,0}
 \definecolor{YELLOW}{cmyk}{0,0,1,0}
}

\usepackage{ae,aecompl}

\makeatother

\usepackage{babel}
\begin{document}

\title{Brownian Dynamics of Confined Rigid Bodies}

\author{Steven Delong}

\affiliation{Courant Institute of Mathematical Sciences, New York University,
New York, NY 10012}

\author{Florencio Balboa Usabiaga}

\affiliation{Courant Institute of Mathematical Sciences, New York University,
New York, NY 10012}

\author{Aleksandar Donev}

\email{donev@courant.nyu.edu}

\selectlanguage{english}%

\affiliation{Courant Institute of Mathematical Sciences, New York University,
New York, NY 10012}
\begin{abstract}
\textcolor{black}{We introduce numerical methods for simulating the
diffusive motion of rigid bodies of arbitrary shape immersed in a
viscous fluid. We parameterize the orientation of the bodies using
normalized quaternions, which are numerically robust, space efficient,
and easy to accumulate. We construct a system of overdamped Langevin
equations in the quaternion representation that accounts for hydrodynamic
effects, preserves the unit-norm constraint on the quaternion, and
is time reversible with respect to the Gibbs-Boltzmann distribution
at equilibrium. We introduce two schemes for temporal integration
of the overdamped Langevin equations of motion, one based on the Fixman
midpoint method and the other based on a random finite difference
approach, both of which ensure the correct stochastic drift term is
captured in a computationally efficient way. We study several examples
of rigid colloidal particles diffusing near a no-slip boundary, and
demonstrate the importance of the choice of tracking point on the
measured translational mean square displacement (MSD). We examine
the average short-time as well as the long-time quasi-two-dimensional
diffusion coefficient of a rigid particle sedimented near a bottom
wall due to gravity. For several particle shapes we find a choice
of tracking point that makes the MSD essentially linear with time,
allowing us to estimate the long-time diffusion coefficient efficiently
using a Monte Carlo method. However, in general such a special choice
of tracking point does not exist, and numerical techniques for simulating
long trajectories, such as the ones we introduce here, are necessary
to study diffusion on long timescales.}
\end{abstract}
\maketitle
\global\long\def\V#1{\boldsymbol{#1}}
\global\long\def\M#1{\boldsymbol{#1}}
\global\long\def\Set#1{\mathbb{#1}}

\global\long\def\D#1{\Delta#1}
\global\long\def\d#1{\delta#1}

\global\long\def\norm#1{\left\Vert #1\right\Vert }
\global\long\def\abs#1{\left|#1\right|}

\global\long\def\grad{\M{\nabla}}
\global\long\def\avv#1{\langle#1\rangle}
\global\long\def\av#1{\left\langle #1\right\rangle }

\global\long\def\P{\mathcal{P}}

\global\long\def\ki{k}
\global\long\def\wi{\omega}

\global\long\def\bu{\V u}
 \global\long\def\bv{\V v}
 \global\long\def\br{\V r}

\global\long\def\sM#1{\M{\mathcal{#1}}}
\global\long\def\Mob{\sM M}
\global\long\def\J{\sM J}
\global\long\def\S{\sM S}
\global\long\def\L{\sM L}

\section{Introduction}

The Brownian motion of rigid bodies suspended in a viscous solvent
is one of the oldest subjects in nonequilibrium statistical mechanics,
and is of crucial importance in a number of applications in chemical
engineering and materials science. Examples include the dynamics of
passive \cite{BoomerangDiffusion,ColloidalClusters_Granick,SphereNearWall,ComplexShapeColloids,RigidRods_BD,AsymmetricBoomerangs}
or active \cite{ActiveSuspensions,Hematites_Science,FlippingNanorods,RotationalDiffusion_Lowen}
particles in suspension, the dynamics of biomolecules in solution
\cite{HYDROPRO,HYDROPRO_Globular,RotationalBD_Torre}, the design
of novel nano-colloidal materials \cite{BiaxialNematic_Boomerang},
and others. At the mesoscopic scales of interest, the erratic motion
of individual molecules in the solvent drives the diffusive motion
of the suspended particles. The number of degrees of freedom necessary
to simulate this motion directly using Molecular Dynamics (MD) is
large enough to make this approach prohibitively expensive. Instead,
the Brownian dynamics approach captures the effect of the solvent
through a mobility operator, and thermal fluctuations are modeled
using appropriate stochastic forcing terms. In previous work \cite{BrownianBlobs},
we used a computational fluid solver and immersed boundary techniques
to simulate the diffusive motion of spherical particles including
hydrodynamic interactions. The fluctuating immersed boundary method
developed in \cite{BrownianBlobs} is suitable for minimally-resolved
computations in which only the translational degrees of freedom are
kept and hydrodynamics is resolved at a far-field level assuming the
particles are spherical. Novel methods are, however, required to model
the behavior of particles with nontrivial shapes such as rigidly-fused
colloidal clusters \cite{ComplexShapeColloids,ColloidalClusters_Granick}
or colloidal boomerangs \cite{BoomerangDiffusion}. In this paper,
we show how to include rotational degrees of freedom in the overdamped
Langevin equations of motion for rigid bodies suspended in a viscous
fluid, develop specialized temporal integrators for these equations,
and apply them to a number of model problems.

One of the important goals of our work is to develop an overdamped
formulation and associated numerical algorithms that apply when the
hydrodynamic mobility (equivalently, resistance) depends strongly
on the configuration. Many previous works have focused on the rotational
diffusion of a single isolated rigid body in an unbounded domain.
However, in practice, rigid particles diffuse either in a suspension,
in which case they interact hydrodynamically with other particles,
or near a boundary such as a microscope slide or the walls of a slit
channel, in which case they interact hydrodynamically with the boundaries.
Here we consider a general case of a rigid body performing translational
and rotational Brownian motion in a confined system, specifically,
we numerically study particles sedimented close to a single no-slip
boundary. This is of particular relevance to recent experimental studies
of the diffusive motion of colloidal particles that are much denser
than water and thus sediment close to the microscope slide (glass
plate) \cite{ColloidalClusters_Granick,SphereNearWall,BoomerangDiffusion}.

When writing the equations of motion for a rigid body one must first
choose how to represent the orientation of the body. For bodies with
a high degree of symmetry one can use simple representations of orientation,
for example, for axisymmetric particles (e.g., rigid rods) in three
dimensions one can use two polar angles or a unit vector to represent
the orientation of the axis of symmetry \cite{BD_Rods_Shaqfeh,RigidRods_BD,SpheroidNearWall}.
More complex (biaxial or skewed) particle shapes \cite{BoomerangDiffusion,AsymmetricBoomerangs},
or asymmetrically patterned particles of symmetric shapes \cite{SphereNearWall},
as common in active particle suspensions \cite{ActiveSuspensions,FlippingNanorods},
require describing the complete orientation of the rigid bodies. Mathematically,
the orientation of a general rigid body in three dimensions is an
element of the rotation group $SO(3)$; the group of unitary $3\times3$
matrices of unit determinant (rotation matrices). This group can be
parameterized in a number of ways, the most fundamental one representing
elements of this group by an orientated rotation angle, represented
as a three-dimensional vector $\V{\phi}$, the direction of which
gives an axes of rotation relative to a reference configuration, and
the magnitude of which gives an angle of rotation around that axes.
Prior work on rotational Brownian motion in the overdamped regime
has considered the use of Euler angles \cite{RotationalDiffusion_Lowen,Overdamped_EulerAngles},
oriented rotation angles \cite{Overdamped_AngleRotation}, as well
as a number of other representations \cite{RotationalBD_Doi,RotationalBD_Torre}.
Each of these representations has its own set of problems, notably,
most of them have singularities or redundancies (which can be avoided
in principle with sufficient care), lead to complex analytical expressions
involving potentially expensive-to-evaluate trigonometric functions,
or require a large amount of storage (e.g., a rotation matrix with
$9$ elements). Furthermore, with the exception of \cite{RotationalDiffusion_Lowen,Overdamped_AngleRotation,Overdamped_EulerAngles},
most prior work on rotational diffusion either assumes that the mobility
does not depend on configuration \cite{RotationalDiffusion_McCammon},
focuses on cases where tracking a single axes is sufficient to describe
the Brownian motion \cite{BD_Rods_Shaqfeh,RigidRods_BD,SphereRotation_Wall},
or is not careful in handling the stochastic drift terms necessary
when the rotational mobility is dependent on the position and orientation
of the body.

In molecular dynamics circles \cite{LangevinRotation_Inertial,LangevinRotation_Inertial1,Event_Driven_HE},
it is well-known that a robust and efficient representation of orientation
is provided by unit quaternions, which are unit vectors in \emph{four}
dimensions (i.e., points on the unit 4-sphere). This representation
contains one redundant degree of freedom (four instead of the minimal
required of three), however, it is free of singularities and thus
numerically robust, and, as we will see, leads to a straightforward
formulation that is simple to work with both analytically and numerically.
In some sense, the quaternion representation is a direct generalization
to bi-axial bodies of the standard representation used in Brownian
Dynamics of uni-axial particles \cite{RigidRods_BD}, namely, a unit
vector in \emph{three} dimensions. That common representation is also
redundant (only two polar angles are required to describe a direction
in three dimensions), however, it offers many advantages over more
compressed representations such as polar angles, and is thus the representation
of choice. Following the submission of this manuscript, we learned
of a very recent work by Ilie \emph{et al} that also uses quaternions
in an overdamped Langevin equation for the motion of a general rigid
body in bulk \cite{RotationalBD_Quaternions}; earlier work \cite{RotationalBD_QuaternionsOld}
has also used quaternions but without carefully considering the required
stochastic drift terms.

We consider the overdamped regime, where the timescale of momentum
diffusion in the fluid is much shorter than the timescale of the motion
of the rigid bodies themselves. Formally, this regime corresponds
to the limit of infinite Schmidt number \cite{StokesEinstein}. Neglecting
inertia, we track only the positions and orientations of the immersed
bodies, deriving evolution equations for the quaternion representation.
This Langevin system exhibits the correct deterministic dynamics and
preserves the Gibbs-Boltzmann distribution in equilibrium, properly
restricted to the unit quaternion 4-sphere. Integrating these equations
proves challenging primarily due to the presence of the stochastic
drift term that arises from the configuration-dependent mobility;
this issue is identified theoretically in Appendix C in \cite{RotationalBD_Quaternions}
but that work is focused on unconfined particles for which a key stochastic
drift term vanishes (see (C21) in \cite{RotationalBD_Quaternions}).
The standard approach to handling the stochastic drift term is Fixman's
method, requiring a costly application of the inverse of the mobility
which in some cases is not directly computable. As an alternative,
we employ a recently-proposed Random Finite Difference (RFD) scheme
\cite{BrownianBlobs,MultiscaleIntegrators} for approximating the
drift; this approach only requires application of the mobility and
its ``square root'' but not the inverse of the mobility.

We perform a number of numerical experiments in which we simulate
the Brownian motion of rigid particles sedimented near a wall in the
presence of gravity, as inspired by recent experimental studies of
the diffusion of asymmetric spheres \cite{SphereNearWall}, clusters
of spheres \cite{ColloidalClusters_Granick,ComplexShapeColloids},
and boomerang colloids \cite{BoomerangDiffusion,AsymmetricBoomerangs}.
In the first example, we study a tetramer formed by rigidly connecting
four colloidal spheres placed at the vertices of a tetrahedron, modeling
colloidal clusters that have been manufactured in the lab \cite{ColloidalClusters_Granick,ComplexShapeColloids,ColloidCluster_Holography,ColloidalCluster_Nano}.
In the second example, we study the rotational and translational diffusion
of an asymmetric colloidal sphere with center of mass displaced from
the geometric center, modeling recently-manufactured ``colloidal
surfers'' \cite{Hematites_Science} in which a dense hematite cube
is embedded in a polymeric spherical particle. In the last example
we study the quasi two-dimensional diffusive motion of a dense boomerang
colloid sedimented near a no-slip boundary, as inspired by recent
experiments \cite{BoomerangDiffusion,AsymmetricBoomerangs,angleBoomerangs}.
We computationally demonstrate the crucial importance of the choice
of tracking point when computing the translational diffusion coefficient.
In particular, we show that with a suitable choice of the origin around
which torques are expressed, one can obtain an approximate but relatively
accurate formula for the effective long-time diffusion coefficient
in the directions parallel to the boundary. However, we are unable
to reach a precise and definite conclusion about the optimal choice
of tracking point even for quasi-two-dimensional diffusion, since
for all shapes studied here and in existing experiments the center
of hydrodynamic stress and the center of mobility are too close to
each other to be distinguished. In the more general case, our results
indicate that there is no exact closed-form expression for the long-time
quasi-two-dimensional coefficient, and numerical methods for simulating
trajectories are necessary in order to study the long-time diffusive
dynamics of even a single rigid body in the presence of confinement.

This paper is organized as follows. In Section \ref{sec:RigidLangevin},
we formulate the equations of motion for rigid bodies with translation
and rotation, giving a brief background on the use of quaternions
to parameterize orientation. Section \ref{sec:TemporalIntegrators}
introduces temporal integrators for these equations, including a Fixman
scheme, as well as a RFD scheme that approximates the stochastic drift
using only applications of the mobility. We perform numerical tests
of our schemes in Section \ref{sec:NumericalResultsRigid} to verify
that we can correctly simulate the dynamics of a rigid body near a
no-slip boundary, and study the influence of the choice of tracking
point on the MSD. Finally, we give concluding thoughts and discuss
future directions in Section \ref{sec:Conclusion}. Technical details
are handled in Appendices.

\section{\label{sec:RigidLangevin}Langevin equations for rigid bodies}

In this section, we formulate Langevin equations for rigid bodies
performing rotational and translational diffusion. We begin by formulating
an overdamped Langevin equation for rotational diffusion using a unit
quaternion representation of rigid-body orientation. For the remainder
of this section, we will assume that we know how to compute the configuration
dependent hydrodynamic mobilities needed for our equations. These
mobility matrices are applied to vectors of forces and torques to
compute the resulting linear and angular velocities of the immersed
rigid bodies. In future work, we will develop algorithms for computing
these objects on the fly using a computational fluid solver as in
the Fluctuating Immersed Boundary (FIB) method \cite{BrownianBlobs},
as we discuss in more detail in Section \ref{sec:Conclusion}.

Our goal is to formulate an equation for the evolution of the orientation
of a rigid body. It is important that the resulting system has the
correct deterministic term, that it is time reversible with respect
to the correct Gibbs-Boltzmann distribution in equilibrium, and that
it preserves the constraint that the quaternion has unit norm. Before
we accomplish this goal, we briefly review some required facts about
quaternions.

\subsection{\label{sub:Quaternions}Quaternions}

Describing the orientation of a rigid body in three dimensions can
be done in many ways. Rotation matrices are perhaps the most straightforward
approach to accomplish this task, but they require the use of 9 floating
point numbers to parameterize a 3 dimensional space. Additionally,
accumulation of numerical errors over many time steps can cause rotation
matrices to lose their orthonormal properties. Euler angles suffer
from gimbal lock, where at certain orientations, two Euler angles
describe rotation about the same axis, and a degree of freedom is
lost. Oriented angles are inconvenient to accumulate (in particular
one cannot simply add oriented angles to represent successive rotations)
and require the evaluation of trigonometric functions. In this work,
we choose to use normalized quaternions, which require 4 floating
point numbers to store, are easy to normalize, can be accumulated
in a convenient manner, and avoid the need for (potentially expensive
to evaluate) trigonometric functions.

A normalized quaternion can be used to represent a finite rotation
relative to a given initial reference frame, and is specified by $\V{\theta}=\{s,\V p\}\in\Set R^{4},$
a combination of a scalar $s$ and a vector $\V p\in\Set R^{3}$ that
satisfy the unit-norm constraint 
\begin{equation}
\norm{\V{\theta}}^{2}=s^{2}+\V p\cdot\V p=1.\label{eq:unit_constraint}
\end{equation}
Quaternions can be combined via the operation of quaternion multiplication,
whereby $\V{\theta}_{3}=\V{\theta}_{1}\V{\theta}_{2}$ is defined
via
\begin{align}
\left[\begin{array}{c}
s_{3}\\
\V p_{3}
\end{array}\right]= & \left[\begin{array}{c}
s_{1}s_{2}-\V p_{1}\cdot\V p_{2}\\
s_{1}\V p_{2}+s_{2}\V p_{1}+\V p_{1}\times\V p_{2}
\end{array}\right],\label{eq:quaternion-mult}
\end{align}
with $\V{\theta}_{i}=\{s_{i},\,\V p_{i}\},\, i=1,2,3$. With this
operation, normalized quaternions form a group with identity $\{1,\,\V 0\}$;
the inverse of a quaternion $\V{\theta}=\{s,\V p\}$ is given by $\V{\theta}^{-1}=\{s,-\V p\}$.

In this work, we will use normalized quaternions to represent the
orientation of a body in three space dimensions. Any finite rotation
can be defined by its oriented angle, a vector $\V{\phi}$, indicating
a turn of $\phi=\norm{\V{\phi}}$ radians counterclockwise (i.e.,
using the right-hand convention) around an axis $\hat{\V{\phi}}=\V{\phi}/\phi$.
This rotation can be associated with the quaternion 
\begin{equation}
\V{\theta}_{\V{\phi}}=\{\cos(\phi/2),\,\sin(\phi/2)\hat{\V{\phi}}\},\label{eq:quaternionAngle}
\end{equation}
i.e., $\V p$ gives the axis of the rotation and the magnitude of
$\V p$ gives the angle of rotation; the inclusion of $s$ and the
normalization constraint is thus not strictly necessary \cite{IsotropicRotationalDiffusion}
but is useful numerically. Note that $\V{\theta}$ and $-\V{\theta}$
correspond to the same physical rotation/orientation %
\footnote{Note that, in principle, the formalism developed here can directly
be applied to two dimensions by replacing quaternions with complex
numbers; a rotation of $\phi$ radians in a counterclockwise direction
is associated with the complex number $\theta_{\phi}=\exp\left(i\phi\right)=\cos\phi+i\sin\phi.$%
}.

Performing a rotation on any three dimensional vector $\V r$ in the
reference frame gives a rotated vector $\V r^{\prime}=\M R(\V{\theta})\V r$,
where the rotation matrix is 
\[
\M R(\V{\theta})=2\left[\V p\V p^{T}+s\M P+\left(s^{2}-\frac{1}{2}\right)\M I\right].
\]
Here $\M P$ is a cross-product $3\times3$ matrix such that $\M P\V r=\V p\times\V r$
for any $\V r$, i.e., $P_{ij}=\epsilon_{ikj}p_{k}$, where $\epsilon$
is the Levi-Civita symbol. Given two normalized quaternions $\V{\theta}_{1}$
and $\V{\theta}_{2}$, their rotation matrices satisfy the condition
\begin{equation}
\M R(\V{\theta}_{1})\M R(\V{\theta}_{2})=\M R(\V{\theta}_{1}\V{\theta}_{2}),\label{eq:RotationHomomorphism}
\end{equation}
that is, successive rotations can be accumulated by multiplying their
associated quaternions. More precisely, if a rotation given by oriented
angle $\V{\phi}$ followed by a rotation $\V{\psi}$ yields a total
rotation $\V{\zeta}$, then it holds that $\V{\theta}_{\V{\zeta}}=\V{\theta}_{\V{\psi}}\V{\theta}_{\V{\phi}}$.

Given an angular velocity $\V{\omega}$, we can write the corresponding
time derivative of orientation as 
\begin{equation}
\dot{\V{\theta}}=\M{\Psi}\V{\omega},\label{eq:thetaDot}
\end{equation}
where $\M{\Psi}\left(\V{\theta}\right)$ is the $4\times3$ matrix
\begin{equation}
\M{\Psi=}\frac{1}{2}\left[\begin{array}{c}
-\V p^{T}\\
s\M I-\M P
\end{array}\right].\label{eq:Psi_def}
\end{equation}
The matrix $\M{\Psi}$ has many properties that will be useful when
we formulate equations of motion for bodies with orientation. First,
it satisfies the property 
\begin{equation}
\M{\Psi}^{T}\V{\theta}=\frac{1}{2}\left(-s\V p+s\V p\right)=\V 0,\label{eq:Psi_theta}
\end{equation}
which together with the relation $\dot{\V{\theta}}=\M{\Psi}\V{\omega}$,
indicates that the deterministic evolution (\ref{eq:thetaDot}) remains
on the constraint (\ref{eq:unit_constraint}). This property is used
in Section \ref{sec:RigidLangevin} to show that the Langevin equations
presented in this work also preserve the constraint. Another useful
relationship is the fact that 
\begin{equation}
\partial_{\V{\theta}}\cdot\M{\Psi}^{T}=0\quad\mbox{i.e.}\quad\partial_{l}\left(\Psi_{lk}\right)=0,\label{eq:div_Psi}
\end{equation}
which is clear because the $j$-th row of $\M{\Psi}$ has no entries
that depend on the $j$-th component of $\V{\theta}$. Here and in
the remainder of this paper we use Einstein's repeated index summation
convention, and denote $\partial_{j}\equiv\partial/\partial\theta_{j}$.

Describing the orientation of a body at several times $t^{n}$ requires
choosing a single initial reference orientation associated with $\V{\theta}^{0}=\left\{ 1,\V 0\right\} $,
and recording the quaternion $\V{\theta}^{n}$ that describes the
rotation from the reference orientation to the orientation at instant
$t^{n}$. Furthermore, if the body undergoes a rotation with constant
angular velocity $\V{\omega}$ from time $t^{n}$ to time $t^{n+1}=t^{n}+\D t,$
we have that $\V{\theta}^{n+1}=\V{\theta}_{\V{\omega}\D t}\V{\theta}^{n}$.
This leads to a natural recipe for tracking orientation using the
Rotate procedure %
\footnote{If the accumulation of numerical errors has caused$\left|\left\Vert \V{\theta}^{n+1}\right\Vert -1\right|>\epsilon$,
for some tolerance $\epsilon$, one should renormalize the quaternion,
$\V{\theta}^{n+1}\leftarrow\V{\theta}^{n+1}/\left\Vert \V{\theta}^{n+1}\right\Vert $.%
}
\begin{equation}
\V{\theta}^{n+1}=\mbox{Rotate(}\V{\theta}^{n},\,\V{\omega}\D t)=\V{\theta}_{\V{\omega}\D t}\V{\theta}^{n}.\label{eq:Rotate_def}
\end{equation}
In constructing numerical schemes in Section \ref{sec:TemporalIntegrators},
it will be necessary to consider the second order expansion of this
rotate procedure 
\begin{align}
\mbox{Rotate(}\V{\theta},\,\V{\omega}\D t)= & \V{\theta}+\M{\Psi}\V{\omega}\D t-\frac{\left(\V{\omega}\cdot\V{\omega}\right)\D t^{2}}{8}\V{\theta}+O\left(\D t^{3}\right),\label{eq:Rotate2ndOrder}
\end{align}
as shown in Appendix (\ref{Add:Quaternions}).

\subsection{\label{sub:RigidLangevinRotationOnly}Rotational Brownian Motion}

For simplicity, we first consider a single rigid body that is free
to rotate but with a reference point $\V q$, around which torques
are measured, that is fixed in space. We let the orientation of this
body (relative to some fixed reference frame) be denoted by the quaternion
$\V{\theta}\left(t\right)$, and we suppose that the body is subjected
to a torque $\V{\tau}$ generated by a given conservative potential
$U(\V{\theta})$. It can be shown (see Appendix \ref{Add:Torques})
the the torque generated by the potential is 
\begin{align}
\V{\tau}= & -\M{\Psi}^{T}\partial_{\V{\theta}}U\label{eq:tau_dU_dq}
\end{align}
In practice, it is not necessary to formulate $U(\V{\theta})$ and
calculate $-\M{\Psi}^{T}\partial U/\partial\V{\theta}$ to obtain
the torque. Often is is much more convenient to calculate torque directly
based on the geometries of the rigid bodies and the forces applied
to them. We will see that (\ref{eq:tau_dU_dq}) will be a convenient
relation for formulating the constrained equations of motion. The
schemes that we develop will be able to simulate the motion of rigid
bodies without direct knowledge of $U(\V{\theta})$; they simply update
the positions and orientations of the bodies based on the total forces
and torques applied to each body.

\subsubsection{Overdamped Langevin Equation}

We introduce the $3\times3$ symmetric positive semidefinite (SPD)
rotational mobility matrix $\M M_{\V{\omega}\V{\tau}}\left(\V{\theta}\right)$,
which acts on torque to produce the resulting angular velocity, $\V{\omega}=\M M_{\V{\omega}\V{\tau}}\V{\tau}$.
Note that the mobility contains all the effects of hydrodynamics,
including the shape of the body, the hydrodynamic interactions with
other bodies or boundaries, far-field boundary conditions, etc. In
this section we will assume this matrix is known, and discuss ways
to obtain it explicitly in Section \ref{sec:NumericalResultsRigid}.
Using (\ref{eq:thetaDot})\textbf{ }and (\ref{eq:tau_dU_dq}), we
can write down a deterministic equation of motion for the rigid body,
\begin{align*}
\frac{d\V{\theta}}{dt}= & \M{\Psi}\M M_{\V{\omega}\V{\tau}}\V{\tau}=-\left(\M{\Psi}\M M_{\V{\omega}\V{\tau}}\M{\Psi}^{T}\right)\partial_{\V{\theta}}U=-\widetilde{\M M}\,\partial_{\V{\theta}}U,
\end{align*}
where we have defined the $4\times4$ mobility matrix $\widetilde{\M M}=\M{\Psi}\M M_{\V{\omega\tau}}\M{\Psi}^{T}$. 

It is now straight forward to formulate an Ito Langevin equation for
the rotational diffusion of the body,
\begin{align}
\frac{d\V{\theta}}{dt}= & -\widetilde{\M M}\partial_{\V{\theta}}U+\sqrt{2k_{B}T}\;\widetilde{\M M}^{\frac{1}{2}}\V{\mathcal{W}}+\left(k_{B}T\right)\partial_{\V{\theta}}\cdot\widetilde{\M M},\label{eq:overdamped_rot}
\end{align}
where $\V{\mathcal{W}}(t)$ is a collection of independent white noise
processes. Here $\widetilde{\M M}^{\frac{1}{2}}=\M{\Psi}\M M_{\V{\omega}\V{\tau}}^{\frac{1}{2}}$,
with the ``square root'' of the mobility $\M M_{\V{\omega}\V{\tau}}^{\frac{1}{2}}$
obeying the fluctuation-dissipation relation $\M M_{\V{\omega}\V{\tau}}^{\frac{1}{2}}\left(\M M_{\V{\omega}\V{\tau}}^{\frac{1}{2}}\right)^{T}=\M M_{\V{\omega}\V{\tau}}$,
for example, it could be the Cholesky factor of $\M M_{\V{\omega}\V{\tau}}$.
Note that in (\ref{eq:overdamped_rot}) the covariance for the noise
satisfies the fluctuation dissipation balance condition $\widetilde{\M M}^{\frac{1}{2}}\left(\widetilde{\M M}^{\frac{1}{2}}\right)^{T}=\widetilde{\M M}.$
The $i$-th component of the stochastic drift term $\partial_{\V{\theta}}\cdot\widetilde{\M M}$
may be written in indicial notation as $\partial_{j}\widetilde{M}_{ji}(\V{\theta})$.

Using Ito's formula, we can show that the overdamped dynamics (\ref{eq:overdamped_rot})
strictly preserves the constraint that $\V{\theta}$ have unit norm,
\[
\frac{d}{dt}\left(\V{\theta}^{T}\V{\theta}\right)=\V{\theta}^{T}\frac{d\V{\theta}}{dt}+\left(k_{B}T\right)\M I:\widetilde{\M M}=\left(k_{B}T\right)\left(\V{\theta}^{T}\left(\partial_{\V{\theta}}\cdot\widetilde{\M M}\right)+\M I:\widetilde{\M M}\right)=\left(k_{B}T\right)\partial_{\V{\theta}}\cdot\left(\V{\theta}^{T}\widetilde{\M M}\right)=0,
\]
where we used (\ref{eq:Psi_theta}) and its consequence $\V{\theta}^{T}\widetilde{\M M}=0$.
Note that the stochastic drift term in (\ref{eq:overdamped_rot})
can be rewritten as (see Appendix \ref{Add:ThermalDriftRigid}), 
\begin{equation}
\partial_{\theta}\cdot\widetilde{\M M}=\partial_{\theta}\cdot\left(\M{\Psi}\M M_{\V{\omega}\V{\tau}}\M{\Psi}^{T}\right)=\M{\Psi}\left(\partial_{\V{\theta}}\M M_{\V{\omega}\V{\tau}}\right):\M{\Psi}^{T}-\frac{1}{4}\text{Tr}\left(\M M_{\V{\omega}\V{\tau}}\right)\V{\theta},\label{eq:rotationDriftExpanded}
\end{equation}
where Tr denotes trace, and colon denotes double contraction; in index
notation $\left(\M{\Psi}\left(\partial_{\V{\theta}}\M M_{\V{\omega}\V{\tau}}\right):\M{\Psi}^{T}\right)_{i}=\Psi_{ij}\partial_{l}\left(M_{\V{\omega}\V{\tau}}\right)_{jk}\Psi_{lk}$
and $\left\{ \text{Tr}\left(\M M_{\V{\omega}\V{\tau}}\right)\V{\theta}\right\} _{i}=\left(M_{\V{\omega}\V{\tau}}\right)_{jj}\theta_{i}$
. We will see that this way of writing the drift is convenient when
we consider numerical methods for integrating (\ref{eq:overdamped_rot})
in Section \ref{sec:TemporalIntegrators}. Note that the stochastic
drift term proportional to $\text{Tr}\left(\M M_{\V{\omega}\V{\tau}}\right)\V{\theta}/4$
can be seen in Eq. (36) in \cite{RotationalBD_Quaternions} to be
related to enforcing the normalization constraint, and it will turn
out we do not need to include it explicitly just as in \cite{RotationalBD_Quaternions}.

In the special case of a free particle with unit mobility, $\M M_{\V{\omega}\V{\tau}}=\M I$,
(\ref{eq:overdamped_rot}) degenerates to the Stratonovich equation
(see (\ref{eq:overdamped_rot_strato})),
\begin{equation}
\dot{\V{\theta}}=\left(2k_{B}T\right)^{\frac{1}{2}}\M{\Psi}\circ\V{\mathcal{W}}.\label{eq:simpleRotation}
\end{equation}
Recall that the infinitesimal change in orientation is given by the
infinitesimal rotation $d\V{\phi}$ in the axes-angle representation,
where the direction of the vector $d\V{\phi}$ is the axes around
which the body is rotated by an angle $d\phi$. Also recall that the
corresponding change in the quaternion is
\[
d\V{\theta}=\M{\Psi}\left(\V{\theta}\right)d\V{\phi},
\]
at least deterministically. Since the standard rules of calculus apply
in the Stratonovich interpretation, (\ref{eq:simpleRotation}) is
equivalent to
\begin{equation}
d\V{\varphi}=\left(2k_{B}T\right)^{\frac{1}{2}}d\V{\mathcal{B}}\label{eq:dphi_isotropic}
\end{equation}
where $\V{\mathcal{B}}\left(t\right)$ is Brownian motion, formally
$\V{\mathcal{W}}\equiv d\V{\mathcal{B}}/dt$. This is a natural definition
of isotropic rotational diffusion \cite{IsotropicRotationalDiffusion}.

We can verify that (\ref{eq:overdamped_rot}) has the correct noise
covariance when $\M M_{\V{\omega}\V{\tau}}$ is not a multiple of
the identity by considering the rotational mean square displacement
at short times. Let us consider a set of orthonormal vectors $\V u_{i}(t)$
which are attached to the rigid body, and define a rotational displacement
following Kraft et al.\textbf{ }\cite{ComplexShapeColloids}, 
\begin{equation}
\D{\hat{\V u}}\left(\D t\right)\equiv\frac{1}{2}\sum_{i=1}^{3}\V u_{i}(0)\times\V u_{i}\left(\D t\right).\label{eq:orientation_increment}
\end{equation}
A straightforward calculation relates this rotational displacement
to the total angle of rotation $\V{\phi}_{\D t}$ relative to the
the initial configuration,
\begin{equation}
\D{\hat{\V u}}\left(\D t\right)=\sin\left(\phi_{\D t}\right)\hat{\V{\phi}}_{\D t}=\V{\phi}_{\D t}+O(\D t^{\frac{3}{2}}),\label{eq:du_phi}
\end{equation}
which shows that the magnitude of the rotational displacement is insensitive
to the choice of the initial triad $\V u_{i}(0)$. If the covariance
of the noise in (\ref{eq:overdamped_rot}) is correct, it should hold
that (c.f. Eqs. (1,2) in Ref. \cite{ComplexShapeColloids})
\begin{align}
\frac{1}{2k_{B}T}\,\lim_{\D t\to0}\frac{\avv{\left(\D{\hat{\V u}}\left(\D t\right)\right)\left(\D{\hat{\V u}}\left(\D t\right)\right)^{T}}}{\D t}=\frac{1}{2k_{B}T}\,\lim_{\D t\to0}\left(\frac{\V{\phi}_{\D t}\V{\phi}_{\D t}^{T}}{\D t}\right)= & \M M_{\V{\omega}\V{\tau}},\label{eq:rotational_diffusion_coeff}
\end{align}
which follows directly from (\ref{eq:overdamped_rot}). This shows
that our equation has the same physical noise covariance as the overdamped
equation in Ref. \cite{ComplexShapeColloids}, only written in a different
representation. In our numerical tests, we will use $\avv{\left(\D{\hat{\V u}}\left(\tau\right)\right)\left(\D{\hat{\V u}}\left(\tau\right)\right)^{T}}$
as a convenient definition of a rotational mean square displacement
(RMSD) at time $\tau$; note that this RMSD is necessarily bounded
and thus must reach a plateau at long times.

\subsubsection{Smoluchowski Equation}

A key property of the overdamped Langevin equation (\ref{eq:overdamped_rot})
is that it is time reversible with respect to the Gibbs-Boltzmann
equilibrium distribution 
\begin{equation}
P_{\text{eq}}\left(\V{\theta}\right)=Z^{-1}\exp\left(-U\left(\V{\theta}\right)/k_{B}T\right)\delta\left(\V{\theta}^{T}\V{\theta}-1\right),\label{eq:GB_rotation}
\end{equation}
with $Z$ a normalization constant. The overdamped equation (\ref{eq:overdamped_rot})
has the familiar structure of a generic Langevin equation (see Section
I.A in Ref. \cite{MultiscaleIntegrators}); however, a crucial difference
is that (\ref{eq:overdamped_rot}) is an SDE on a manifold, namely,
the unit 4-sphere, rather than an SDE in Eucledian space. A discussion
of overdamped Langevin equations constrained on a manifold can be
found in Ref. \cite{ConstrainedStochasticDiffusion}. As explained
there, for general curved manifolds one has to carefully construct
the stochastic drift terms in order to ensure consistency with the
desired equilibrium distribution. Note that the original (true or
physical) dynamics is unconstrained, and could, in principle, be described
using a non-redundant parameterization of the rotation group such
as Euler angles \cite{RotationalDiffusion_Lowen}; the unit norm constraint
implicit in (\ref{eq:overdamped_rot}) arises because it is mathematically
simpler to embed the unit 4-sphere in a four dimensional Euclidean
space than to parameterize it directly. The geometric matrix $\M{\Psi}\left(\V{\theta}\right)$
plays the role of the projection operator in Ref. \cite{ConstrainedStochasticDiffusion},
but unlike a projection operator, $\M{\Psi}$ is not square and projects
from the original (physical) three-dimensional space of angular velocity
to the tangent space of the unit 4-sphere.

To see that (\ref{eq:GB_rotation}) is indeed the equilibrium distribution
let us consider the case of a freely-rotating particle, $U\left(\V{\theta}\right)=0$,
which should correspond to uniform probability of all orientations.
The uniform distribution over the space of orientations of a rigid
body in three dimensions is the so-called Haar measure over the group
$SO(3)$, and has been the subject of mathematical study \cite{rummler2002distribution,miles1965random}.
It is known that in the quaternion representation this Haar measure
corresponds to a \emph{constant} probability density over the surface
of the unit 4-sphere, i.e., the Hausdorff measure on the unit 4-sphere
\cite{rummler2002distribution,prentice1978invariant,IsotropicRotationalDiffusion};
generating random uniformly-distributed orientations amounts to simply
generating a point uniformly sampled on the unit 4-sphere %
\footnote{Numerically, a uniformly-distributed unit 4-vector can be sampled
by generating a vector of 4 standard Gaussian random variables and
normalizing the result; to see this observe that the resulting distribution
must be uniform by virtue of the rotational invariance of the multivariate
Gaussian distribution.%
}. This uniform distribution over the unit quaternion sphere is captured
in (\ref{eq:GB_rotation}) by the term $\delta\left(\V{\theta}^{T}\V{\theta}-1\right)$,
and the additional prefactor $\exp\left(-U\left(\V{\theta}\right)/k_{B}T\right)$
captures the standard Gibbs-Boltzmann weighting of the configurations
based on their potential energy.

Note that more generally, for a manifold $\Sigma$ defined by the
scalar constraint $g(\V{\theta})=0,$ the Hausdorff measure $d\sigma_{\Sigma}$
on the the surface contains a metric factor relative to the Lebesque
measure $d\V{\theta}$ in unconstrained coordinates, as given by the
co-area formula \cite{ConstrainedStochasticDiffusion},
\[
d\sigma_{\Sigma}(\V{\theta})=\delta\left(g(\V{\theta})\right)\norm{\grad g(\V{\theta)}}_{2}\, d\V{\theta}.
\]
In our case, however, $g(\V{\theta})=\V{\theta}^{T}\V{\theta}-1$
and $\norm{\grad g(\V{\theta)}}_{2}=\norm{\V{\theta}}_{2}=1$ is constant
over the surface of the unit 4-sphere, and the metric factor can be
absorbed into the normalization factor $Z$. The fact that no metric
factors appear in the quaternion representation simplifies the equations;
in other representations such as Euler angles or rotation angles the
Gibbs-Boltzmann distribution is \emph{not} uniform even in the absence
of external potentials, and therefore ``metric forces'' need to
be included in the stochastic drift terms to ensure the correct equilibrium
distribution \cite{Overdamped_AngleRotation,Overdamped_EulerAngles}.
This subtle point has been missed in a number of prior works even
though the concept of metric forces is well understood for rather
general constrained Langevin equations \cite{ConstrainedBD}.

To demonstrate that (\ref{eq:GB_rotation}) is the equilibrium distribution
(invariant measure) for (\ref{eq:overdamped_rot}), we examine the
Fokker-Planck equation (FPE) for the probability density $P\left(\V{\theta},t\right)$,
\begin{align}
\partial_{t}P & =\partial_{i}\left\{ \widetilde{M}_{ij}\left[\left(\partial_{j}U\right)P+\left(k_{B}T\right)\partial_{j}P\right]\right\} .\label{eq:RotationFPE}
\end{align}
When $P$ is the Gibbs-Boltzmann distribution (\ref{eq:GB_rotation}),
we formally obtain
\begin{align*}
\left[\left(\partial_{j}U\right)P_{\mbox{eq}}+\left(k_{B}T\right)\partial_{j}P_{\mbox{eq}}\right]\sim & \exp\left(-U\left(\V{\theta}\right)/k_{B}T\right)\delta^{\prime}\left(\V{\theta}^{T}\V{\theta}-1\right)\theta_{j}.
\end{align*}
We can then use the fact that $\M{\Psi}^{T}\V{\theta}=0$ to see that
at thermodynamic equilibrium the thermodynamic driving force inside
the square brackets in (\ref{eq:RotationFPE}) vanishes, which implies
that the Gibbs-Boltzmann distribution is an equilibrium distribution;
using standard tools combined with reasonable assumptions on $U(\V{\theta})$,
it can also be shown that (\ref{eq:GB_rotation}) is the \emph{unique}
invariant measure \cite{ConstrainedStochasticDiffusion}. Note that
the calculation above is formal, but one can make a more precise argument
by considering the backward Kolmogorov equation applied to $\mathbb{E}\left[f\right]$
for an arbitrary well behaved function $f$ and expressing expectation
values as integrals over the unit 4-sphere, similar to the approach
taken in \cite{ConstrainedStochasticDiffusion}. A similar calculation
can be used to show that the generator of the Markov diffusion process
(\ref{eq:overdamped_rot}) is self-adjoint with respect to a dot product
weighted by the invariant measure (\ref{eq:GB_rotation}), which proves
that the overdamped dynamics is time reversible with respect to (\ref{eq:GB_rotation}).

We can compare the Eq. (\ref{eq:RotationFPE}) with the FPE derived
for rigid rods in Ref. \cite{RigidRods_BD}. A rigid rod can be parameterized
with a unit 3-vector $\V{\psi}$ indicating the orientation of the
rod. If we expand (\ref{eq:RotationFPE}) and use the property (\ref{eq:div_Psi}),
we can rewrite the FPE in the form
\begin{align*}
\partial_{t}P= & \partial_{i}\left\{ \Psi_{ik}\left(M_{\V{\omega}\V{\tau}}\right)_{kl}\Psi_{jl}\left[\left(\partial_{j}U\right)P+\left(k_{B}T\right)\partial_{j}P\right]\right\} \\
= & \Psi_{ik}\partial_{i}\left\{ \left(M_{\V{\omega}\V{\tau}}\right)_{kl}\left(\Psi_{jl}\left(\partial_{j}U\right)P+\left(k_{B}T\right)\Psi_{jl}\partial_{j}P\right)\right\} .
\end{align*}
This FPE has exactly the same form as the rotational part of Eq. (4.149)
in \cite{RigidRods_BD}, with the crucial difference that for rods
$\M{\Psi}$ is the cross product matrix corresponding to the direction
$\V{\psi}$. We see that (\ref{eq:RotationFPE}) is a natural generalization
of the standard Smoluckowski equation for uniaxial bodies to biaxial
bodies.

\subsection{Rotation-Translation Coupling}

In order to describe Brownian motion of freely suspended particles,
it is necessary to also include translation in our model of rigid
body motion. We first consider tracking both the location and orientation
of a single rigid body. To do this, we introduce a variable $\V q\left(t\right)$
for the Cartesian coordinates of a chosen tracking point fixed in
the body frame. We assume that we are given hydrodynamic information
in the form of a known \emph{grand mobility} matrix $\M N\left(\V q,\V{\theta}\right)$,
which is the linear mapping from given force $\V F$ and torque $\V{\tau}$
(about $\V q$) to the resulting velocity $\V u\equiv\dot{\V q}$
and angular velocity $\V{\omega}$, 
\begin{align}
\left[\begin{array}{c}
\V u\\
\V{\omega}
\end{array}\right]=\M N\left[\begin{array}{c}
\V F\\
\V{\tau}
\end{array}\right]= & \left[\begin{array}{cc}
\M M_{\V u\V F} & \M M_{\V u\V{\tau}}\\
\M M_{\V{\omega}\V F} & \M M_{\V{\omega}\V{\tau}}
\end{array}\right]\left[\begin{array}{c}
\V F\\
\V{\tau}
\end{array}\right],\label{eq:N_block}
\end{align}
where $\M M_{\V u\V{\tau}}=\M M_{\V{\omega}\V F}^{T}$ is the translation-rotation
coupling tensor, and $\M M_{\V u\V F}$ is the translation-translation
mobility familiar from Brownian Dynamics of spherical particles.

Let us suppose that the torque and force are generated from a conservative
potential $U(\V q,\V{\theta})$. Then using the fact that $\dot{\V q}=\V u$,
along with (\ref{eq:thetaDot}) and (\ref{eq:tau_dU_dq}) we can write
the overdamped Langevin equation including translation as the Ito
SDE, 
\begin{align}
\V{\upsilon}=\frac{d\V x}{dt}= & -\widetilde{\M N}\partial_{\V x}U+\sqrt{2k_{B}T}\,\widetilde{\M N}^{\frac{1}{2}}\V{\mathcal{W}}+\left(k_{B}T\right)\partial_{\V x}\cdot\widetilde{\M N}\label{eq:LangevinWithTranslation}\\
= & -\left(\M{\Xi}\M N\M{\Xi}^{T}\right)\partial_{\V x}U+\sqrt{2k_{B}T}\,\M{\Xi}\M N^{\frac{1}{2}}\V{\mathcal{W}}+\left(k_{B}T\right)\partial_{\V x}\cdot\left(\M{\Xi}\M N\M{\Xi}^{T}\right),\nonumber 
\end{align}
where $\V x=\left(\V q,\V{\theta}\right)^{T}$ and $\V{\upsilon}=\left(\V u,\dot{\V{\theta}}\right)^{T}$
are composite vectors of the translational and rotational variables
(and their velocities), and we have introduced the block matrix 
\begin{equation}
\M{\Xi}=\left[\begin{array}{cc}
\M I & \M 0\\
\M 0 & \M{\Psi}
\end{array}\right].\label{eq:Sigma_Psi}
\end{equation}
The ``square root'' of the mobility $\M N^{\frac{1}{2}}$ satisfies
the fluctuation-dissipation relation $\M N^{\frac{1}{2}}\left(\M N^{\frac{1}{2}}\right)^{T}=\M N$.
A similar computation to that mentioned in Section \ref{sub:RigidLangevinRotationOnly}
shows that (\ref{eq:LangevinWithTranslation}) is time reversible
with respect to the Gibbs-Boltzmann distribution \cite{ConstrainedStochasticDiffusion},
\begin{align}
P_{\mbox{eq}}(\V q,\V{\theta})= & Z^{-1}\exp\left(-U\left(\V q,\V{\theta}\right)/k_{B}T\right)\delta\left(\V{\theta}^{T}\V{\theta}-1\right).\label{eq:GibbsBoltzmannWithLocation}
\end{align}

\section{\label{sec:TemporalIntegrators}Temporal Integrators}

In this section we introduce temporal integrators for the overdamped
equations of motion of rigid bodies immersed in fluid, as formulated
in Section \ref{sec:RigidLangevin}. We update the quaternion representation
of orientation using the Rotate procedure (\ref{eq:Rotate_def}) introduced
in Section \ref{sub:Quaternions}, preserving the unit-norm constraint
to numerical precision. The stochastic drift term in (\ref{eq:LangevinWithTranslation})
is approximated in two ways, using a Fixman midpoint scheme and a
Random Finite Difference (RFD) scheme, see Section I.B in Ref. \cite{MultiscaleIntegrators}
for a comparison of the two approaches in the context of unconstrained
overdamped Langevin equations.

\subsection{\label{sub:EMRotation}Euler-Maruyama scheme}

For illustration purposes, we begin by considering a naive Euler-Maruyama
(EM) scheme applied to an incorrect variant of (\ref{eq:overdamped_rot}),
in which we do not carefully handle the stochastic drift term $\left(k_{B}T\right)\partial_{\V{\theta}}\cdot\widetilde{\M M}$.
In the EM scheme, we advance the configuration from time level $n$
to time level $n+1$ with the time step
\begin{align}
\V{\omega}^{n}= & -\M M_{\V{\omega}\V{\tau}}^{n}\V{\tau}^{n}+\left(\frac{2k_{B}T}{\D t}\M M_{\V{\omega}\V{\tau}}^{n}\right)^{\frac{1}{2}}\V W^{n}\label{eq:EM_rotation}\\
\V{\theta}^{n+1}= & \text{Rotate}\left(\V{\theta}^{n},\,\V{\omega}^{n}\D t\right),\nonumber 
\end{align}
where a superscript denotes the point in time at which a particular
quantity is evaluated, e.g. $\V{\theta}^{n}\approx\V{\theta}(n\D t)$
and $\M M_{\V{\omega}\V{\tau}}^{n}=\M M_{\V{\omega}\V{\tau}}\left(\V{\theta}^{n}\right)$,
and the Rotate procedure is defined by (\ref{eq:Rotate_def}). Here
$\V W^{n}$ is a collection of i.i.d. standard (i.e., mean zero and
unit variance) Gaussian variates generated using a pseudo-random number
generator. Here and henceforth, we have used (\ref{eq:tau_dU_dq})
to express the updates directly in terms of torque $\V{\tau}(\V{\theta})$.
While the scheme (\ref{eq:EM_rotation}) is not actually consistent
with (\ref{eq:overdamped_rot}), it makes a natural starting point
when discussing temporal integrators for (\ref{eq:overdamped_rot}).

Note that because we use the Rotate procedure (\ref{eq:Rotate_def}),
this update actually moves along the unit norm constraint of normalized
quaternions, as opposed to stepping off of the constraint and then
projecting back onto it \cite{ConstrainedStochasticDiffusion}. This
is a natural way to update orientation multiplicatively while still
being consistent with the additive Langevin equations formulated in
Section \ref{sec:RigidLangevin}. In the alternative approach followed
in \cite{RotationalBD_Quaternions} one has to solve a quadratic equation
(c.f. (15) in \cite{RotationalBD_Quaternions}) for a Lagrange multiplier
to enforce the normalization constraint; while this avoids the use
of trigonometric functions, it is difficult to make such methods second-order
accurate. We can expand the Rotate procedure using the Taylor series
(\ref{eq:Rotate2ndOrder}) and truncate the result at first order
in $\D t$, to obtain an expression for the leading order change in
$\V{\theta}$,
\begin{align*}
\frac{\V{\theta}^{n+1}-\V{\theta}^{n}}{\D t}\approx & \M{\Psi}^{n}\V{\omega}^{n}-\D t\frac{\left(\V{\omega}^{n}\cdot\V{\omega}^{n}\right)}{8}\V{\theta}^{n}.\\
= & \M{\Psi}^{n}\left(-\M M_{\V{\omega}\V{\tau}}^{n}\V{\tau}^{n}+\left(2k_{B}T\,\M M_{\V{\omega}\V{\tau}}^{n}\right)^{\frac{1}{2}}\V W^{n}\right)\\
 & -\left(k_{B}T\right)\frac{\left(\V W^{n}\right)^{T}\M M_{\V{\omega}\V{\tau}}^{n}\V W^{n}}{4}\V{\theta}^{n}+O\left(\D t\right).
\end{align*}
Note that the last term is equal in expectation to $-k_{B}T\left(\text{Tr}\left(\M M_{\V{\omega}\V{\tau}}^{n}\right)/4\right)\V{\theta}^{n}$,
which gives us the second term in the stochastic drift on the right
hand side of (\ref{eq:rotationDriftExpanded}). Therefore, when constructing
temporal integrators that are actually consistent with the correct
dynamics (\ref{eq:overdamped_rot}), we see that we only need to add
terms in the orientational update $\V{\theta}^{n+1}-\V{\theta}^{n}$
that will generate the remaining stochastic drift $k_{B}T\left(\M{\Psi}\left(\partial_{\V{\theta}}\M M_{\V{\omega}\V{\tau}}\right):\M{\Psi}^{T}\right)^{n}$.
Fortunately, adding this term to the orientation looks to first order
like a Rotate procedure with angular velocity $k_{B}T\left(\left(\partial_{\V{\theta}}\M M_{\V{\omega}\V{\tau}}\right):\M{\Psi}^{T}\right)^{n}$.
With this in mind, we now construct first order weakly accurate temporal
integrators for (\ref{eq:overdamped_rot}).

\subsection{\label{sub:RigidFixman}Midpoint Scheme}

The standard approach to handling the stochastic drift in overdamped
Langevin equations is to use Fixman's midpoint scheme \cite{BD_Fixman,BD_Hinch}.
Henceforth we consider the full equations (\ref{eq:LangevinWithTranslation})
including translation and rotational diffusion. To apply the Fixman
method to (\ref{eq:LangevinWithTranslation}) we rewrite (\ref{eq:LangevinWithTranslation})
in a split Stratonovich-Ito form,
\begin{equation}
\frac{d\V x}{dt}=-\left(\M{\Xi}\M N\M{\Xi}^{T}\right)\frac{\partial U}{\partial\V x}+\left(2k_{B}T\right)^{\frac{1}{2}}\M{\Xi}\M N\circ\M N^{-\frac{1}{2}}\V{\mathcal{W}},\label{eq:overdamped_rot_strato}
\end{equation}
where the terms after $\circ$ are evaluated at the beginning of the
time interval in the spirit of the Ito interpretation, while the terms
before $\circ$ are evaluated at the midpoint of the time interval,
in the spirit of the Stratonovich interpretation. Here $\M N^{-\frac{1}{2}}$
satisfies $\M N^{-\frac{1}{2}}\left(\M N^{-\frac{1}{2}}\right)^{T}=\M N^{-1}$;
the term $\M N^{-\frac{1}{2}}\V{\mathcal{W}}$ can be thought of as
a ``random force and torque'' \cite{ForceCoupling_Fluctuations}
and is equivalent in law to $\M N^{-1}\M N^{\frac{1}{2}}\V{\mathcal{W}}$.
We demonstrate that (\ref{eq:overdamped_rot_strato}) is equivalent
to (\ref{eq:overdamped_rot}) in section \ref{sub:FixmanRotation}
of the Appendix.

Note that the Fixman scheme can be seen as a direct application of
the Euler-Heun %
\footnote{The Euler-Heun method is the natural generalization of the Euler-Maruyuama
method to SDEs with Stratonovich noise \cite{EulerHeun}.%
} predictor-corrector method \cite{EulerHeun} to the split Ito-Stratonovich
form (\ref{eq:overdamped_rot_strato}). We also ensure that the scheme
is weakly second-order accurate for the linearized Langevin equations
(i.e., for additive noise) by following the predictor-corrector approach
described in detail in Ref. \cite{MultiscaleIntegrators}, giving
our midpoint predictor-corrector Fixman-like temporal integrator,
\begin{align}
\V{\upsilon}^{n}=\left(\V u^{n},\V{\omega}^{n}\right)^{T}= & \left(\M N\left[\begin{array}{c}
\V F\\
\V{\tau}
\end{array}\right]\right)^{n}+\sqrt{\frac{2k_{B}T}{\D t/2}}\left(\M N^{\frac{1}{2}}\right)^{n}\V W^{n,1}\label{eq:Fixman}\\
\V q^{p,n+\frac{1}{2}}= & \V q^{n}+\frac{\D t}{2}\V u^{n}\nonumber \\
\V{\theta}^{p,n+\frac{1}{2}}= & \mbox{Rotate}\left(\V{\theta}^{n},\,\frac{\D t}{2}\V{\omega}^{n}\right)\nonumber \\
\V{\upsilon}^{p,n+\frac{1}{2}}= & \left(\M N\left[\begin{array}{c}
\V F\\
\V{\tau}
\end{array}\right]\right)^{p,n+\frac{1}{2}}+\sqrt{\frac{k_{B}T}{\D t}}\M N^{p,n+\frac{1}{2}}\left(\M N^{-\frac{1}{2}}\right)^{n}\left(\V W^{n,1}+\V W^{n,2}\right)\nonumber \\
\V q^{n+1}= & \V q^{n}+\D t\,\V u^{p,n+\frac{1}{2}}\nonumber \\
\V{\theta}^{n+1}= & \mbox{Rotate}\left(\V{\theta}^{n},\,\D t\V{\omega}^{p,n+\frac{1}{2}}\right).\nonumber 
\end{align}
We show that this scheme produces the correct stochastic drift in
Appendix \ref{Add:ThermalDriftRigid}, more precisely, the scheme
(\ref{eq:Fixman}) is a first-order weak integrator for (\ref{eq:LangevinWithTranslation}). 

The Fixman scheme requires the application of $\M N^{-\frac{1}{2}}$,
or, equivalently, of $\M N^{-1}$; this is computationally expensive
in cases when only $\M N$ is easy to compute, and it is prohibitive
in cases when only the application of $\M N$ and $\M N^{\frac{1}{2}}$
to a vectors can be computed. In the examples we study here these
matrices will be small and thus easy to compute using direct linear
algebra, but this approach does not extend easily to suspensions of
many rigid particles. In the next section, we show how to avoid using
$\M N^{-\frac{1}{2}}$ or $\M N^{-1}$ by using a random finite difference
(RFD) approach.

It is important to observe that the Fixman scheme (\ref{eq:Fixman})
is unaffected by the change of representations of orientations. All
that needs to be changed to use other representations of orientations
is to simply change the Rotate procedure. This point has already been
intuited in prior works, where the standard Fixman method has been
used for non-spherical bodies, such as, for example, work on Brownian
dynamics for rigid rods \cite{BD_Rods_Shaqfeh}. The analytical simplicity
of the quaternion representation makes it straightforward for us to
prove first order weak accuracy for the Fixman scheme in the general
case (see Appendix \ref{Add:ThermalDriftRigid}), although the key
idea is in fact to write the dynamics in the split Ito-Strato form
(\ref{eq:overdamped_rot_strato}).

\subsection{\label{sub:RigidRFD}Random Finite Difference scheme}

To avoid the computation of $\M N^{-\frac{1}{2}}$ or $\M N^{-1}$,
we formulate a random finite difference (RFD) scheme \cite{BrownianBlobs,MultiscaleIntegrators}
by expanding the stochastic drift term into pieces (see Appendix \ref{Add:ThermalDriftRigid}),
\begin{align}
\left\{ \partial_{\V x}\cdot\left(\M{\Xi}\M N\M{\Xi}^{T}\right)\right\} _{i}= & \Xi_{im}\left(\partial_{n}N_{mp}\right)\Xi_{np}+\left[\begin{array}{c}
0\\
\left(\partial_{s}\Psi_{it}\right)\left(M_{\V{\omega}\V{\tau}}\right)_{tu}\Psi_{su}
\end{array}\right],\label{eq:translational_drift_pieces_indices}
\end{align}
where $i,n,m,$ and $p$ represent any component of $\V x$, and $s,t$,
and $u$ represent indices that range over only the orientation components,
i.e. components of $\V{\theta}$. 

An Euler-Maruyama scheme such as (\ref{eq:EM_rotation}) will, in
expectation, produce the second term on the right-hand side of (\ref{eq:translational_drift_pieces_indices}),
as we saw in Section \ref{sub:EMRotation}. The remaining term $\M{\Xi}\left(\partial_{\V x}\left(\M N\right):\M{\Xi}^{T}\right)$
can be approximated in expectation using an RFD correction \cite{BrownianBlobs,MultiscaleIntegrators}
to the velocity as follows, 
\begin{align}
\tilde{\V{\upsilon}}= & \left(\tilde{\V u}^{n},\,\tilde{\V{\omega}}^{n}\right)^{T}=\widetilde{\V W}^{n}\label{eq:RFDForTranslation}\\
\tilde{\V x}= & \left(\tilde{\V q},\,\tilde{\V{\theta}}\right)=\left(\V q^{n}+\delta\tilde{\V u}^{n},\;\mbox{Rotate}\left(\V{\theta}^{n},\,\delta\tilde{\V{\omega}}^{n}\right)\right)\nonumber \\
\V{\upsilon}^{n}=\left(\V u^{n},\V{\omega}^{n}\right)= & \left(\M N\left[\begin{array}{c}
\V F\\
\V{\tau}
\end{array}\right]\right)^{n}+\sqrt{\frac{2k_{B}T}{\D t}}\left(\M N^{\frac{1}{2}}\right)^{n}\V W^{n}+\frac{k_{B}T}{\delta}\left(\M N\left(\tilde{\V x}\right)-\M N^{n}\right)\widetilde{\V W}^{n}\nonumber \\
\V q^{n+1}= & \V q^{n}+\D t\V u^{n}\nonumber \\
\V{\theta}^{n+1}= & \mbox{Rotate}\left(\V{\theta}^{n},\D t\V{\omega}^{n}\right),\nonumber 
\end{align}
where $\widetilde{\V W}^{n}$ is a collection of i.i.d. standard normal
variates generated independently at each time step. Here $\delta$
is a small parameter that should be chosen to minimize roundoff errors
in the finite difference \cite{MultiscaleIntegrators}. Observe that
the RFD scheme only requires the application of $\M N^{n}$, $\M N^{\frac{1}{2}}$
, and $N(\tilde{\M x})$, which can be a considerable advantage over
the Fixman scheme in the case when $\M N$ is expensive to invert.
In Appendix \ref{Add:ThermalDriftRigid} we show that this RFD approach
does in fact generate the correct drift terms, and the scheme (\ref{eq:RFDForTranslation})
is a first-order weak integrator for (\ref{eq:LangevinWithTranslation}).
Observe that the RFD scheme is also invariant under changes of representation
for the orientation of the body; all that is required is an appropriate
Rotate procedure.

\subsection{\label{sub:RigidSuspensions}Suspensions of Rigid Bodies}

The temporal integration schemes presented above straightforwardly
generalize to suspensions of more than one rigid body. The overdamped
Langevin equation (\ref{eq:LangevinWithTranslation}) continues to
hold, but now $\V x$ collects the positions and orientations of all
bodies, and $\V{\upsilon}$ collects the linear and angular velocities
of all bodies, and $\V{\Xi}$ is a block-diagonal matrix with one
diagonal block (\ref{eq:Sigma_Psi}) per body. We assume here that
we can compute the grand mobility tensor $\M N$ for all of the bodies,
which maps the forces and torques applied on the bodies to the resulting
linear and angular velocities, and accounts for the hydrodynamic interactions
among the bodies. 

The deterministic and stochastic terms are handled in a straightforward
way; we accumulate deterministic velocities and angular velocities
on each body using the grand mobility tensor, and the random velocities
and angular velocities that the bodies experience are given by $\sqrt{2k_{B}T/\D t}\,\M N^{\frac{1}{2}}\V W$
where $\M N^{\frac{1}{2}}\left(\M N^{\frac{1}{2}}\right)^{T}=\M N$.
Note that the direct computation of $\M N^{\frac{1}{2}}$ can be expensive
in the multi-body setting; a generalization of the fluctuating immersed
boundary method \cite{BrownianBlobs} or the fluctuating force coupling
method \cite{ForceCoupling_Fluctuations} can, however, generate the
stochastic forcing in essentially linear time by using a fluctuating
hydrodynamic solver.

We focus here on generalizing the Fixman and RFD approximations of
the stochastic drift term. The grand mobility $\M N$ consists of
blocks $\M N_{AB}$ which take forces and torques on body $B$ and
produce the resulting velocities and angular velocities on body $A$
(which can be the same as body $B$). We consider now the stochastic
drift for a given body $A$, denoting the position and orientation
of body $A$ with $\V x_{A}=\left\{ \V q_{A},\V{\theta}_{A}\right\} $,
\begin{align}
\frac{d\V x_{A}}{dt}\mbox{ drift}= & \left(k_{B}T\right)\sum_{B}\partial_{\V x_{B}}\cdot\left(\V{\Xi}_{A}\M N_{AB}\M{\Xi}_{B}^{T}\right),\nonumber \\
= & \left(k_{B}T\right)\sum_{B}\left[\left(\partial_{\V x_{B}}\V{\Xi}_{A}\right):\left(\M N_{AB}\M{\Xi}_{B}^{T}\right)+\V{\Xi}_{A}\partial_{\V x_{B}}\left(\M N_{AB}\M{\Xi}_{B}^{T}\right)\right]\nonumber \\
= & \left(k_{B}T\right)\left(\partial_{\V x_{A}}\V{\Xi}_{A}\right):\left(\M N_{AA}\M{\Xi}_{A}^{T}\right)+\left(k_{B}T\right)\sum_{B}\V{\Xi}_{A}\left(\partial_{\V x_{B}}\M N_{AB}\right):\M{\Xi}_{B}^{T}.\label{eq:A_drift}
\end{align}
where the sums range over all bodies $B$, and we used the fact that
$\partial_{\V x_{B}}\V{\Xi}_{A}$ is nonzero only when $A=B$. The
first term term on the right hand side of (\ref{eq:A_drift}) is a
local term that does not contain any many-body effects, and can therefore
be approximated by using the Rotate procedure, as for a single body.
The second term on the right hand side of (\ref{eq:A_drift}) can
be approximated using a random finite difference or Fixman approach
in the same way as for a single body. This second term contains many-body
interactions which are captured in the computation of $\M N^{-\frac{1}{2}}$
in the Fixman approach, and in the RFD approach they are captured
by randomly displacing \emph{all} bodies together (rather than one
by one).

\section{\label{sec:NumericalResultsRigid}Diffusion Along a No-Slip Boundary}

In this section we apply the Fixman and RFD temporal integrators to
several model examples of a single rigid body immersed in a viscous
fluid. Since we want to focus on examples with configuration-dependent
mobilities, we examine rigid bodies confined to be in the vicinity
of a no-slip boundary. Specifically, we simulate the diffusive motion
of a tetramer of colloidal spheres (Section \ref{sub:Tetrahedron}),
an asymmetric sphere (Section \ref{sub:AsymmetricSphere}), and a
colloidal boomerang (Section \ref{sub:Boomerang}), in the presence
of gravity and a no-slip wall located at the plane $z=0$. For the
tetramer we validate our new methods by comparing to the FIB method
\cite{BrownianBlobs} with stiff springs used to keep the tetramer
nearly rigid. The Python codes used to produce the results reported
here are available as open source at \url{https://github.com/stochasticHydroTools/RotationalDiffusion}.

There are two main types of quantities that we examine in these simulations,
the first static and the second dynamic. The first type of quantities
are various moments of the equilibrium distribution for the position
and orientation of the rigid bodies, which we compare to moments of
the expected Gibbs-Boltzmann distribution. The second type of quantities
we study are components of the mean square displacement of the rigid
bodies, as we now explain in more detail.

\subsection{\label{sub:MSDTheory}Mean Square Displacement}

We define the total mean square displacement (MSD) at time $\tau$
as the outer product 
\begin{align}
\M D(\tau;\V x)= & \avv{\D{\V X}(\tau;\V x)\left(\D{\V X}(\tau;\V x)\right)^{T}}=\left[\begin{array}{cc}
\M D_{t} & \M D_{c}\\
\M D_{c}^{T} & \M D_{r}
\end{array}\right](\tau;\V x),\label{eq:MSDPoint}
\end{align}
where $\D X(\tau;\V x)=(\D{\V q}(\tau;\V x),\,\D{\hat{\V u}}(\tau;\V x)$),
with position increment $\D{\V q}(\tau)=\V q(\tau)-\V q(0)$ and orientation
increment $\D{\hat{\V u}(\tau)}$ as defined in (\ref{eq:orientation_increment}).
The average in (\ref{eq:MSDPoint}) is taken over trajectories started
at $\V x=\left(\V q\left(0\right),\V{\theta}\left(0\right)\right)$.
The short-time diffusion coefficient is given by the mobility in agreement
with the Stokes-Einstein (SE) relation
\begin{equation}
\frac{1}{2k_{B}T}\,\lim_{\tau\to0}\frac{\M D(\tau;\V x)}{\tau}=\M N(\V x),\label{eq:MSDMobility}
\end{equation}
where the grand mobility tensor $\M N$ is the block matrix (\ref{eq:N_block}).
Our overdamped Langevin equation (\ref{eq:LangevinWithTranslation})
is consistent with the SE relation (\ref{eq:MSDMobility}), specifically,
using (\ref{eq:du_phi}) for the rotational component $\M D_{r}$,
we obtain
\begin{equation}
\M D(\tau;\V x)=\av{\left[\begin{array}{cc}
\left(\V d_{\tau}\V d_{\tau}^{T}\right) & \left(\V d_{\tau}\V{\phi}_{\tau}^{T}\right)\\
\left(\V{\phi}_{\tau}\V d_{\tau}^{T}\right) & \left(\V{\phi}_{\tau}\V{\phi}_{\tau}^{T}\right)
\end{array}\right]}+O\left(\tau^{\frac{3}{2}}\right),\label{eq:MSD_phi}
\end{equation}
where $\V d_{\tau}=\D{\V q}(\tau;\V x)$ is the translational displacement
and $\V{\phi}_{\tau}$ is the angle of rotation over the short time
interval $\tau$. The SE formula (\ref{eq:MSDMobility}) follows directly
from (\ref{eq:MSD_phi}) the noise term in (\ref{eq:LangevinWithTranslation})
and (\ref{eq:rotational_diffusion_coeff}).

We can further define the \emph{equilibrium} MSD via the ergodic average
\begin{align}
\M D(\tau)= & \avv{\M D\left(\tau;\V x\right)},\label{eq:MSD}
\end{align}
where the average is taken over $\V x$ distributed according to the
Gibbs-Boltzmann distribution (\ref{eq:GibbsBoltzmannWithLocation}).
In practice, we estimate $\M D(\tau)$ from our simulations by taking
a time average over one long trajectory (using the ergodic property)
with the initial condition distributed according to (\ref{eq:GibbsBoltzmannWithLocation}),
as generated using an accept/reject Monte Carlo method. We estimate
error bars by running an ensemble of statistically independent trajectories. 

The Stokes-Einstein relation (\ref{eq:MSDMobility}) gives the \emph{short-time}
mean square displacement, which can be used to define a short-time
translational diffusion tensor
\begin{equation}
\M{\chi}_{st}=\frac{1}{2}\,\lim_{\tau\to0}\frac{\av{\M D_{t}(\tau;\V x)}}{\tau}=k_{B}T\av{\M M_{\V u\V F}\left(\V x\right)}.\label{eq:chi_st}
\end{equation}
In general, it is much harder to characterize the \emph{long-time}
diffusion coefficient
\begin{equation}
\M{\chi}_{lt}=\frac{1}{2}\,\lim_{\tau\to\infty}\frac{\av{\M D_{t}(\tau;\V x)}}{\tau},\label{eq:chi_lt}
\end{equation}
in the presence of confinement, even for a single body. The only simple
case is when the MSD is \emph{strictly} linear with time so that the
long and short time diffusion coefficients are equal and one can just
average the mobility over the Gibbs-Boltzmann distribution in order
to obtain $\M{\chi}_{lt}$ using (\ref{eq:chi_st}).

Observe that the long-time diffusion coefficient is independent of
the choice of point to track on the body, i.e., the choice of the
point around which torques are expressed, however, the short-time
one does depend on the choice. Our goal will therefore be to identify
a tracking point that makes the MSD as close to linear as possible,
so that the short-time diffusion coefficient provides a good estimate
of the long-time one. If this can be accomplished, then the long-time
diffusion coefficient can be estimated from the purely equilibrium
calculation (\ref{eq:chi_st}), \emph{without} requiring us to simulate
long trajectories and use (\ref{eq:chi_lt}).

\subsubsection{Choice of tracking point}

The choice of the origin around which torques are expressed, which
is the point on the body whose position we track when computing the
translational MSD, strongly affects the short-time MSD. Given two
fixed tracking points $\V q_{1}$ and $\V q_{2}$, it is straightforward
to derive the following relationship between the blocks of $\M N^{1}$
calculated using origin $\V q_{1}$, and $\M N^{2}$ calculated with
origin $\V q_{2}$ \cite{bernal1980transport},
\begin{align}
\M M_{\V{\omega}\V{\tau}}^{2}= & \M M_{\V{\omega}\V{\tau}}^{1}\label{eq:MSDChangeOfPoint}\\
\M M_{\V{\omega}\V F}^{2}= & \M M_{\V{\omega F}}^{1}+\M M_{\V{\omega\tau}}^{1}\times\V r_{12}\nonumber \\
\M M_{\V{uF}}^{2}= & \M M_{\V{uF}}^{1}-\V r_{12}\times\left(\M M_{\V{\omega\tau}}^{1}\times\V r_{12}\right)+\left(\M M_{\V{\omega F}}^{1}\right)^{T}\times\V r_{12}-\V r_{12}\times\M M_{\V{\omega F}}^{1},\nonumber 
\end{align}
where $\V r_{12}=\V q_{2}-\V q_{1}$.\textbf{ }Cross-products between
vectors and tensors are defined in Eqs. (4,5) in \cite{bernal1980transport},
with $\M A\times\V b$ corresponding to taking cross products between
rows of $\M A$ and $\V b$, in index notation,
\begin{equation}
\left(\M A\times\V b\right)_{ij}=\left(\M A_{i,:}\times\V b\right)_{j}=\epsilon_{jkl}A_{ik}b_{l},\label{eq:A_cross_b}
\end{equation}
where\textbf{ }$\M{\epsilon}$ is the Levi-Civita tensor, and similarly,
\begin{equation}
\left(\V b\times\M A\right)_{ij}=\left(\V b\times\M A_{:,j}\right)_{i}=\epsilon_{ikl}b_{k}A_{lj}.\label{eq:b_cross_A}
\end{equation}

In general, the cross-coupling (translation-rotation) mobility tensors
$\M M_{\V{\omega}\V F}=\M M_{\V u\V{\tau}}^{T}$ are not symmetric.
However, it can be shown that for any body shape, there exists a unique
point in the body called the center of diffusion or center of mobility
(CoM), such that, when that point is taken as the origin, the coupling
tensors are symmetric, $\M M_{\V{\omega}\V F}^{T}=\M M_{\V{\omega}\V F}=\M M_{\V u\V{\tau}}=\M M_{\V u\V{\tau}}^{T}$.
The location of the CoM can be found by solving for $i\ne j$ the
linear system 
\begin{equation}
\left[\epsilon_{ikl}\left(M_{\boldsymbol{\omega\boldsymbol{\tau}}}\right)_{jk}-\epsilon_{jkl}\left(M_{\boldsymbol{\omega\boldsymbol{\tau}}}\right)_{ik}\right]r_{l}^{CoM}=\left(\M M_{\V{\omega}\V F}\right)_{ij}-\left(\M M_{\V{\omega}\V F}\right)_{ji},\label{eq:CoM_system}
\end{equation}
where the mobilities are evaluated at an arbitrary origin, and $\V r^{CoM}$
goes from the origin to the CoM. It is very important to note that
in the presence of confinement the location of the CoM is \emph{not
fixed} relative to the body but changes with the position and orientation
of the body relative to the boundaries. Therefore, one should consider
the CoM computed in the absence of confinement only as an \emph{approximate}
CoM. For some bodies of sufficient symmetry, there exists a point
called the center of hydrodynamic stress (CoH) \cite{BrennerBook},
where the cross-coupling vanishes, $\M M_{\V{\omega}\V F}=\M M_{\V u\V{\tau}}=0$;
note that if a CoH exists it is also the CoM. A CoH always exists
in two dimensions, however, for general skew bodies in three dimensions
a CoH does not exist \cite{BrennerBook}. In two dimensions the CoH
is the origin for which a torque out of the plane does not induce
any translational motion in the plane, and its position can be found
from %
\footnote{For three dimensional particles diffusing parallel to the $x-y$ plane,
one can define a quasi-two-dimensional CoH as an origin for which
a torque around the $z$ axis does not induce any translational motion
in the $x-y$ plane; the set of such points is a line parameterized
by a parameter $s$, 
\[
\V r^{CoH}(s)=(-(M_{\omega_{z}F_{y}}-M_{\omega_{z}\tau_{x}}s)/M_{\omega_{z}\tau_{z}},\;(M_{\omega_{z}F_{x}}+M_{\omega_{z}\tau_{y}}s)/M_{\omega_{z}\tau_{z}},\; s).
\]
The two-dimensional result (\ref{eq:CoH_system}) is a special case
of this more general formula for $s=0$.%
}
\begin{equation}
\V r^{CoH}=(-M_{\omega F_{y}}/M_{\omega\tau},\; M_{\omega F_{x}}/M_{\omega\tau}).\label{eq:CoH_system}
\end{equation}
Experimental investigations have lead to the suggestion that for planar
particles confined to perform essentially quasi two-dimensional diffusion,
the point (\ref{eq:CoH_system}) should be tracked \cite{BoomerangDiffusion,AsymmetricBoomerangs}.

\subsubsection{Free isotropic diffusion}

For a freely-diffusing rigid body in an unbounded fluid in the absence
of any external forces and torques, all orientations are equally likely.
It is well-known that in the oriented angle representation the Haar
measure over the rotation group corresponds to $\hat{\V{\phi}}$ uniformly
distributed over the unit 3-sphere, and a probability density $P(\phi)=\left(2/\pi\right)\sin^{2}\left(\phi/2\right)$
for the angle of rotation (see (14) in \cite{miles1965random}). Combined
with (\ref{eq:du_phi}) this shows that for free isotropic rotational
diffusion the asymptotic long-time value of the rotational MSD is
finite,
\begin{align}
\lim_{\tau\rightarrow\infty}\M D_{r}(\tau)= & \lim_{\tau\rightarrow\infty}\avv{\left(\D{\hat{\V u}}\left(\tau\right)\right)\left(\D{\hat{\V u}}\left(\tau\right)\right)^{T}}=\av{\hat{\V{\phi}}\hat{\V{\phi}}^{T}}\,\frac{2}{\pi}\int_{0}^{\pi}\sin^{2}\left(\phi/2\right)\sin^{2}\left(\phi\right)d\phi\label{eq:asymptotic_rot_msd}\\
= & \frac{1}{3}\M I\cdot\frac{1}{2}=\frac{1}{6}\M I\approx0.167\,\M I,\nonumber 
\end{align}
independent of the shape of the body.

When the CoM is used as the tracking point, the translational MSD
for free isotropic diffusion is strictly linear in time,
\begin{equation}
\M D_{t}^{CoM}(\tau)=\left(2\chi\tau\right)\cdot\M I,\label{eq:MSD_CoM}
\end{equation}
where the average (short- and long-time) diffusion coefficient is
$\chi=\left(k_{B}T\right)\,\mbox{Tr}\left(\M M_{\V u\V F}\right)/d$,
with $d$ being the dimensionality and the orientation of the body
used to evaluate $\M M_{\V u\V F}$ being arbitrary. See discussion
around Eq. (46) in \cite{RotationalBD_Doi} for more details, and
note that the mathematical reason behind this result is the fact that
for a symmetric cross-coupling mobility $\M M_{\V u\V{\tau}}$, the
translational component of the thermal drift term $\left(k_{B}T\right)\partial_{\V{\theta}}\cdot\left(\M M_{\V u\V{\tau}}\M{\Psi}^{T}\right)$
vanishes identically for a freely-diffusing body, see (31,32) in \cite{RotationalBD_Doi}.

\subsubsection{Confined Diffusion}

Our primary focus in this work is diffusion of rigid bodies in the
presence of confinement and gravity, specifically, we consider diffusion
in the vicinity of a no-slip boundary \cite{StokesianDynamics_Wall}.
In a number of experiments, the Brownian particles being tracked are
substantially denser than the solvent and thus sediment close to the
bottom microscope slide due to gravity \cite{ColloidalClusters_Granick,SphereNearWall},
or, the particles are confined in a narrow slit channel \cite{BoomerangDiffusion,AsymmetricBoomerangs}.
In both cases the boundaries strongly modify the hydrodynamic response.
Notably, the CoM will depend on the position of the body relative
to the boundary, and for non-skew particles, there may not be a CoH
in the presence of a boundary even if there is one in an unbounded
domain. Note that in the presence of gravity the typical height of
a rigid body above a plane wall is on the order of the gravitational
height $h_{g}\sim k_{B}T/\left(m_{e}g\right)$, where $m_{e}$ is
the excess mass of the particle relative to the solvent, and $g$
is the gravitational acceleration. The value of the gravitational
height varies widely in experiments from tens of nanometers to tens
of micrometers, depending on the size and density of the colloidal
particles. 

In the numerical studies that follow we examine the MSD of isolated
rigid particles sedimented near a wall in the presence of gravity.
We orient our coordinate system so that the $x$ and $y$ axes are
parallel to the wall and the $z$ axis is perpendicular to the wall
located at $z=0$. In experiments based on confocal or optical microscopy,
only the motion of the particle parallel to the wall can be observed
and measured, in particular, what is measured is the parallel mean
square displacement
\begin{align*}
D_{\parallel}(\tau)= & D_{xx}(\tau)+D_{yy}(\tau).
\end{align*}
In our simulations, we apply no forces in the $x$ and $y$ directions,
and at large times we expect that $D_{\parallel}(\tau)$ will grow
linearly with slope proportional to the long-time quasi-two-dimensional
diffusion coefficient $\chi_{2D}$ which can be measured from simulations
or experiments,
\[
D_{\parallel}(\tau)\sim4\chi_{2D}\tau\quad\mbox{ at long times}.
\]

In general, we do not expect $D_{\parallel}(\tau)$ to be strictly
linear in time. However, if it is, then the long-time diffusion coefficient
is the same as the short-time diffusion coefficient, and can be obtained
by averaging the parallel translational mobility over the equilibrium
Gibbs-Boltzmann distribution (\ref{eq:GibbsBoltzmannWithLocation}),
\begin{equation}
\chi_{2D}=k_{B}T\av{M_{\parallel}}=k_{B}T\av{M_{F_{x}F_{x}}}=k_{B}T\av{M_{F_{y}F_{y}}}.\label{eq:chi2D_theory}
\end{equation}
If a CoH exists, and is \emph{independent} of the configuration, then
translational and rotational motion will decouple and (\ref{eq:chi2D_theory})
will be exact. In general, however, a CoH does not exist in three
dimensions even in the absence of confinement. However, as argued
in Refs. \cite{BoomerangDiffusion,AsymmetricBoomerangs}, if the diffusion
is strongly confined to be effectively two-dimensional, either because
of strong gravity or because of the presence of two tightly-spaced
confining walls, an approximate CoH should exist and therefore the
parallel MSD will be approximately linear in time. We will examine
these claims numerically in Section \ref{sub:Boomerang}.

We also investigate the perpendicular mean square displacement for
the height above the plane wall,
\begin{align*}
D_{\perp}(\tau)= & D_{zz}(\tau),
\end{align*}
which we expect to reach a finite asymptotic value at large times
due to the presence of gravity,
\begin{align}
\lim_{\tau\to\infty}D_{\perp}(\tau)= & \avv{\left(z_{1}-z_{2}\right)^{2}}_{\left(z_{1},z_{2}\right)}.\label{eq:D_perp_asympt}
\end{align}
where $z_{1}$ and $z_{2}$ are the heights of the tracking point
for two configurations sampled uniformly and randomly from the Gibbs-Boltzmann
distribution (\ref{eq:GibbsBoltzmannWithLocation}). A good generalization
of the concept of a gravitational height for nonspherical particles
is
\begin{equation}
h_{g}=\left(\frac{1}{2}\lim_{\tau\to\infty}D_{\perp}(\tau)\right)^{\frac{1}{2}}\sim f_{g}\left(\frac{k_{B}T}{m_{e}g}\right),\label{eq:h_g_def}
\end{equation}
where $f_{g}$ is a geometric factor that is hard to compute analytically
for a general body but can be computed using a Monte Carlo algorithm
from (\ref{eq:D_perp_asympt}); the factor of $1/2$ is chosen so
that $h_{g}=k_{B}T/\left(m_{e}g\right)$ for a point particle. Note,
however, that $f_{g}$ depends on the choice of the tracking point,
and should therefore be associated with a particular fixed point on
the body. Also note that $h_{g}$ measures the \emph{relative} displacement
of the particle in the vertical direction rather than the distance
to the plane; it may therefore be more appropriate to think of it
as gravitational \emph{thickness} rather than \emph{height}.

For rotation, we examine the diagonal components of the RMSD $\M D_{r}(\tau)$,
which must reach a finite asymptotic limit at large times since the
rotational displacements are bounded,
\begin{equation}
\lim_{\tau\to\infty}\M D_{r}(\tau)=\avv{\left(\D{\hat{\V u}}\left(\V{\theta}_{1},\V{\theta}_{2}\right)\right)\left(\D{\hat{\V u}}\left(\V{\theta}_{1},\V{\theta}_{2}\right)\right)^{T}}_{\left(\V{\theta}_{1},\V{\theta}_{2}\right)},\label{eq:D_rot_asympt}
\end{equation}
where $\D{\hat{\V u}}\left(\V{\theta}_{1},\V{\theta}_{2}\right)$
is the rotational displacement (\ref{eq:orientation_increment}) between
two random orientations $\V{\theta}_{1}$ and $\V{\theta}_{2}$ sampled
uniformly from the Gibbs-Boltzmann distribution (\ref{eq:GibbsBoltzmannWithLocation}).

In order to compute averages over the Gibbs-Boltzmann distribution,
which includes the effects of gravity and steric repulsion from the
wall, we use a Monte Carlo method to generate random samples distributed
according to (\ref{eq:GibbsBoltzmannWithLocation}). The simplest
way to do this is an accept-reject method in which we first generate
a random position $\V q$ with uniform height and a uniform random
orientation $\V{\theta}$ of the body, and accept the random configuration
$\V x$ with probability $\exp\left(-U(\V x)/k_{B}T\right)$. Note
that in principle the Gibbs-Boltzmann distribution is unbounded in
the $z>0$ direction, and cannot be captured exactly by such an accept-reject
method with height distributed in a finite interval. However, since
the probability decays monotonically in the tail as $\sim\exp\left(-m_{e}gh/k_{B}T\right)$
where $h$ is the height of the tracking point of the body, we can
adjust the upper bound of the uniformly distributed height empirically
to ensure that we are only neglecting an insignificant portion of
the distribution. One can avoid this bias by using an exponential
distribution as a proposal density in the accept/reject process, or
by using Markov Chain Monte Carlo to generate samples from the Gibbs-Boltzmann
distribution; we have found this to produce indistinguishable results
for our purposes, while being significantly slower. We estimate asymptotic
values of the MSD from (\ref{eq:D_perp_asympt}) and (\ref{eq:D_perp_asympt})
by using the Monte Carlo method to generate a large number of pairs
of samples from the equilibrium distribution, calculating the mean
square displacement between each pair, and averaging this value over
all of the pairs. 

When using discrete time steps with stochastic forcing, it is possible
for unphysical configurations to occur; this leads to a finite-time
breakdown of explicit integrators \cite{AdaptiveEM_SDEs,MetropolizedBD}
such as our Fixman and RFD schemes. Specifically, in our numerical
tests, it is possible for the stochastic terms to ``kick'' some
part of the body through the wall; this invalidates the hydrodynamic
calculations used to compute the mobility, or makes the mobility not
positive-semidefinite. To handle this possibility in our simulations,
after each configurational update (including the predictor step to
the midpoint in the Fixman scheme), we check whether any part of the
body overlaps the wall, and if the new configuration is not valid,
we start again at state $\V x^{n}$ and repeat the time step, drawing
new random numbers. This procedure is repeated until a valid new state
is found (note that it is possible for multiple rejections to occur
in one time interval). Because this rejection of invalid states changes
the dynamics, and therefore the statistics of the system, we ensure
that the number of rejections is very low compared to the total number
of steps taken. In the results presented in this section, the rejection
rate (number of rejections divided by number of attempted steps) is
never greater than $5\times10^{-5}$; in most cases it is zero.

\subsection{Rigid multiblob models for hydrodynamic calculations}

\begin{figure}
\begin{centering}
\includegraphics[height=4cm]{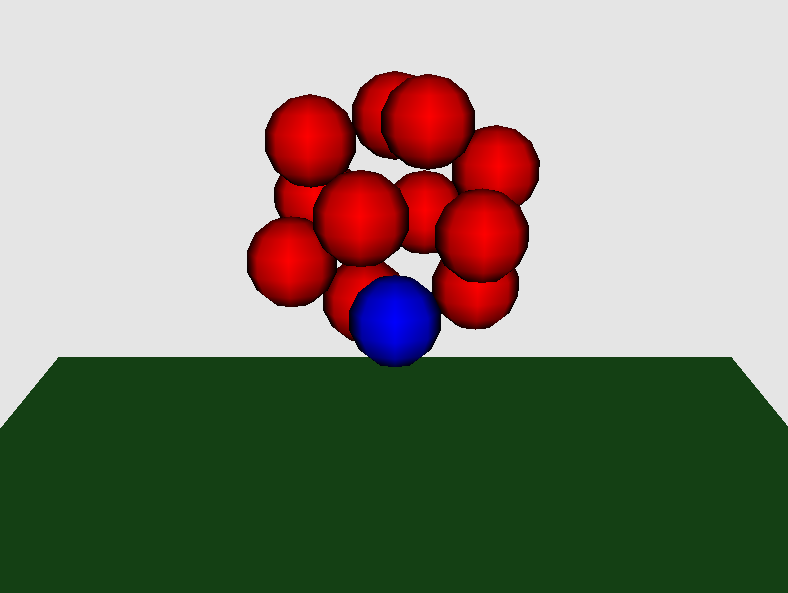}\hspace{0.25cm}\includegraphics[height=4cm]{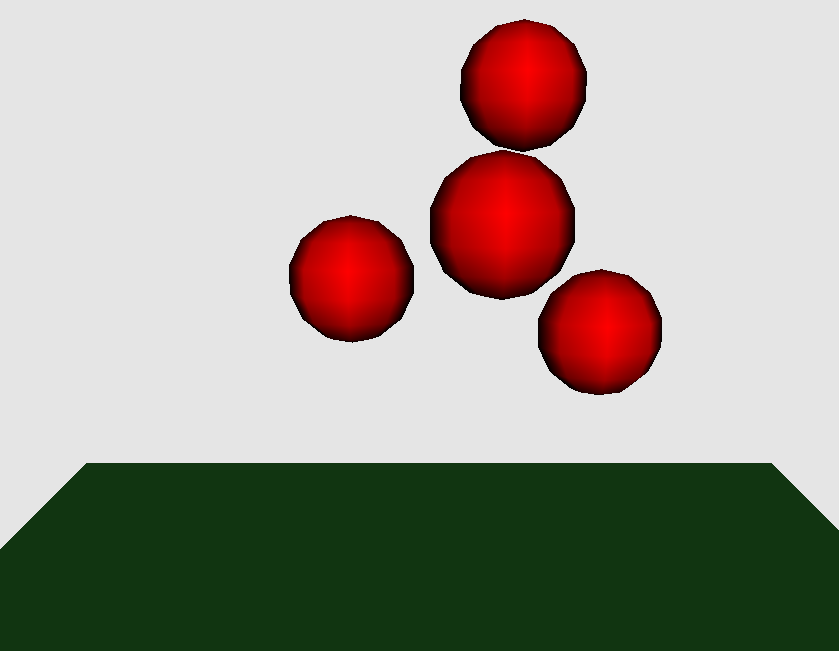}\hspace{0.25cm}\includegraphics[height=4cm]{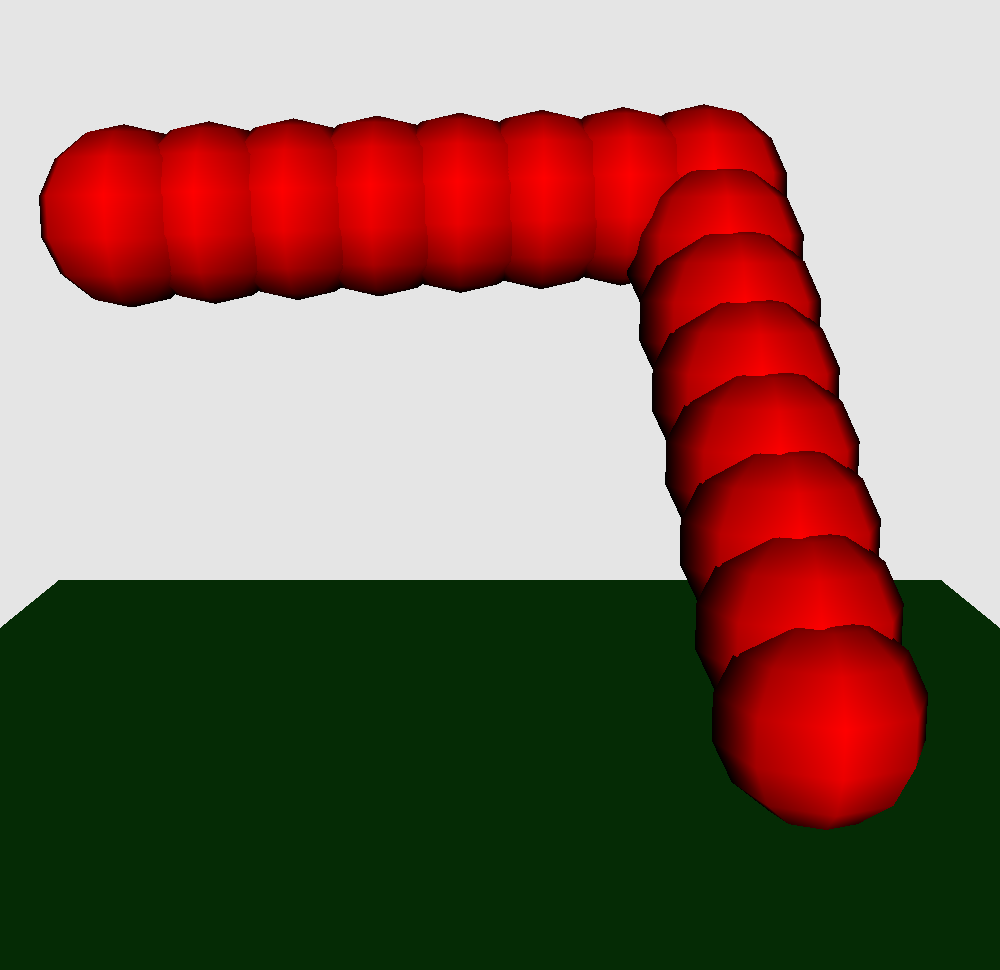}
\par\end{centering}

\caption{\label{fig:BlobModels}Rigid multiblob models of three types of particles
studied in this work. The blobs are shown as red spheres of radius
equal to the blob hydrodynamic radius, and the no-slip bottom wall
is shown as a green plane. (\emph{Left}) A spherical colloidal ``surfer''
that has a much denser metallic cube of hematite embedded in it, taken
from the work of Palacci et al. \cite{Hematites_Science}. In our
computer simulations we model this as an icosahedron of rigidly-connected
blobs, one of which (indicated by a blue sphere) models the dense
hematite and holds all of the mass of the particle. (\emph{Middle})
A tetramer formed by connecting four colloidal particles using DNA
bonding into a tetrahedron, as in the work of Kraft et al. \cite{ComplexShapeColloids}.
The multiblob model has four blobs rigidly placed at the vertices
of a tetrahedron. (\emph{Right}) \label{fig:BlobBoomerang}A right-angle
boomerang colloid manufactured using lithography and studied in a
slit channel formed by two microscope slides by Chakrabarty et al.
\cite{BoomerangDiffusion}, modeled here using a 15-blob approximation.}
\end{figure}

\begin{figure*}
\begin{centering}
\includegraphics[width=0.32\textwidth]{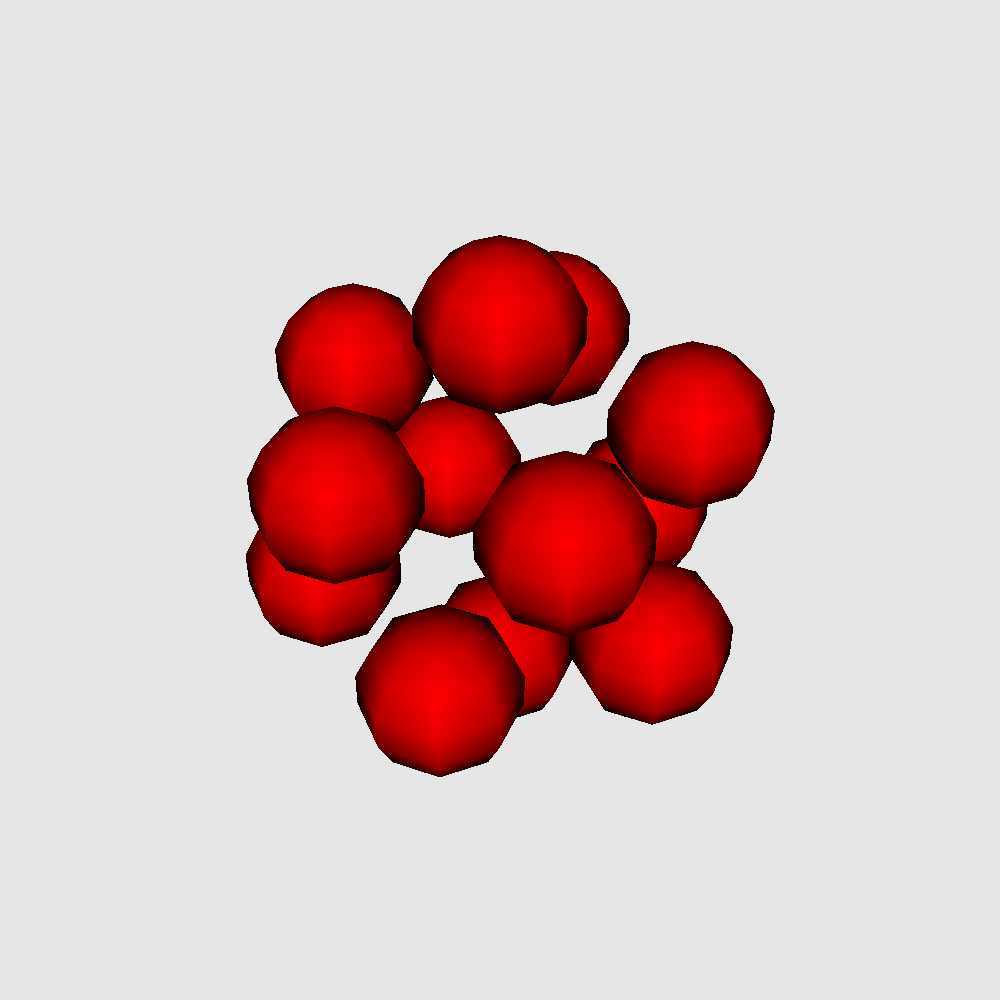}\hspace{0.05cm}\includegraphics[width=0.32\textwidth]{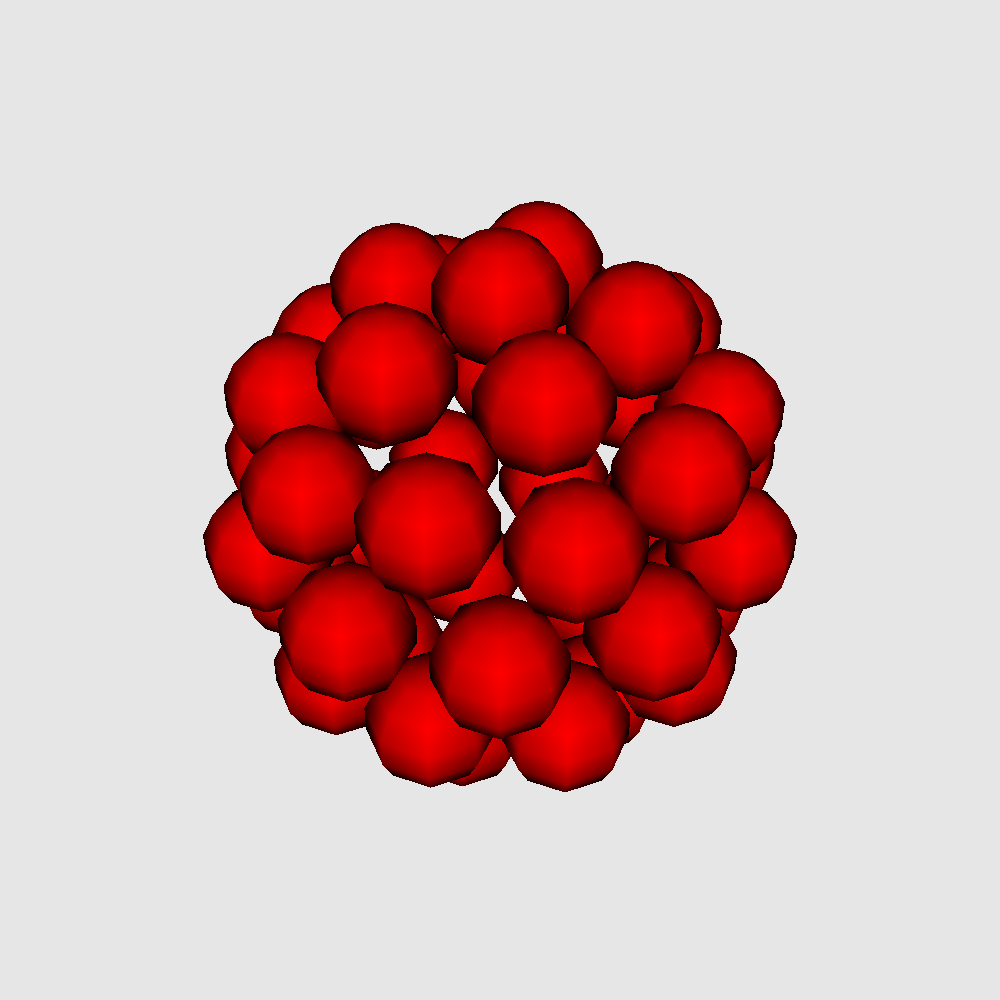}\hspace{0.05cm}\includegraphics[width=0.32\textwidth]{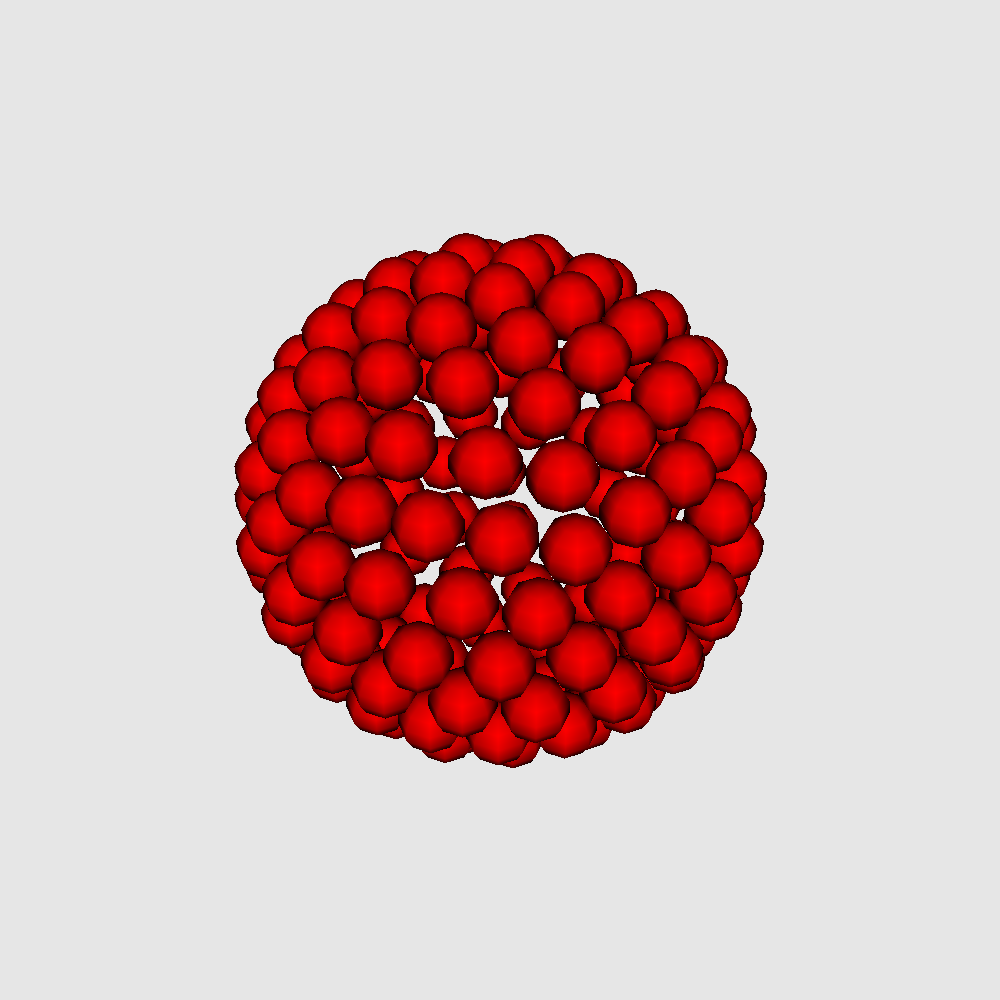}
\par\end{centering}

\caption{\label{fig:BlobSpheres}Rigid multiblob or ``raspberry'' models
of a rigid sphere, containing 12 blobs (left), 42 blobs (center),
or 162 blobs (right) placed on the surface of a sphere using a recursive
triangulation procedure starting from an icosahedron (left-most panel).
The radius of the red spheres is equal to the effective hydrodynamic
radius of a blob; the hydrodynamic radius of the resulting rigid multiblob
sphere is computed numerically \cite{MultiblobSprings,RigidIBM} and
is generally larger than the geometric radius of the sphere used for
the recursive triangulation.}
\end{figure*}

For the purposes of hydrodynamic calculations, we discretize rigid
bodies by constructing them out of multiple rigidly-connected spherical
``blobs'' of hydrodynamic radius $a$. These blobs can be thought
of as hydrodynamically minimally-resolved spheres forming a rigid
conglomerate that approximates the hydrodynamics of the actual rigid
object being studied. Examples of such ``multiblob'' \cite{MultiblobSprings}
models of several types of rigid bodies studied in recent experiments
are given in Fig. \ref{fig:BlobModels}. As Fig. \ref{fig:BlobSpheres}
illustrates for a rigid sphere, the hydrodynamic fidelity of \emph{rigid
multiblob} \cite{MultiblobSprings} models can be refined by increasing
the number of blobs (and decreasing their hydrodynamic radius $a$
accordingly); of course, increasing the resolution comes at a significant
increase in the computational cost of the method. Similar ``bead''
or ``raspberry'' models appear in a number of studies of hydrodynamics
of particle suspensions \cite{HYDROPRO,HYDROPRO_Globular,RotationalBD_Torre,Raspberry_MPCD,Raspberry_LBM,SPM_Rigid,StokesianDynamics_Rigid,HYDROLIB,SphereConglomerate,RigidBody_SD,IBM_Sphere,MultiblobSprings},
with the blobs or beads being either connected rigidly as we do here,
or connected via stiff springs; in some models the fluid or particle
inertia is included also. Since in this work we focus on the long-time
diffusive dynamics it is crucial to use rigid rather than stiff springs,
and to eliminate inertia in the spirit of the overdamped approximation,
in order to allow for a sufficiently large time step to reach physical
time scales of interest (seconds to minutes in actual experiments).

After discretizing a rigid body using $n$ blobs, we write down a
system of equations that constrain the blobs to move rigidly. These
intuitive equations are written in a large number of prior works \cite{HYDROPRO,HYDROPRO_Globular,RotationalBD_Torre,StokesianDynamics_Rigid,HYDROLIB,SphereConglomerate,RigidBody_SD,RegularizedStokeslets}
but we refer to \cite{StokesianDynamics_Rigid} for a clear yet detailed
exposition; the authors also provide associated computer codes (not
used in this work) in the supplementary material. Letting $\V{\lambda}=\left\{ \V{\lambda}_{1},\ldots,\V{\lambda}_{n}\right\} $
be a vector of forces (Lagrange multipliers) that act on each blob
to enforce the rigidity of the body, we have the linear system for
$\V{\lambda}$, $\V u$ and $\V{\omega}$,
\begin{align}
\sum_{j}\left(\M M_{B}\right)_{ij}\V{\lambda}_{j}= & \V u+\V{\omega}\times(\V r_{i}-\V q),\quad\forall i\label{eq:rigidSystem}\\
\sum_{i}\V{\lambda}_{i}= & \V F\nonumber \\
\sum_{i}(\V r_{i}-\V q)\times\V{\lambda}_{i}= & \V{\tau},\nonumber 
\end{align}
where $\V u$ is the velocity of the tracking point $\V q$, $\V{\omega}$
is the angular velocity of the body around $\V q$, $\V F$ is the
total force applied on the body, $\V{\tau}$ is the total torque applied
to the body about point $\V q$, and $\V r_{i}$ is the position of
blob $i$. 

Here the blob-blob translational mobility $\M M_{B}$ describes the
hydrodynamic relations between the blobs, accounting for the influence
of the boundaries. The $d\times d$ block of the translational mobility
$\left(\M M_{B}\right)_{ij}$ computes the velocity of blob $i$ given
forces on blob $j$, neglecting the presence of the other blobs in
a \emph{pairwise} approximation. In the presence of a single wall,
an analytic approximation to $\left(\M M_{B}\right)_{ij}$ is given
by Swan and Brady \cite{StokesianDynamics_Wall}, as a generalization
of the Rotne-Prager (RP) tensor \cite{RotnePrager} to account for
the no-slip boundary using Blake's image construction. In this work
we utilize the translation-translation part of the Rotne-Prager-Blake
mobility given by Eqs. (B1) and (C2) in \cite{StokesianDynamics_Wall}
to compute $\M M_{B}$, ignoring the higher order torque and stresslet
terms in the spirit of the minimally-resolved blob model \cite{BrownianBlobs}.
The self-mobilities for a single blob are given in (\ref{eq:SingleWallBlob})
in Appendix \ref{add:SphereMob}. Note that in a suitable limit of
infinitely many blobs of appropriate radius, solving (\ref{eq:rigidSystem})
computes the exact grand mobility for the rigid body (or a collection
of bodies), even though only a low-order RP-like approximation is
used for $\M M_{B}$ \cite{RigidIBM,RegularizedStokeslets}. To see
this, note that blob methods can be considered as a discretization
of a regularized first-kind integral equation \cite{RegularizedStokeslets}
for the Stokes flow around the rigid bodies. In a recent experimental
and computational study \cite{TetheredNanowires}, the mobility of
a rigid rod tethered to a hard wall was computed using the multiblob
approach as we do here, however, in that work the wall was created
from (many) blobs instead of using the known generalization of the
RP tensor to a single-wall geometry \cite{StokesianDynamics_Wall}
as we do here.

The solution of the linear system (\ref{eq:rigidSystem}) defines
a linear mapping from applied force and torque to body velocity and
angular velocity, and thus gives the grand mobility $\M N$ (for explicit
formulas, see \cite{StokesianDynamics_Rigid}). Observe that generalizing
the system (\ref{eq:rigidSystem}) to a collection of rigid bodies
is trivial. In the examples considered in the work, the number of
blobs is small and the system (\ref{eq:rigidSystem}) can easily be
solved by computing the Schur complement \cite{StokesianDynamics_Rigid}
and inverting it directly with dense linear algebra. The use of dense
linear algebra allows us to focus our attention on the temporal integrators
for the overdamped dynamics and not on linear algebra or hydrodynamics
issues. In principle, our temporal integrators can be used with a
variety of methods for computing the hydrodynamic mobility of suspensions
of rigid bodies, for example, boundary-integral or boundary-element
methods can be used to compute the (action of the) grand mobility
with higher accuracy.

We compute the square root $\M N^{\frac{1}{2}}$ by performing a dense
Cholesky factorization on $\M N$. It is important to note that if
$\M M_{B}$ is SPD, the grand mobility $\M N$ computed by solving
(\ref{eq:rigidSystem}) is also SPD. Note that the Swan-Brady approximation
to $\M M_{B}$ \cite{StokesianDynamics_Wall} used here is based on
the Rotne-Prager tensor and is thus only guaranteed to be positive
definite when the blobs do not overlap each other or the wall, i.e.,
when no two blobs are closer than a distance $2a$ and the distance
of all blob centroids to the wall is greater than $a$. It is possible
to generalize the Rotne-Prager-Yamakawa tensor to confined systems
\cite{RPY_Shear_Wall}, thus guaranteeing an SPD $\M M_{B}$ even
when blobs overlap each other or the wall (but of course their centroids
must remain above the wall), but we know of no published explicit
formula that accomplishes this even for the case of a single wall.
Fortunately, for our model of boomerang-shaped particles we numerically
observe an SPD mobility when the blobs do not overlap the wall even
though blobs overlap each other.

\subsection{\label{sub:Tetrahedron}Colloidal tetramer: Tetrahedron}

In this section we study a tetramer formed by rigidly connecting four
colloidal spheres at the vertices of a tetrahedron \cite{ComplexShapeColloids,ColloidalClusters_Granick},
diffusing near a single no-slip boundary. The tetrahedron is discretized
in a minimally-resolved way using 4 blobs, one at each vertex of a
regular tetrahedron, as illustrated in the pictured in the left panel
of Fig. \ref{fig:BlobModels}. In some arbitrary units, each blob
is a distance $d=2$ away from all of the others and has a hydrodynamic
radius of $a=0.5$; this somewhat arbitrary choice makes the tetrahedron
hydrodynamically sufficiently different from a sphere to require resolving
the orientation of the body as well.

To avoid symmetries and make the test more general, we assume each
of the four spheres to have a different density; the gravitational
forces on the vertices in the negative $z$ direction are set to $F_{1}=0.15$,
$F_{2}=0.1$, $F_{3}=0.3$ and $F_{4}=0.05$ in units of $k_{B}T/a$.
To prevent the tetrahedron from passing through the wall, we include
a repulsive potential (\ref{eq:Yukawa_blob}) between each of the
blobs and the wall, based on an ad-hoc combination of a Yukawa and
a hard-sphere-like divergent potential,
\begin{align}
U_{\mbox{wall}}(h;\, a)= & \frac{\epsilon a}{h-a}\,\exp\left(-\frac{h-a}{b}\right),\label{eq:Yukawa_blob}
\end{align}
where $h$ is the height of the center of the blob above the wall,
$\epsilon=20k_{B}T$ is the repulsion strength, and $b=0.5a$ to be
the Debye length (these values are selected somewhat arbitrarily).
The total force and torque on the rigid tetramer is the sum of the
forces and torques on the individual blobs. The above choice of parameters
gives the center of the tetrahedron a gravitational height (\ref{eq:h_g_def})
of $h_{g}\approx1.75a$.

For comparison and validation, we also construct an approximation
to the freely-moving rigid tetrahedron using four blobs connected
by stiff springs, and then employ the FIB method \cite{BrownianBlobs}
to simulate the diffusive motion of the almost-rigid tetramer; the
same parameters are used in both simulations.\textbf{ }The FIB simulation
was performed in a domain of $64\times64\times64$ grid cells of width
$\D x=0.796a$ using the 4-point Peskin kernel, which ensures that
the effective hydrodynamic radius of the blobs is $a$ \cite{BrownianBlobs}.
The boundaries on the top and bottom of the domain are both no-slip
walls, which differs from the domain for the rigid body simulations,
but since the center of the tetrahedron almost never goes past $20\%$
of the channel width (see figure \ref{fig:TetrahedronEqDist}), the
effect of the top wall is relatively minor. The spring stiffness was
set to $k=200\, k_{B}T/a$ to keep the deformations of the tetrahedron
small; this imposes a stringent limit on the time step size $\D t$.

\subsubsection{Equilibrium Distribution}

\begin{figure}
\centering{}\includegraphics[width=0.75\textwidth]{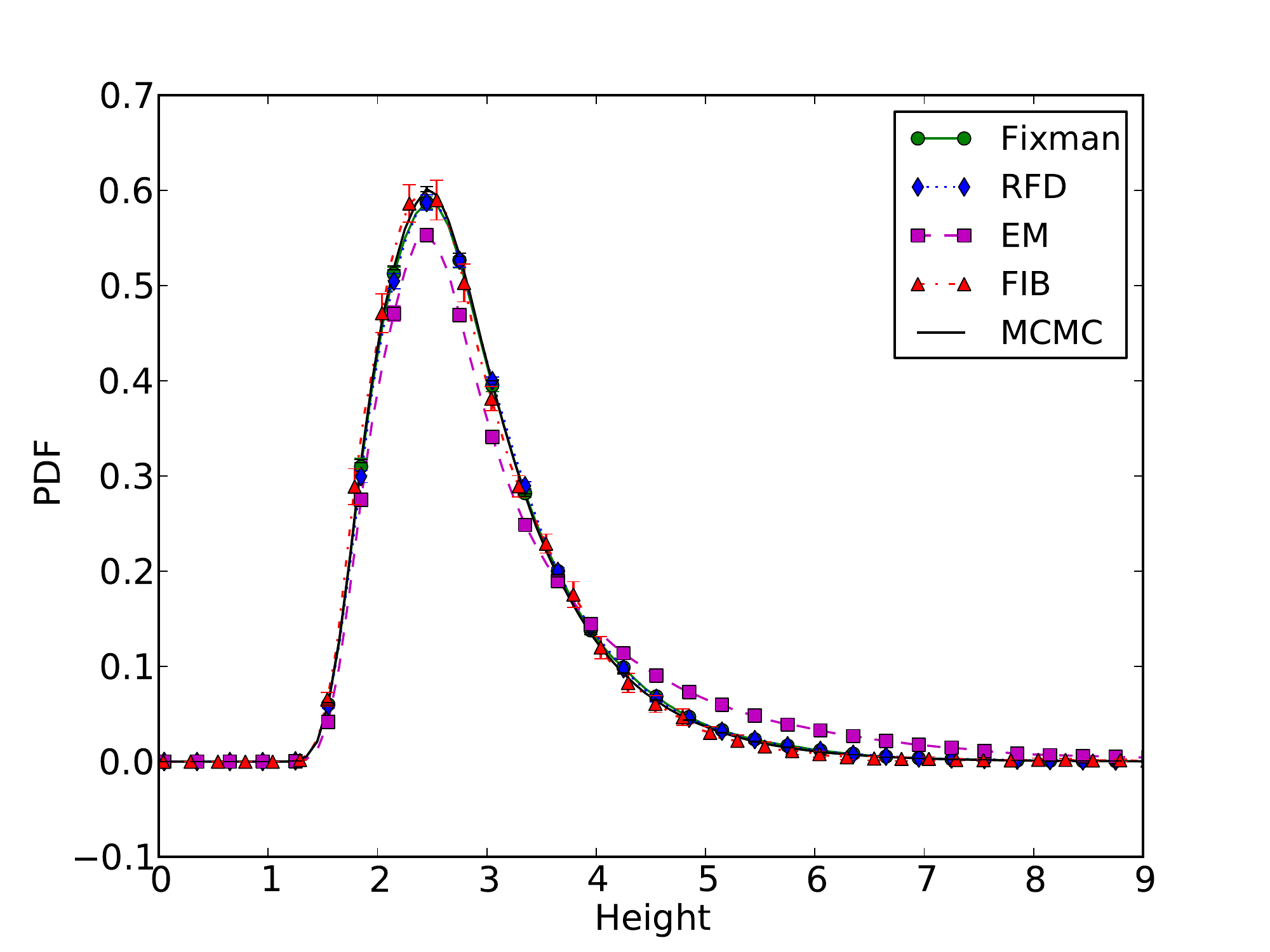}\caption{\label{fig:TetrahedronEqDist}Equilibrium distribution for the height
of the geometric center of the tetramer colloid pictured in the left
panel of Fig. \ref{fig:BlobModels}. The results obtained by using
the FIB method and the Fixman (Section \ref{sub:RigidFixman}) and
RFD (Section \ref{sub:RigidRFD}) integratrors agree with the Gibbs-Boltzmann
distribution generated using Monte Carlo sampling. The results obtained
by using the inconsistent EM scheme (\ref{eq:EM_rotation}) demonstrate
that neglecting the stochastic drift term yields an incorrect equilibrium
distribution. This plot is based on 16 runs of $3\cdot10^{5}$ time
steps with a small time step size of $\D t\approx0.0653\,\left(6\pi\eta a^{3}/k_{B}T\right)$;
no rejections were needed for this small time step in any of the runs.}
\end{figure}

In this section we examine the equilibrium distribution for the colloidal
tetramer. We use a Monte Carlo method to generate the marginal Gibbs-Boltzmann
distribution for the height of the geometric center of the tetrahedron,
and compare to our numerical results. We see in Fig. \ref{fig:TetrahedronEqDist}
that the Fixman (\ref{eq:Fixman}) and RFD schemes (\ref{eq:RFDForTranslation})
are in good agreement with the Gibbs-Boltzmann distribution. The Euler-Maruyama
scheme (\ref{eq:EM_rotation}), with the obvious additions to include
translation, however, neglects parts of the stochastic drift and generates
an equilibrium distribution which has clear errors that do not vanish
as the time step size is refined (not shown).

\subsubsection{Mean Square Displacement}

\begin{figure}
\centering{}\includegraphics[width=0.49\textwidth]{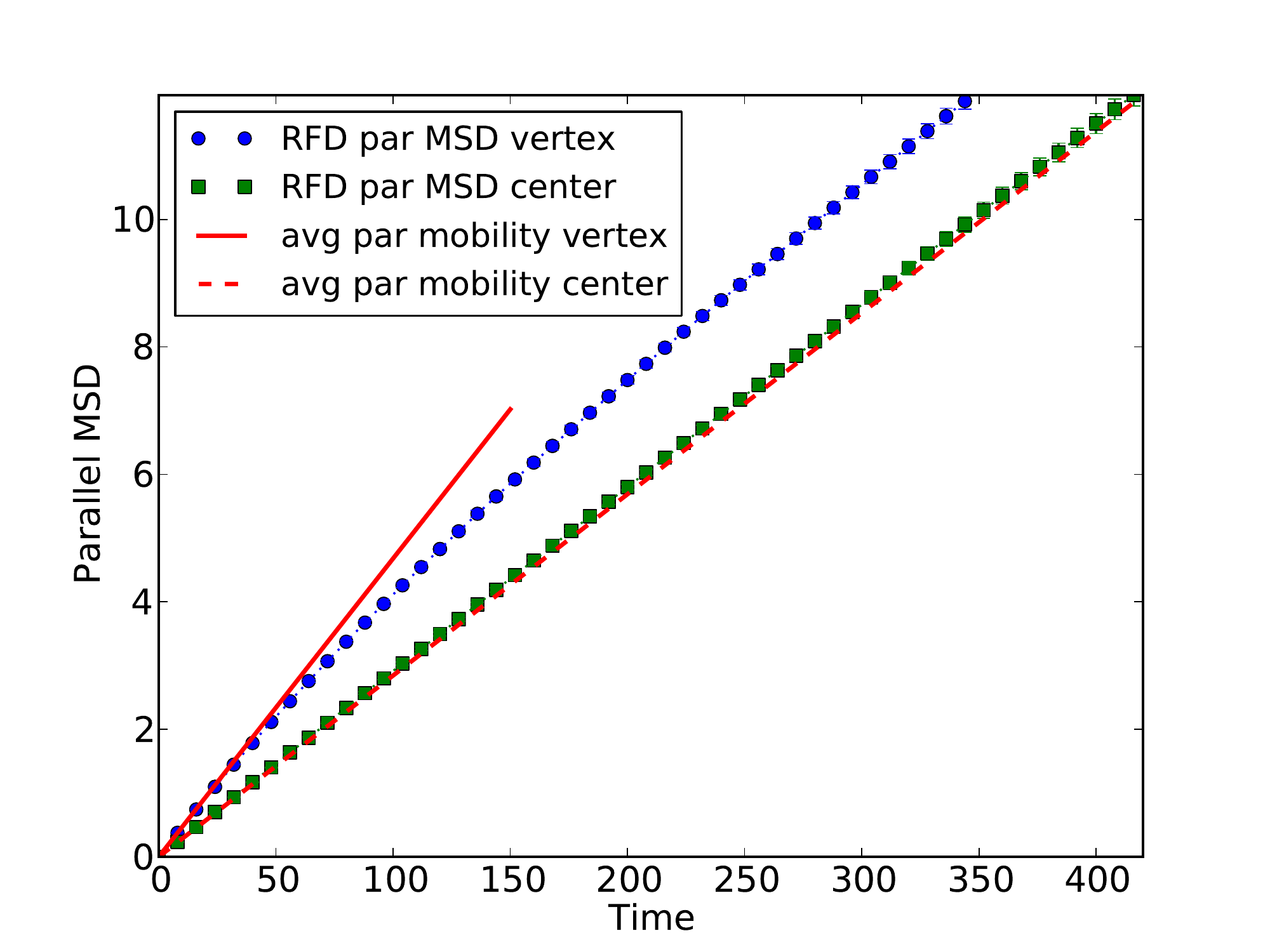}\includegraphics[width=0.49\textwidth]{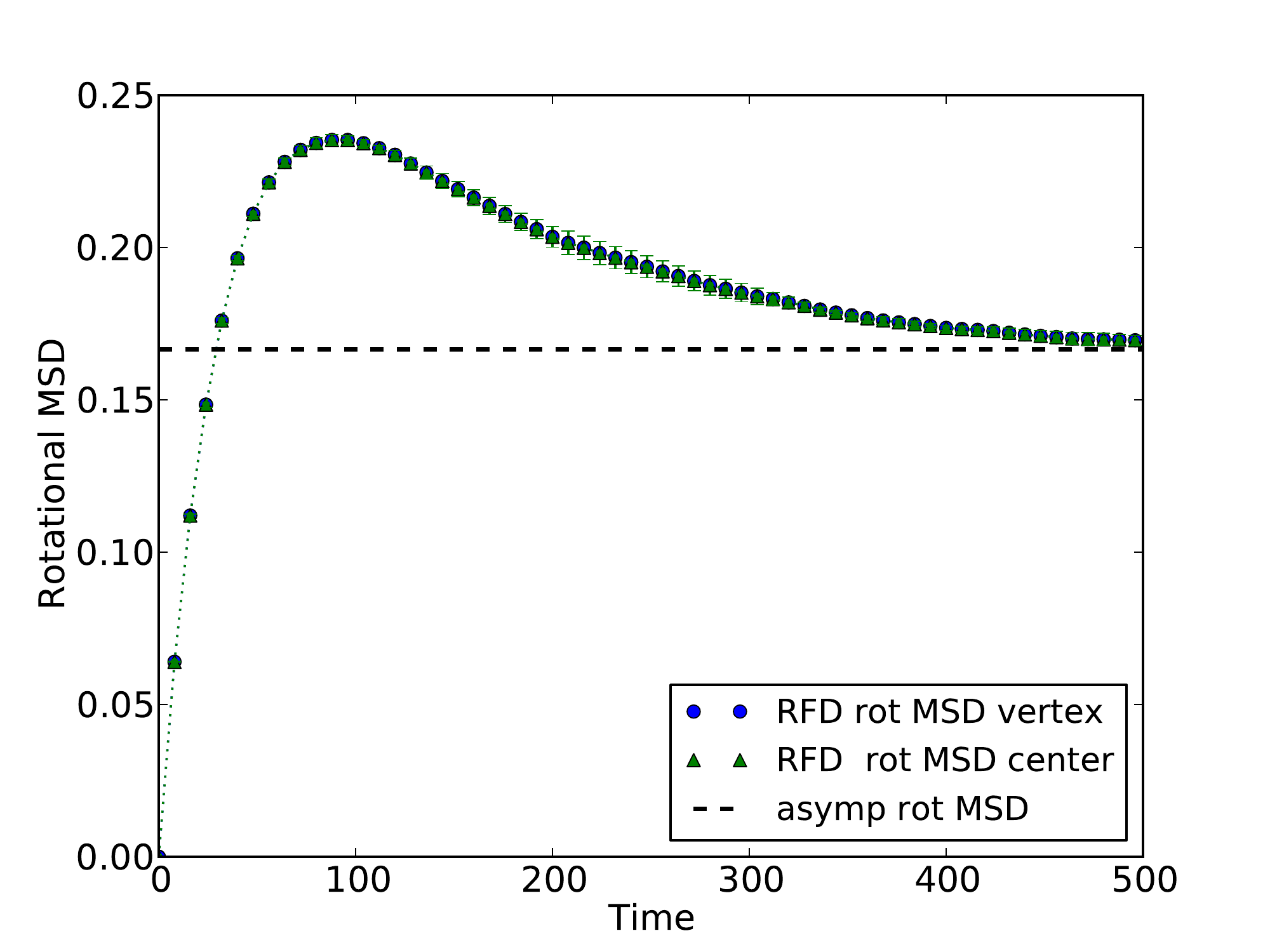}\caption{\label{fig:TetrahedronVertexComparison}\textit{\emph{Mean square
displacement for a colloidal tetramer sedimented near a bottom wall.
}}The data for both figures is generated from 4 independent runs of
$3\cdot10^{5}$ time steps with a time step size of $\D t\approx0.136\,\left(6\pi\eta a^{3}/k_{B}T\right)$;
the highest rejection rate was $2.33\times10^{-5}$ (a total of 7
rejections). The MSDs for each tracking point are calculated from
the same trajectories.\textit{\emph{ }}\textit{(Left)} Comparison
of parallel translational MSD $D_{\parallel}\left(\tau\right)$ when
tracking the geometric center (green squares), versus when tracking
the fourth vertex (blue squares). We see that at short times the numerical
slope agrees with (\ref{eq:MSDMobility}), shown as a red dashed line
of slope $1.34(k_{B}T/6\pi\eta a)\approx2.8\cdot10^{-2}$ for the
geometric center, and as a red solid line of slope $2.21(k_{B}T/6\pi\eta a)\approx4.7\cdot10^{-2}$
for the vertex. \textit{(Right)} Comparison of the parallel ($(x-x)$
or $(y-y)$) component of the rotational mean square displacement
using the two choices of tracking point. The asymptotic rotational
MSD predicted by (\ref{eq:D_rot_asympt}) is shown as a dashed line.}
\end{figure}

In this section, we examine the translational mean square displacement
of the tetrahedron. In the left panel of Fig. \ref{fig:TetrahedronVertexComparison}
we examine the effect of the choice of tracking point on the parallel
mean square displacement by comparing $D_{\parallel}\left(\tau\right)$
when tracking the geometric center of the tetrahedron, versus tracking
one of the four vertices. In both cases (\ref{eq:MSDMobility}) gives
the initial slope of the MSD as it must, and these slopes are clearly
different. Since at long times the slopes of the parallel MSD is independent
of the choice of tracking point, the MSD cannot be linear at all times
for both choices of tracking point. Indeed, the results in Fig. \ref{fig:TetrahedronVertexComparison}
show that the parallel MSD is linear to within statistical and numerical
truncation errors only when the geometric center is tracked. By contrast,
the rotational MSD is insensitive to the choice of tracking point,
as seen in the right panel of Fig. \ref{fig:TetrahedronVertexComparison}.

We note that far from the wall, torques applied about the center of
the tetrahedron generate no translation, indicating that in the absence
of confinement the geometric center is both the CoM and the exact
three dimensional CoH (which does not exist for general rigid bodies).
In the presence of the boundary, this is not strictly the case, but
we nonetheless observe in Fig. \ref{fig:TetrahedronVertexComparison}
that the average parallel mobility evaluated using the center of the
tetrahedron as an origin gives a good approximation to the long time
quasi two-dimensional diffusion coefficient $\chi_{2D}$. This is
perhaps not surprising due to the high symmetry of a tetrahedron,
as the geometric center is the ``obvious'' point to track. 

\begin{figure}
\centering{}\includegraphics[width=0.49\textwidth]{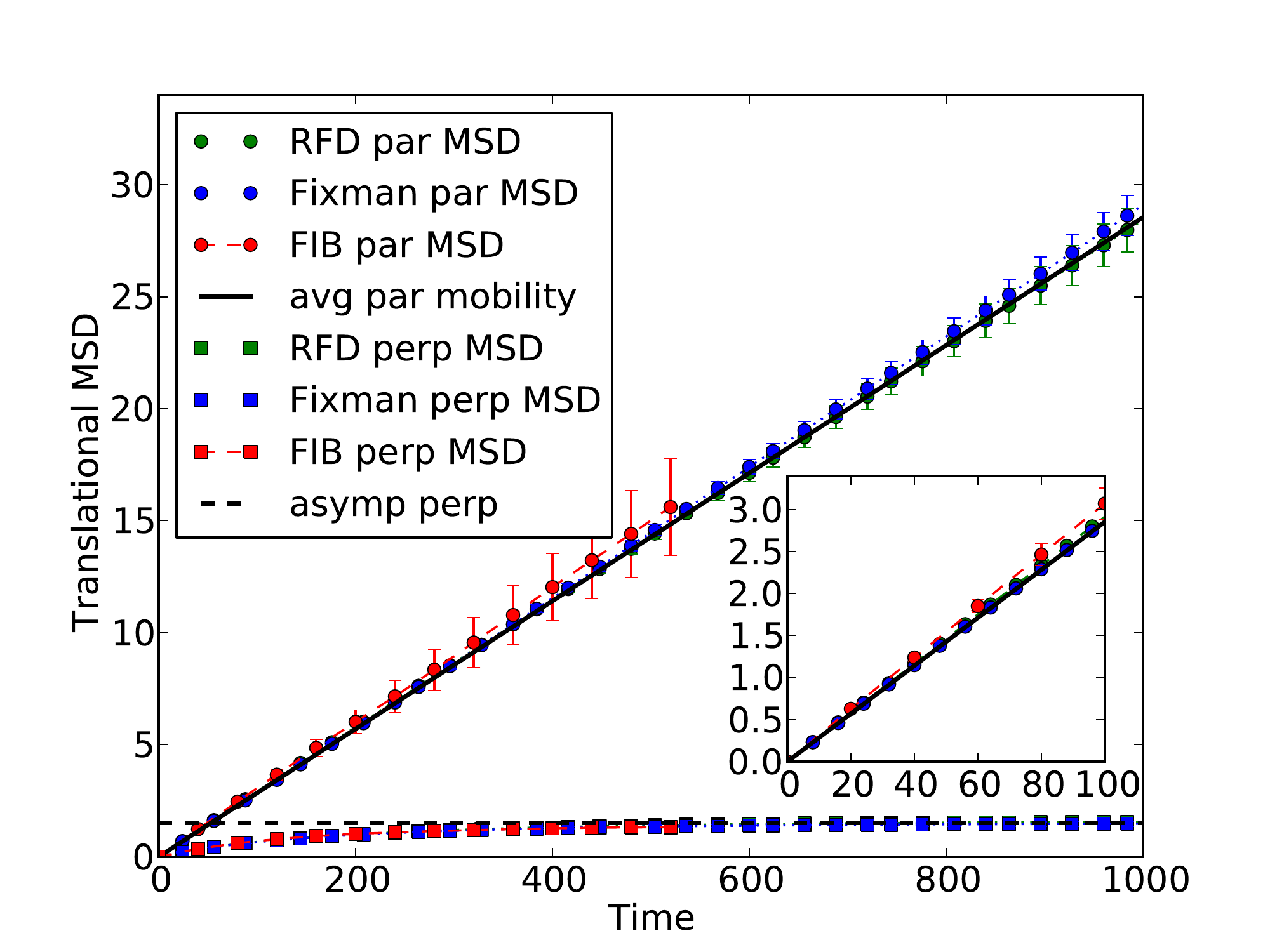}\includegraphics[width=0.49\textwidth]{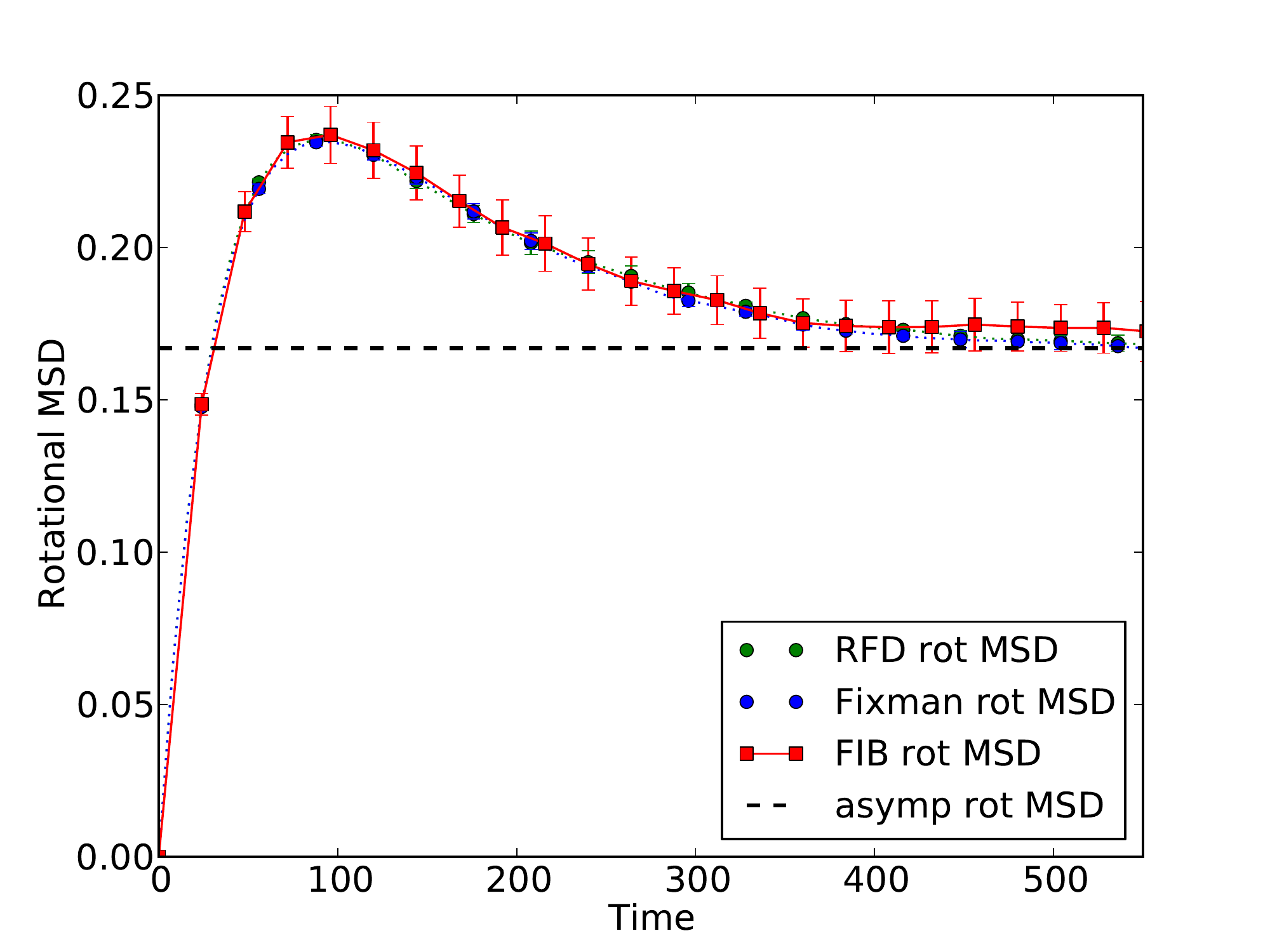}\caption{\label{fig:TetrahedronIBAMRComparison}\textit{\emph{Comparison of
the mean square displacement for a colloidal tetramer sedimented near
a bottom wall,}} obtained by treating the body as rigid using the
RFD and Fixman methods developed here (see caption of Fig. \ref{fig:TetrahedronVertexComparison}),
versus keeping it nearly rigid with stiff springs and using the FIB
method \cite{BrownianBlobs}. For the FIB runs we used 32 simulations
of $10^{5}$ time steps each, with a time step size 32 times smaller
than in the rigid-body simulations. \textit{(Left)} Parallel translational
MSD when tracking the geometric center of the tetrahedron. The inset
focuses on the short time diffusion, and shows a slight hydrodynamic
difference between the rigid and semi-rigid models that is due to
the different methods used to handle the hydrodynamics, as well as
the slight flexibility of the tetrahedron in the FIB simulations.
\textit{(Right) }The parallel ($(x-x)$ or $(y-y)$) component of
the rotational MSD (\ref{eq:rotational_diffusion_coeff}).}
\end{figure}

In Figure \ref{fig:TetrahedronIBAMRComparison} we compare results
for the MSD obtained using the overdamped rigid-body integrators from
Section \ref{sec:TemporalIntegrators} to results obtained using the
FIB method and stiff springs. We examine the mean square translational
displacement parallel and perpendicular to the wall, as well as the
rotational MSD, and find that the behavior of the tetrahedron is the
same for both the stiff and rigid simulations; this provides a validation
of our rigid-body methods and our codes. However, due to the presence
of the stiff springs, using the FIB method to simulate a rigid body
requires a time step size that is 32 times shorter. Due to the small
time step size required for the tetrahedron constructed using rigid
springs, and the high cost of numerically solving a Stokes problem
each time step, it is computationally impractical to study the long
time diffusion coefficient using the FIB method. The time step size
for the rigid-body method could in principle be even larger and still
resolve the dynamics of the body, but it is limited by the stiff potential
used to repel the particle from the wall; we keep $\D t$ sufficiently
small to strictly control the number of rejections of unphysical states
where a blob gets too close to or passes through the wall. In Section
\ref{sec:Conclusion} we discuss some ideas that may allow for the
use of larger time step sizes even in the presence of steep repulsive
forces.

\subsection{\label{sub:AsymmetricSphere}Asymmetric sphere: Icosahedron}

In this section we examine the diffusive motion of a rigid sphere
whose center of mass is displaced away from the geometric center,
in the presence of gravity and a bottom wall (no-slip boundary). This
models recently manufactured colloidal ``surfers'' that become active
when the particles sediment to a microscope slide \cite{Hematites_Science};
here we consider a passive particle in the absence of chemical driving
forces. Diffusive and rotational dynamics of a symmetric patterned
(Janus) sphere near a boundary has been studied experimentally by
Anthony et al. \cite{SphereNearWall}, and can be described well by
theoretical approximations for the mobility of a rigid sphere near
a planar wall \cite{BrennerBook}.

We construct a hydrodynamic model of an asymmetric rigid sphere of
radius $a_{I}$ by rigidly constraining 12 blobs at the vertices of
an icosahedron; a similar blob model of a sphere was used in Ref.
\cite{MultiblobSprings} but was based on (stiff) penalty springs
rather than rigid-body constraints. Note that more accurate results
can be obtained by using more blobs to construct the spherical shell
\cite{RigidIBM}. Each blob has a hydrodynamics radius of $a=0.175$
and is located a distance $2.5a$ from the center of the icosahedron,
so that the minimal distance between two blobs is about $2.63a$.
These parameters are chosen so that the icosahedron is hydrodynamically
nearly rotationally invariant, and has an effective translational
hydrodynamic radius (computed numerically) in bulk (i.e., far from
the wall) of 
\begin{align*}
a_{I}= & \frac{1}{6\pi\eta\left(\mbox{Tr}\left(\M M_{\V u\V F}\right)/3\right)}\approx2.86a=0.5,
\end{align*}
in some arbitrary units. A gravitational force of $F=0.5=1.25\, k_{B}T/a_{I}$
is applied to one of the 12 blobs, which represents the dense hematite
cube embedded in the nearly spherical colloidal surfers of Palacci
et al. \cite{Hematites_Science}. Gravity therefore generates a torque
around the center of the sphere and causes the icosahedron to prefer
orientations where the heavy blob is facing down. A short-ranged repulsive
force $U\left(h;\, a_{I}\right)$ given by (\ref{eq:Yukawa_blob})
is added to keep the icosahedron from overlapping the wall, where
now $h$ is the distance from the center of the icosahedron to the
wall, the repulsion strength is $\epsilon=20k_{B}T$, and the Debye
length is (arbitrarily) set to $b=a_{I}$. This choice of parameters
gives the center of the icosahedron a gravitational height (\ref{eq:h_g_def})
of $h_{g}\approx0.96a_{I}$. Note that in this example the icosahedron
is considered to be a hydrodynamic \emph{approximation} of a physical
sphere and therefore the repulsive force acts on the center of the
sphere (thus not generating any torque), rather than acting on each
of the 12 blobs individually (which would generate some small spurious
torque).

\subsubsection{Equilibrium Distribution}

\begin{figure}
\begin{centering}
\includegraphics[width=0.49\textwidth]{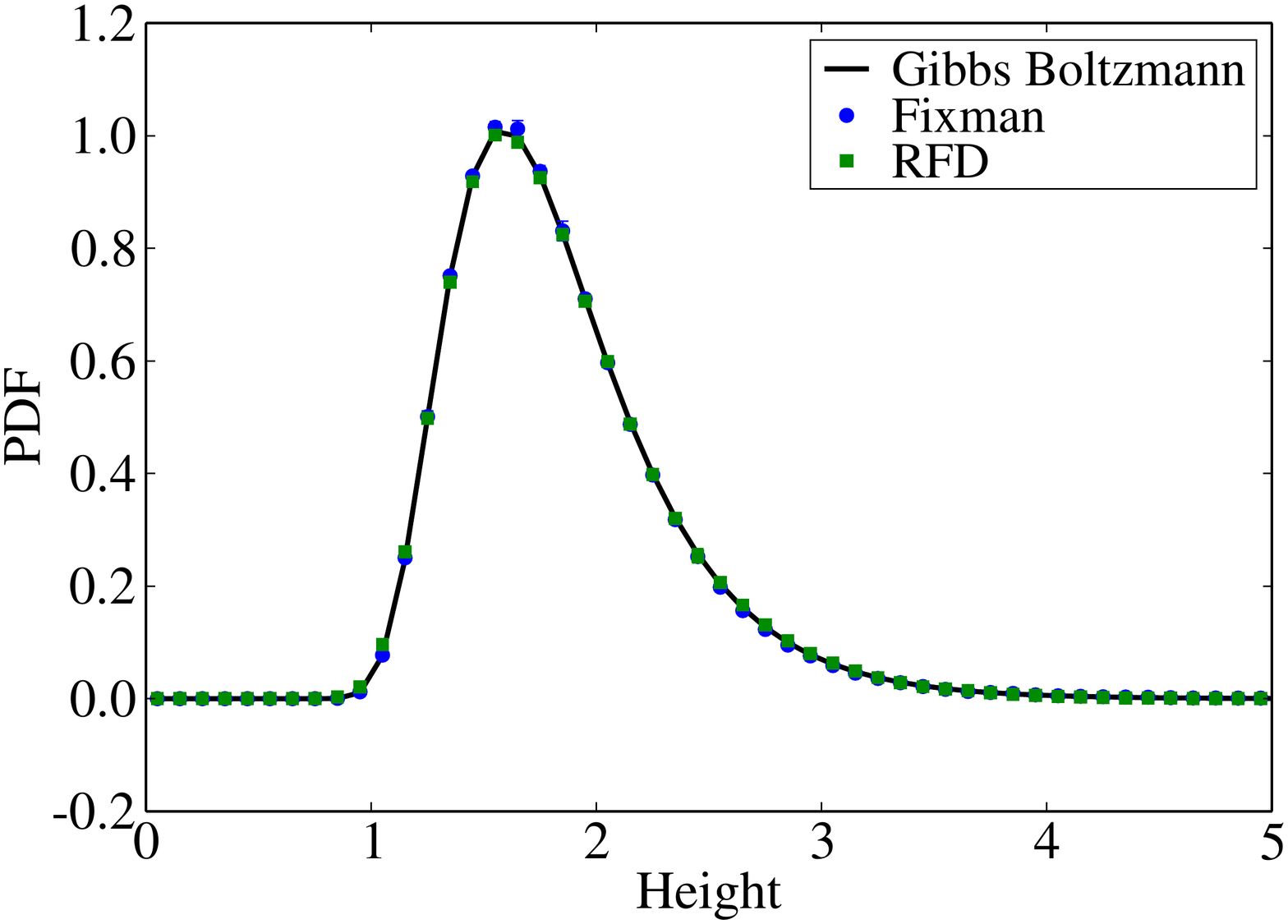}\includegraphics[width=0.49\textwidth]{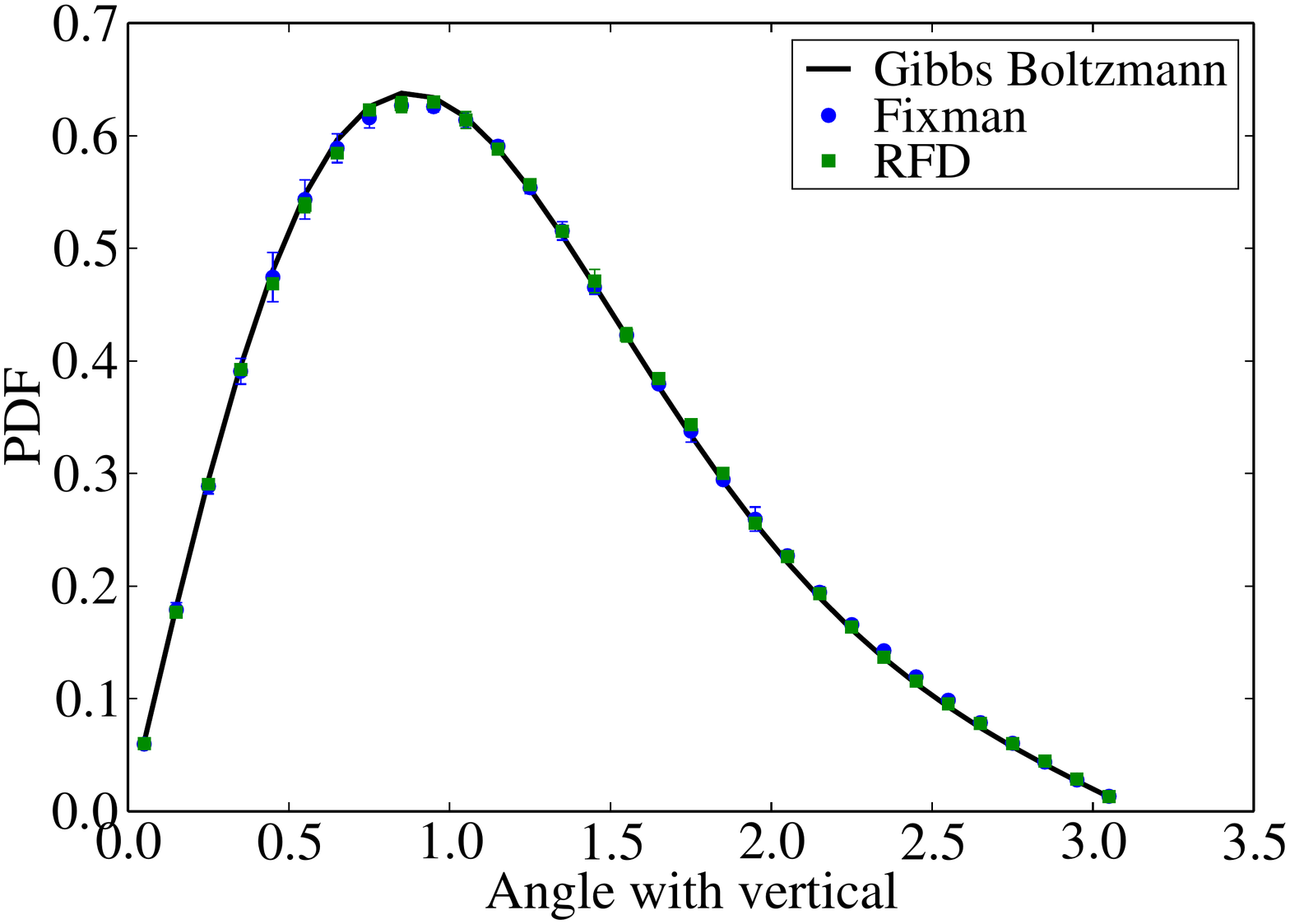}
\par\end{centering}

\caption{\label{fig:IcosahedronDistribution}Equilibrium distribution for a
rigid icosahedron of blobs compared to analytic expressions for the
Gibbs-Boltzmann distribution. These figures were created using data
from 6 independent runs of $4\cdot10^{5}$ time steps with a small
time step size of $\D t=0.04\,(6\pi\eta a_{I}^{3})/k_{B}T$ to avoid
rejections; no rejections occurred during these simulations. \textit{(Left)
}Equilibrium distribution of the height $h$, the distance from the
center of the icosahedron to the wall. \textit{(Right) }Equilibrium
distribution of the angle $\theta$, where $\theta=0$ indicates that
the heavy blob is at the bottom of the icosahedron, and $\theta=\pi$
indicates that it is at the top. As expected, we see the distribution
skewed towards smaller values of $\theta$ due to the gravitational
force.}
\end{figure}

We first investigate the equilbrium distribution $P_{eq}\left(\V q,\V{\theta}\right)$,
examining the marginal distributions of height $h$ and orientation
angle $\theta$, which is the angle between the $z$ axis and the
vector connecting the center of the icosahedron to the blob to which
we apply the gravitational force. In this simple example, we can compute
the marginals of the equilibrium Gibbs-Boltzmann distribution analytically
for both $h$ and $\theta$, and they are compared to numerical results
in Fig. \ref{fig:IcosahedronDistribution}. We see that the RFD and
Fixman schemes agree with each other and with theory. Due to the nonuniform
gravitational forcing on the icosahedron, it prefers orientations
with $\theta$ closer to zero, but the thermal fluctuations causes
it to explore all orientations.

\subsubsection{Mean Square Displacement}

\begin{figure}
\centering{}\includegraphics[width=0.49\textwidth]{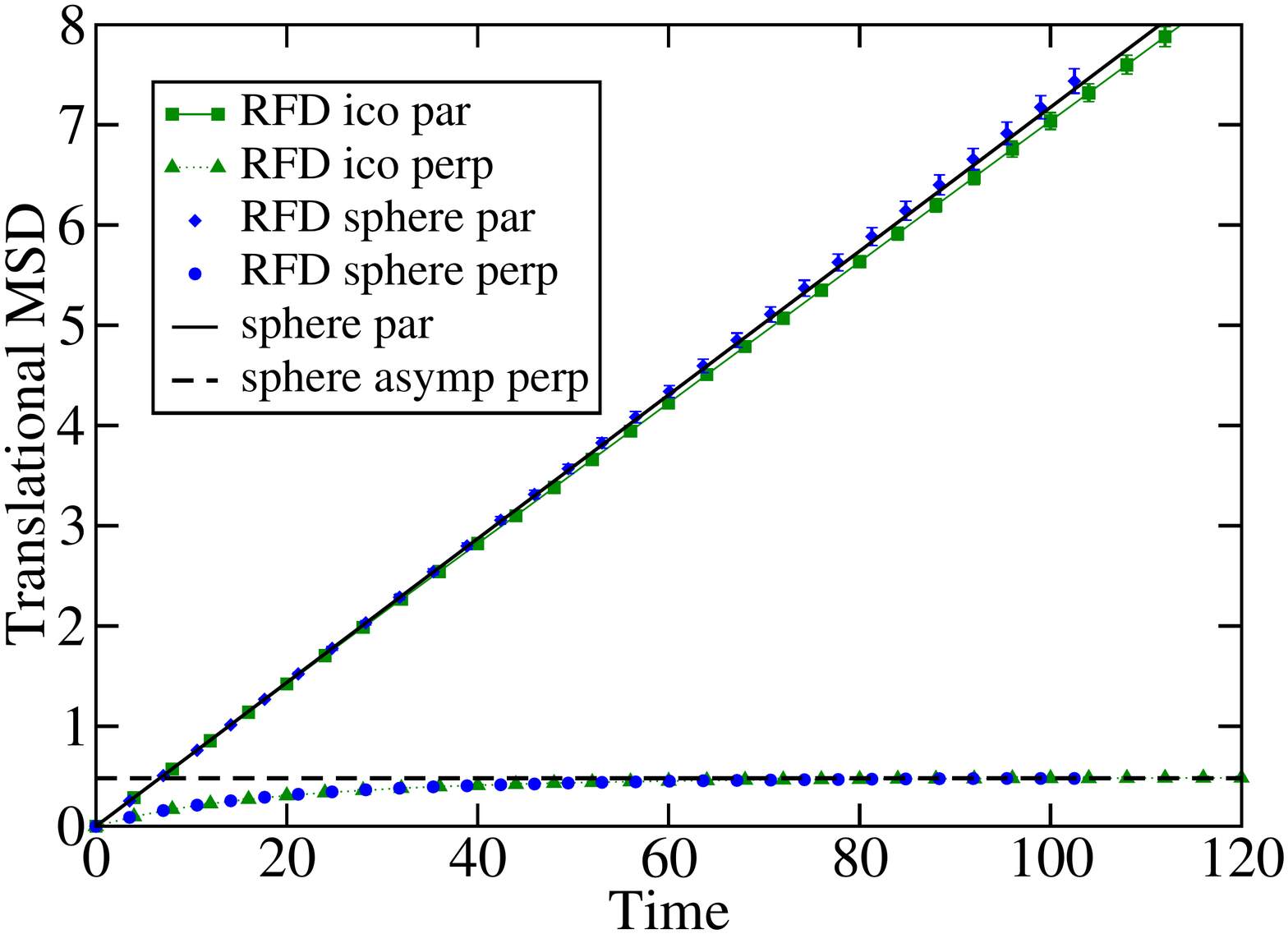}\includegraphics[width=0.49\textwidth]{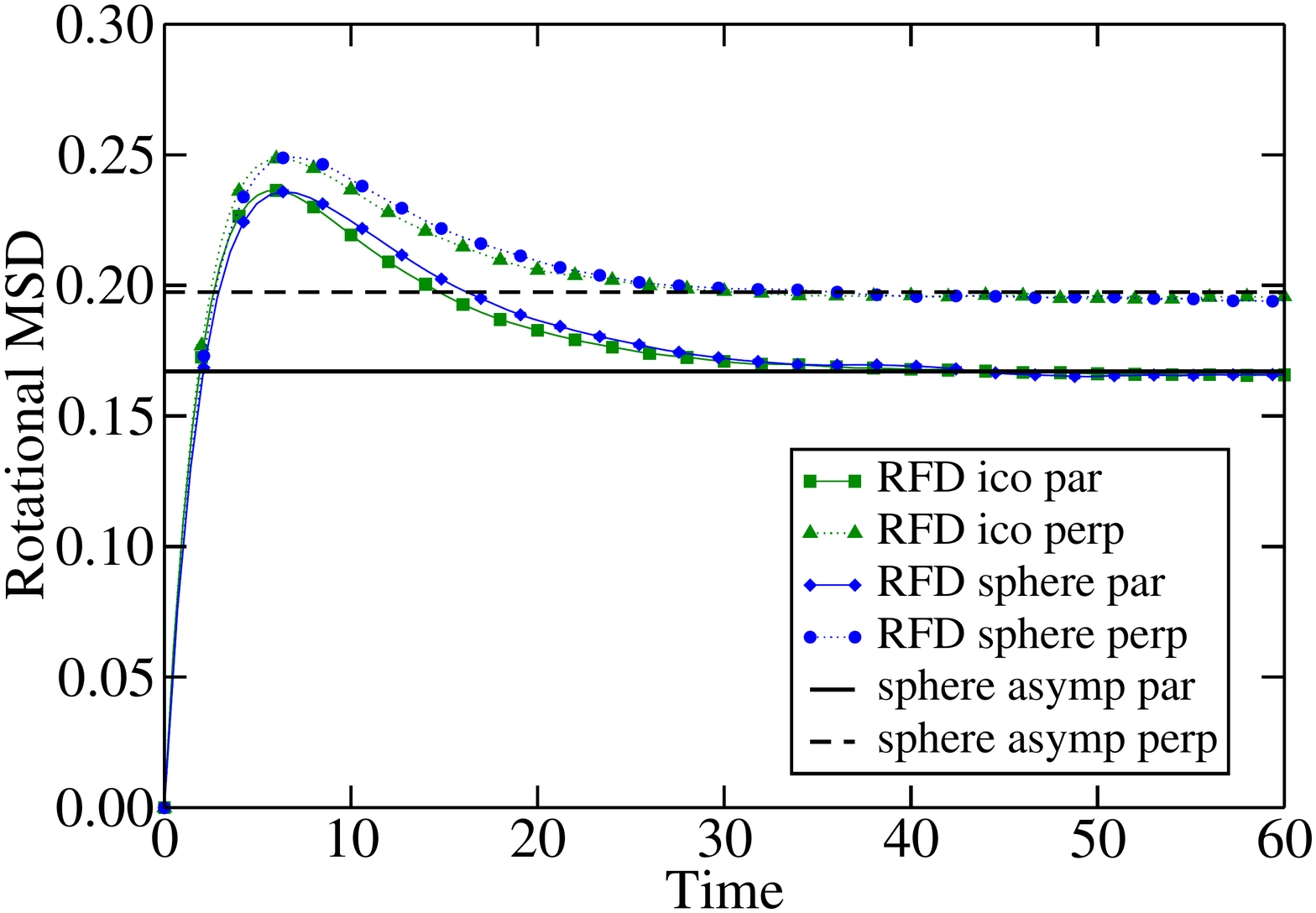}\caption{\label{fig:IcosahedronMSD}Mean square displacements for a sphere
with nonuniform mass distribution diffusing near a planar boundary.
Symbols show numerical results obtained by processing long equilibrium
trajectories, while lines show theoretical predictions. These figures
were generated using data from 16 independent trajectories of length
$5\cdot10^{5}$ time steps, using the RDF temporal integrator with
a small time step size of $\D t=0.004\:(6\pi\eta a_{I}^{3})/k_{B}T$
to eliminate rejections. \textit{(Left)} Parallel ($D_{\parallel}(\tau)$)
and perpendicular ($D_{\perp}(\tau)$) translational MSD of a rigid
icosahedron of blobs. The solid black line shows the theoretical parallel
translational MSD predicted by (\ref{eq:chi2D_theory}) for a rigid
sphere using the best-known approximations to the hydrodynamic mobility
(\ref{eq:SingleWallHardSpherePerp},\ref{eq:SingleWallHardSphereLub},\ref{eq:SingleWallHardSphereParallel}),
while the dashed black line shows the asymptotic perpendicular translational
MSD (\ref{eq:D_perp_asympt}). As expected, the icosahedron behaves
like a sphere with equal effective hydrodynamic radius \cite{MultiblobSprings}.
\textit{(Right)} Parallel ($x-x$ or $y-y$) as well as perpendicular
($z-z$) components of the rotational mean square displacement (\ref{eq:MSD}).
The dashed line shows the asymptotic rotational MSD (\ref{eq:D_rot_asympt}).
We see that the rotational dynamics of the rigid icosahedron and a
true sphere are also in good agreement.}
\end{figure}

To validate how well our scheme captures the dynamics of the system,
we examine the mean square displacement of the geometric center of
the icosahedron. In Fig. \ref{fig:IcosahedronMSD} we compare our
results to the mean square displacement of an actual hard sphere with
hydrodynamic radius $a=0.5.$ We apply torques and forces to the sphere
that are identical to those applied to the icosahedron, but for the
hydrodynamic mobility of the sphere we use the most accurate theoretical
expressions available in the literature, see (\ref{eq:SingleWallHardSpherePerp},\ref{eq:SingleWallHardSphereLub},\ref{eq:SingleWallHardSphereParallel})
in Appendix \ref{add:SphereMob}, instead of relying on the blob approximation
to a sphere (\ref{eq:SingleWallBlob}), even though in this specific
case (\ref{eq:SingleWallBlob}) is sufficiently accurate. This tests
allows us to both evaluate our temporal integration method, as well
as to examine how well the 12-bead model approximates a single spherical
particle.

The results shown in Fig. \ref{fig:IcosahedronMSD} demonstrate that
the dynamics of the icosahedral rigid multiblob is essentially identical
to that of an actual sphere. Note that for a sphere the mobility does
not depend on the orientation of the sphere. Furthermore, by symmetry,
the gravitational force (perpendicular to the wall) cannot induce
rotation of the sphere, and by symmetry, a torque cannot introduce
vertical displacements. Because of these special symmetries the parallel
MSD is linear for all times and therefore (\ref{eq:chi2D_theory})
gives the long-time quasi two-dimensional diffusion coefficient $\chi_{2D}$;
this has in fact been confirmed experimentally with relatively good
accuracy for spheres whose center of mass is very close to their geometric
center \cite{SphereNearWall}.

\subsection{Colloidal Boomerang\label{sub:Boomerang}}

The authors of reference \cite{BoomerangDiffusion} perform a detailed
experimental study of the quasi-two-dimensional translational and
rotational diffusion of lithographed symmetric right-angle boomerang
colloids (see the right panel of Fig. \ref{fig:BlobSpheres}) confined
between two closely-spaced microscope slides. Subsequently this work
was extended to asymmetric (L-shaped) right-angle boomerangs \cite{AsymmetricBoomerangs}
as well as non-right-angle boomerangs \cite{angleBoomerangs}. Some
theoretical analysis is also performed assuming that the overdamped
dynamics of the particles is strictly two-dimensional. Of course,
the actual dynamics of the particles is three dimensional, and a complete
theoretical or numerical analysis of the diffusive dynamics requires
the complete formalism developed in this paper.

In this section we examine a single symmetric right-angle boomerang
near a single no-slip boundary (bottom wall) in the presence of gravity.
We choose to study a single boundary rather than a slit channel as
done in the experiments in order to simplify the hydrodynamic calculations
of mobilities \cite{StokesianDynamics_Wall}; in principle one can
construct tabulated approximations of self and pairwise mobilities
in a slit channel but this is quite complex and expensive \cite{StokesianDynamics_Slit}.
While we cannot make direct comparisons with the experimental values
reported in Ref. \cite{BoomerangDiffusion} in this work, we can still
address the fundamental questions about differences between fully
three-dimensional and quasi two-dimensional diffusion. Specifically,
by enlarging the gravitational force we apply to the boomerang (i.e.,
increasing its effective density mismatch with the solvent), we can
cause the motion to be more or less confined to a two dimensional
plane parallel to the bottom wall. In this section we use microns
as the unit of length, seconds as the unit of time, and milligrams
as the unit of mass.

For hydrodynamic calculations, we construct a blob model of a boomerang
and try to match the physical parameters in the experiments \cite{BoomerangDiffusion}
as close as possible. Our model of the boomerang particle is constructed
by rigidly connecting 15 blobs, one at the cross point, and 7 for
each arm, as illustrated in the right panel in Fig. \ref{fig:BlobBoomerang}.
Prior investigations in the context of the immersed boundary method
\cite{IBM_Sphere}, which we have also confirmed independently by
using the Rotne-Prager tensor as the pairwise blob mobility, have
shown that to construct a good hydrodynamic approximation of a rigid
cylinder of radius $r$ using blobs, one should set the effective
hydrodynamic radius of each blob to $a\approx\sqrt{3/2}\, r$, and
place the blobs centers on a line at a distance of around $a$ (the
precise value does not matter much). Following these recommendations,
we set the blob radius to $a=0.325$, which gives an effective cylinder
radius of 0.265, and the blobs are spaced a distance 0.3 apart. Note
that in this minimally-resolved blob model the cross-section of the
arms of the boomerang is cylindrical rather than square, as would
be more realistic for modeling the lithographed particles. We have,
however, compared to a more resolved 120-blob model constructed from
the initial boomerang by replacing each of its 15 blobs by 8 smaller
blobs of radius $0.1625$ placed at the vertices of a rectangular
prism of size $0.15\times0.285\times0.245$ centered at the location
of the original blob. We find only minor differences with the minimally-resolved
model, for example, in bulk (without confinement) the diffusion coefficients
in the plane of the boomerang are computed to be (in units of $\mu m^{2}/s$)
$0.243$ and $0.283$ for the 15-blob model, and $0.245$ and $0.291$
for the 120-blob model.

For a free boomerang far away from boundaries, there is a unique CoM
that, due to symmetry, must lie on the the line that bisects the boomerang
arms. Also, there must be a unique point on the bisector for which
there is no coupling between torque applied out of the plane of the
boomerang and the translational motion in the plane of the boomerang.
We can consider this point as the CoH for quasi-two-dimensional diffusion
\cite{BoomerangDiffusion}, although, as already explained, this point
is not a CoH in the strict sense for three-dimensional diffusion.
The locations of the bulk CoM and the bulk quasi-two-dimensional CoH,
which we shall henceforth imprecisely refer to as just CoM and CoH,
can be computed from (\ref{eq:CoM_system}) and (\ref{eq:CoH_system}),
respectively. For our blob model, we compute the CoH to be is about
1.08 microns away from the cross point (center of the intersection
blob), and the CoM is 0.96 microns from the cross point; we get the
same estimates from the more refined 120-blob models. These numbers
compare favorably to the experimental findings in \cite{BoomerangDiffusion},
where the CoH is estimated to be a distance of 1.16 microns from the
cross point; the CoM is not mentioned in the experimental works on
boomerang particles. The difference between the CoH and CoM is too
small for this specific particle shape for us to be able to tell the
difference to within statistical errors; in future work we will look
for other planar particle shapes for which the difference may be more
significant and measurable in both simulations and experiments.

The total gravitational force applied to the body is $0.18\times g\,\left(k_{B}T/a\right)$
where $g$ is a parameter that we vary; we split the gravitational
force evenly among the 15 blobs. Here $g=1$ gives a rough approximation
of the gravitational binding experienced by the actual lithographed
particles, which have a density of $1.2\,\mbox{g}/\mbox{cm}^{3}$.
Each blob is also repelled from the wall using the potential (\ref{eq:Yukawa_blob})
with screening length $b=0.5a$ and strength $\epsilon=23.08kT$.
The gravitational height (\ref{eq:h_g_def}) for one of the two (equivalent)
tips of the boomerang are shown in Table \ref{tab:Boomerang2DDiffusion}
for several values of $g$. Since the tips are the points that are
most likely to venture further from the wall, these values give an
indication of how close to two-dimensional the dynamics of the boomerang
is.

\subsubsection{Translational Diffusion}

\begin{figure}
\centering{}\includegraphics[width=0.49\textwidth]{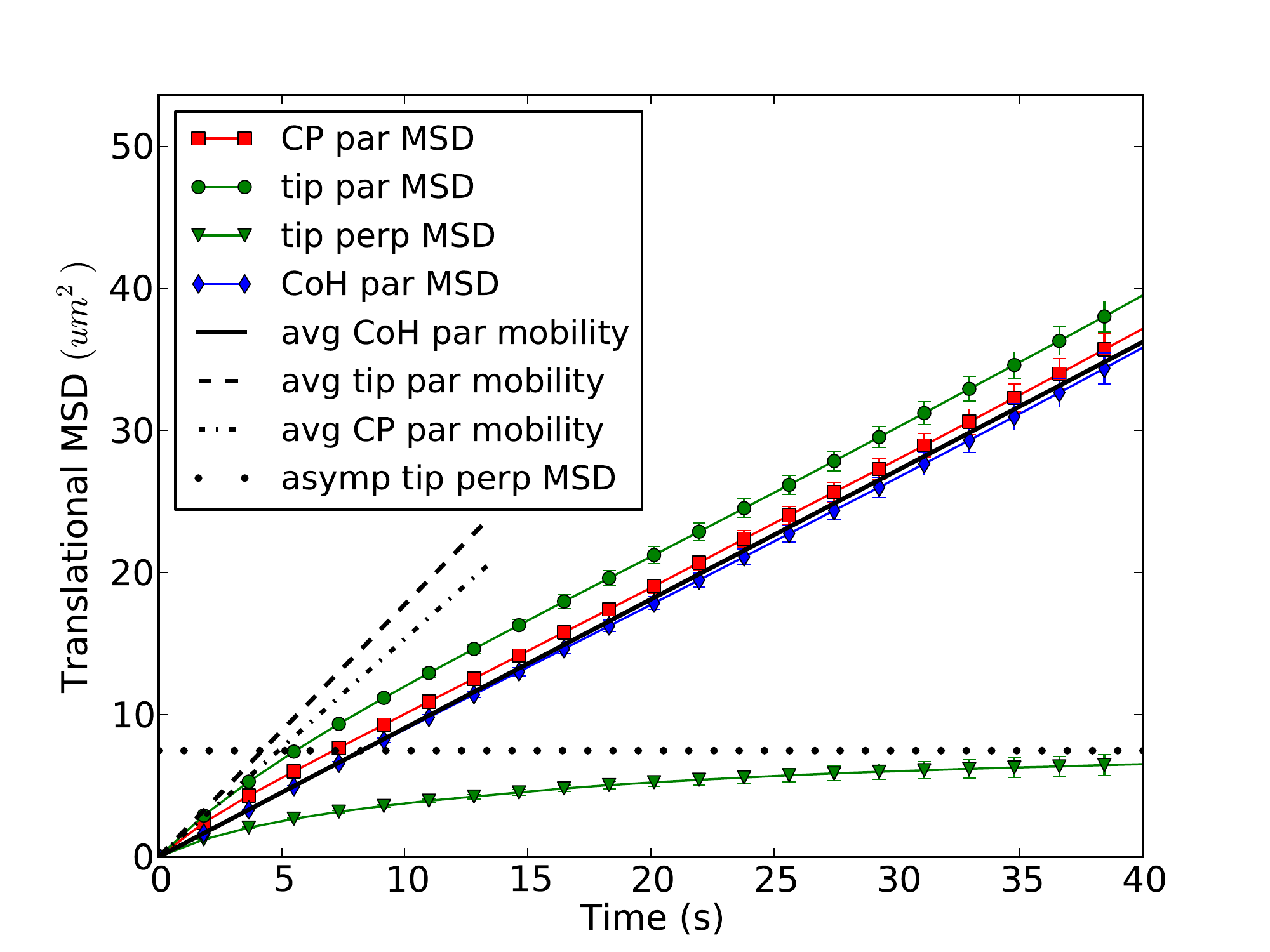}\includegraphics[width=0.49\textwidth]{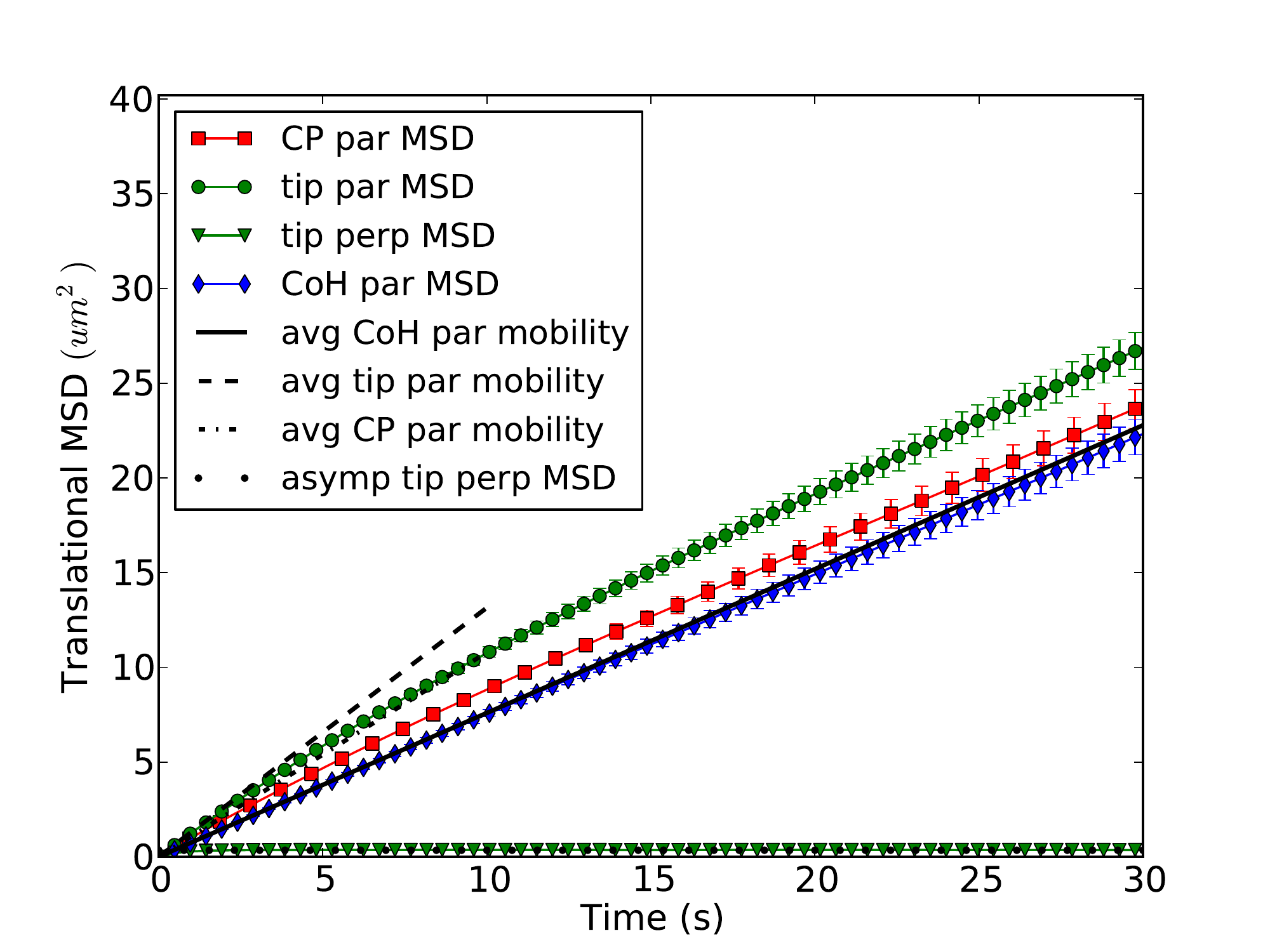}\caption{\label{fig:BoomerangMSD}Translational MSD of a right-angle symmetric
boomerang for gravity $g=1$ (left panel) and $g=20$ (right panel)
parallel and perpendicular to the wall, see legend. The same trajectories
are used but the parallel MSD is computed using three different choices
of the tracking point: cross point (CP) at the corner of the right
angle, (one of the) tip(s) of the boomerang, and the center of hydrodynamic
stress (CoH), which we note is essentially indistinguishable for this
purpose from the center of mobility (CoM). For the perpendicular MSD
we only show the results for the tip and show the expected asymptotic
value with a dotted line. The slopes of the parallel MSD at the origin
are shown with lines (see legend) and estimated using Monte Carlo
sampling via (\ref{eq:chi2D_theory}). We see that by calculating
the MSD tracking the CoH/CoM, the MSD is essentially linear with time,
and therefore the short time diffusion coefficient (\ref{eq:chi2D_theory})
matches the (unique) long time diffusion coefficient $\chi_{2D}$.
This is not the case when tracking the cross point or the tip of one
of the arms.}
\end{figure}

In Fig. \ref{fig:BoomerangMSD} we show the parallel and perpendicular
translational MSDs of the boomerang for a weak ($g=1$, left panel)
and a strong gravitational sedimentation ($g=20$, right panel), where
strong here means that the gravity is sufficiently strong to keep
the boomerang essentially flat against the surface. For the parallel
MSD, we show results based on three different choices of the tracking
point: 1) the CoH, or in this example, equivalently the CoM; 2) the
center of the blob at the tip of one of the arms; and 3) the cross
point (CP), which is the center of the blob at the cross point where
the arms meet. For an unconfined boomerang, we expect that the parallel
MSD measured using the CoM will be strictly linear in time. We also
expect that for the boomerang confined to a plane by strong gravity
and exhibiting quasi two-dimensional diffusion, the MSD will be linear
in time when tracking the CoH \cite{BoomerangDiffusion,AsymmetricBoomerangs}.
However, in our simulations, we find that due to the close proximity
of the CoH and the CoM, the MSD is identical to within statistical
error independent of which of these two points was tracked. For clarity,
we only include the MSD calculated from the CoH in our results, with
the understanding that the CoM is indistinguishable at this level
of accuracy. We see from the figure that by choosing the CoH as the
tracking point, we obtain a MSD that is linear over all times up to
statistical accuracy for both gravities. This means that we can get
an accurate estimate of the long-time diffusion coefficient $\chi_{2D}$
by using equation (\ref{eq:chi2D_theory}) over a broad range of gravities.
This statement should, however, be checked for other particle shapes
for which the CoH and CoM are sufficiently far apart, before drawing
broad conclusions.

\begin{table}
\centering{}%
\begin{tabular}{|c|c|c|c|c|c|}
\hline 
$g$ &
$\chi_{2D}(\mu m^{2}/s)$ &
$\chi_{2D}/\chi_{3D}$ &
$\chi_{\theta}\left(\mbox{rad}^{2}/s\right)$ &
$h_{g}(\mu m)$ &
$\chi_{2D}/\chi_{3D}$ sphere\tabularnewline
\hline 
\hline 
$1$ &
0.226 &
0.834 &
1.42 &
1.77 &
0.837\tabularnewline
\hline 
$10$ &
0.194 &
0.716 &
0.79 &
0.605 &
0.724\tabularnewline
\hline 
$20$ &
0.185 &
0.683 &
0.22 &
0.419 &
0.680\tabularnewline
\hline 
\end{tabular}\caption{\label{tab:Boomerang2DDiffusion}Long-time quasi two-dimensional diffusion
coefficient $\chi_{2D}$ for the boomerang colloid at $g=1,\,10,$
and $20$ estimated using (\ref{eq:chi2D_theory}) and tracking the
CoH. The diffusion coefficient for a free boomerang in an unbounded
fluid $\chi_{3D}$ is computed using (\ref{eq:MSD_CoM}). The rotational
diffusion coefficient $\chi_{\theta}$ is calculated using (\ref{eq:chi_theta_theory}).
Comparing the effective gravitational heights (\ref{eq:h_g_def}),
calculated for the tip of the boomerang, and the boomerang's arm length
($2.1\mu m$) or the blob radius ($0.325\mu m$) gives an indication
of how flat the boomerang is against the wall. In the last column
we estimate the reduction in mobility relative to bulk for a sphere
of the same radius as the blobs and at the same gravitational height
as the tip of the boomerang.}
\end{table}

In Table \ref{tab:Boomerang2DDiffusion} we show the estimated long-time
parallel diffusion coefficient $\chi_{2D}$ obtained from (\ref{eq:chi2D_theory})
for different strengths of the gravitational sedimentation. We find
that, perhaps surprisingly, the presence of the boundary does not
strongly reduce the effective short-time diffusion coefficient compared
to bulk, except at the largest gravity. In the last column of the
table we give the corresponding reduction in mobility for a sphere
of the same radius as the blob radius, as obtained from the theoretical
estimate (\ref{eq:SingleWallHardSphereParallel}) averaged against
a Gibbs-Boltzmann distribution $\sim\exp\left(-h/h_{g}\right)$ for
the height above the wall; a remarkable agreement is observed despite
the significant difference in the particle shape. The value of the
quasi two dimensional diffusion coefficient is measured experimentally
in Ref. \cite{BoomerangDiffusion} for the case of a boomerang particle
confined between two microscope slides a distance $2\mu m$ apart,
and a value of $\chi_{2D}^{\mbox{exp }}=0.054\,\mu m^{2}/s$ is reported.
This is lower than the values calculated here, which we expect is
largely due to the absence of the top wall in our simulations, which
will significantly increase the drag on the boomerang for such strong
confinement.%

\subsubsection{Rotational Diffusion}

\textcolor{black}{}
\begin{figure}
\centering{}\textcolor{black}{\includegraphics[width=0.75\textwidth]{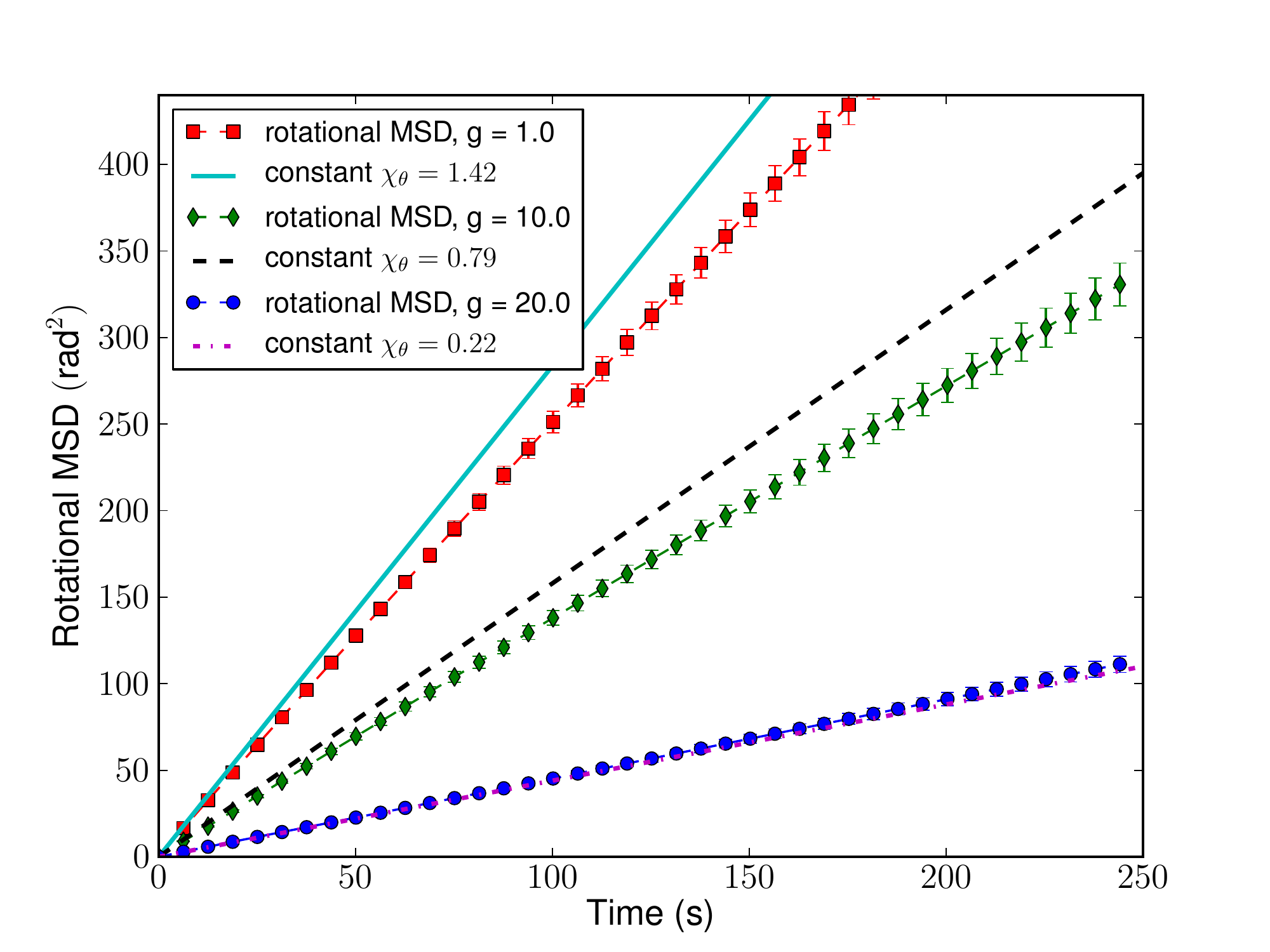}\caption{\label{fig:RotationalBoomerang}Planar rotational mean squared displacement
(\ref{eq:2d_rot_msd}) of the colloidal boomerang for three different
strengths of the gravitational confinement.}
}
\end{figure}

To estimate the quasi-two-dimensional rotational diffusion coefficient
$\chi_{\theta}$, measured experimentally in Ref. \cite{BoomerangDiffusion},
we project the bisector of the boomerang arms into the $x-y$ plane,
and define $\theta(t)$ to be the angle of rotation around the $z$
axis between this projected vector at time $t$ and the initial projected
bisector. We count each full counterclockwise rotation as an addition
of $2\pi$, and similarly we subtract $2\pi$ for each full clockwise
rotation, allowing the value of $\theta$ to take values in all of
$\mathbb{R}$. Truly two-dimensional rotational diffusion (where the
colloid stays in the $x-y$ plane) corresponds to $\theta(t)$ being
standard Brownian motion. We define a planar rotational mean square
displacement from increments of the angle $\theta$, 
\begin{equation}
D_{\theta}\left(\tau\right)=\av{\D{\theta}^{2}}=\av{\left(\theta(t+\tau)-\theta(t)\right)^{2}}.\label{eq:2d_rot_msd}
\end{equation}
In Fig. \ref{fig:RotationalBoomerang} we show numerical results for
$D_{\theta}\left(\tau\right)$. We see that for $g=20$, when the
boomerang is most flat, the quantity $D_{\theta}$ is linear in time
for all times to within statistical error bars, while for lower gravities
we see some deviations from linearity, as expected since the definition
of $\theta$ assumes the diffusion is essentially two-dimensional.

We estimated the short-time planar rotation coefficient
\[
\chi_{\theta}=\lim_{\D t\to0}\,\frac{\avv{\D{\theta}^{2}}}{2\D t}
\]
using Monte Carlo averaging based on (\ref{eq:chi_theta_theory})
from Appendix \ref{Add:RotationalDiff}, and tabulate the computed
values in Table \ref{tab:Boomerang2DDiffusion}. Note, however, that
the trajectory $D_{\theta}\left(\tau\right)$ is only continuous if
the boomerang never flips, i.e., the bisector is never nearly perpendicular
to the wall; note that we \emph{do} observe flips for two lower gravities.
We see that $\chi_{\theta}$ is much larger for lower gravities, both
because the boomerang diffuses more rapidly far from the wall, and
also because in low gravity, the boomerang is not confined to the
$x-y$ plane, and hence small changes in orientation can lead to large
changes in our calculated two-dimensional angular displacement. The
rotational diffusion coefficient measured experimentally in Ref. \cite{BoomerangDiffusion}
for a boomerang confined between two microscope glass slips is $\chi_{\theta}=0.044\,\mbox{rad}^{2}/s$,
which is much smaller than our result for $g=20$. This is most likely
in large part due to the absence of drag from the top wall. Additionally,
without this second boundary, our simulated boomerang is able to rotate
out of the $x-y$ plane more easily, reaching configurations where
a small change in orientation can lead to a large change in $\theta(t)$.

\section{Conclusion\label{sec:Conclusion}}

In this paper, we studied the Brownian motion of rigid bodies of arbitrary
shape immersed in a viscous fluid in the overdamped regime, in the
presence of confinement and gravity. We parameterized the orientation
of the rigid bodies with normalized quaternions, which offer several
advantages over other previously-used representations. Furthermore,
we do not assume any particular symmetry for the rigid bodies, and
we account carefully for the fact that the hydrodynamic mobility $\M N$
depends on the configuration due to confinement or hydrodynamic interactions
with other particles. We derived the appropriate form of the overdamped
Langevin equations of motion, including all of the stochastic drift
terms required to give the correct Gibbs-Boltzmann distribution in
equilibrium, and to preserve the unit norm constraint of the quaternions. 

In section \ref{sec:TemporalIntegrators} we developed temporal integrators
for the rigid-body overdamped Langevin system and presented two ways
to handle the stochastic drift term. The first approach is a generalization
of the well known midpoint Fixman scheme \cite{BD_Fixman,BD_Hinch},
which generates the drift terms using a midpoint predictor step but
requires a costly application or factorization of $\M N^{-1}$. The
second approach employs a Random Finite Difference approach to generate
the drift terms using only applications of $\M N$ and $\M N^{\frac{1}{2}}$,
making it an appealing choice. The RFD approach is especially promising
in situations where the action of the mobility and the stochastic
terms are generated by using a fluctuating hydrodynamics fluid solver,
as in the fluctuating force coupling method (FCM) \cite{ForceCoupling_Fluctuations},
or extensions of our fluctuating immersed boundary (FIB) method \cite{BrownianBlobs}
to include rotlet (and possible also stresslet) terms in the minimally-resolved
blob model.

In Section \ref{sec:NumericalResultsRigid} we performed several numerical
simulations of the Brownian motion of rigid particles diffusing near
a wall in the presence of gravity, motivated by a number of recent
experiments studying the diffusion of asymmetric spheres \cite{SphereNearWall},
clusters of spheres \cite{ColloidalClusters_Granick,ComplexShapeColloids},
and boomerang colloids \cite{BoomerangDiffusion,AsymmetricBoomerangs}.
First, we examined the behavior of a tetramer formed by rigidly connecting
four colloidal spheres together, modeling colloidal clusters that
have been manufactured in the lab \cite{ColloidalClusters_Granick,ComplexShapeColloids}.
Second, we studied the rotational and translational diffusion of a
colloidal sphere with nonuniform density, modeling recently-manufactured
``colloidal surfers'' \cite{Hematites_Science} in which a dense
hematite cube is embedded in a polymeric spherical particle. Finally,
we investigated the quasi two-dimensional diffusive motion of a dense
boomerang colloid sedimented near a no-slip boundary, inspired by
recent experimental studies of lithographed boomerang-shaped particles
\cite{BoomerangDiffusion,AsymmetricBoomerangs}.

We demonstrated that the choice of tracking point is crucial when
computing the translational diffusion coefficient, as already observed
and explained in Refs. \cite{BoomerangDiffusion,AsymmetricBoomerangs}.
In particular, we demonstrate that in some cases there exists a suitable
choice of the origin (around which torques are expressed) which can
be used to obtain an approximate but relatively accurate formula for
the effective \emph{long-time} diffusion coefficient in the directions
parallel to the boundary. For highly symmetric shapes with a clear
geometric center it turned out that the ``obvious'' tracking point
is the best one to use. However, for the boomerang shapes studied
here we found that the CoH and CoM are so close to each other that
we cannot numerically distinguish between them. Therefore, it remains
to be confirmed whether the CoH, rather than the CoM, is the correct
point to track in quasi-two-dimensional confinement as claimed in
Refs. \cite{BoomerangDiffusion,AsymmetricBoomerangs}. Ideally one
would find a planar shape for which these two points are far apart;
unfortunately our calculations indicate that all of the boomerang
shapes studied in published experiments have a CoH and a CoM that
are too close to each other to be distinguished to within experimental
and statistical accuracy. Additional investigations of other particle
shapes are necessary to reach more definitive conclusions about diffusion
in quasi-two-dimensional (strong) confinement.

In many practical situations only part of diffusing particle may be
tracked, for example, a unit of a protein may be labeled by a fluorescent
dye. In such cases, one must be very careful in interpreting the results
for translational diffusion as if the particle were spherical and
the center of the sphere were tracked. Furthermore, there are many
particle shapes for which there is no obvious geometric center and
it is then not trivial to determine what the best point to track is,
even if one can track an arbitrary point on the body. In general,
we find that there is no exact closed-form expression for the long-time
quasi-two-dimensional coefficient; it appears necessary to perform
numerical simulations in order to study the long-time diffusive dynamics
of even a single rigid body in the presence of confinement. Our temporal
integrators can easily be extended to study quasi-two-dimensional
suspensions of passive or active particles sedimented near a boundary,
which is quite relevant in practice since active particles often have
metallic components and are therefore much denser than the solvent
\cite{FlippingNanorods,Hematites_Science}.

In our simulations, the time step size was strongly restricted in
order to keep the rigid body from passing through the wall. To this
end, we rejected steps that encountered an unphysical state (e.g.,
a configuration where the computed mobility matrix is not positive
semi-definite). This naive approach modifies the dynamics in a way
that violates ergodicity and detailed balance, and we reduced our
time step size to avoid performing a significant number of rejections.
Several more sophisticated approaches exist that may solve this problem,
including Metropolization \cite{MetropolizedBD}, adaptive time-stepping
\cite{AdaptiveEM_SDEs}, or continious-time discretizations \cite{ContTimeDiscSpace_SDE}.
Employing these techniques in our integrators remains an area of future
exploration.

Recently, the Brownian motion of a spheroid (an axisymmetric particle)
near a single no-slip wall has been studied \cite{SpheroidNearWall}
by using a finite element method for pre-computing the hydrodynamic
mobility over many positions of the particle relative to the wall,
and using the RFD approach to compute the divergence of the mobility
in expectation. The strategy of Ref. \cite{SpheroidNearWall} of pre-computing
the mobility does not extend to suspensions of particles, and constructing
body-fitted finite element meshes and solving the resulting Stokes
equations is rather computationally intensive. In this work we relied
on a simple rigid multiblob approach for computing the hydrodynamic
mobilities \cite{StokesianDynamics_Rigid}, using direct dense linear
algebra to compute inverses and Cholesky factorizations. This was
useful for validating our methods, but it does not scale well with
increasing numbers of rigid bodies or blobs per rigid body. Furthermore,
the analytical approximation we used for the blob mobility is valid
only for the case of a single no slip boundary \cite{StokesianDynamics_Wall},
and even in that case it is not guaranteed to lead to a symmetric
positive-definite grand mobility for all configurations. The RDF scheme
developed in this work can be coupled with a computational fluid solver,
similarly to the approach taken in the FIB method \cite{BrownianBlobs},
in a way that will allow us to do simulations in more complex geometries
such as slit or square channels or chambers, and scale to large numbers
of blobs. The required rigid-body immersed boundary method has recently
been developed \cite{RigidIBM}, and in the future the temporal integrators
developed in this work will be employed to account for the Brownian
motion of the rigid particles.
\begin{acknowledgments}
We thank Qi-Huo Wei for discussions and shared data regarding the
experiments on boomerang colloids. We also thank Eric Vanden-Eijnden
and Miranda Holmes-Cerfon for numerous stimulating and informative
discussions regarding SDEs on manifolds. A. Donev and F. Balboa were
supported in part by the Air Force Office of Scientific Research under
grant number FA9550-12-1-0356. Partial support for A. Donev and S.
Delong was provided by the National Science Foundation under award
DMS-1418706.
\end{acknowledgments}
\newpage

\section*{Appendix}

\appendix

\section{\label{Add:Rotation}Quaternions and Rotation}

In this appendix we derive some relations regarding the quaternion
representation of orientations, as used in the main text.

\subsection{\label{Add:Quaternions}Rotating a body}

In this section, we derive eq. (\ref{eq:Rotate2ndOrder}). We proceed
by first writing the Rotate procedure (\ref{eq:Rotate_def}) using
its definition, and then expand the trigonometric functions to second
order. Letting $\V{\theta}=\left\{ s,\V p\right\} $, and $\norm{\V{\omega}}=\omega$,
we have
\begin{align*}
\mbox{Rotate(}\V{\theta},\V{\omega}\D t)= & \left[\begin{array}{c}
s\cos\left(\frac{\omega\D t}{2}\right)-\V p\cdot\sin\left(\frac{\omega\D t}{2}\right)\V{\omega}/\omega\\
s\sin\left(\frac{\omega\D t}{2}\right)\V{\omega}/\omega+\cos\left(\frac{\omega\D t}{2}\right)\V p+\sin\left(\frac{\omega\D t}{2}\right)\V{\omega}\times\V p/\omega
\end{array}\right]\\
= & \left[\begin{array}{c}
s(1-\frac{\omega^{2}\D t^{2}}{8})-\V p\cdot\V{\omega}\frac{\D t}{2}\\
s\V{\omega}\frac{\D t}{2}+\left(1-\frac{\omega^{2}\D t^{2}}{8}\right)\V p+\V{\omega}\times\V p\frac{\D t}{2}
\end{array}\right]+O\left(\D t^{3}\right)\\
= & \left[\begin{array}{c}
s-\V p\cdot\V{\omega}\frac{\D t}{2}\\
\V p+s\V{\omega}\frac{\D t}{2}-\M P\V{\omega}\frac{\D t}{2}
\end{array}\right]-\frac{\omega^{2}\D t^{2}}{8}\left[\begin{array}{c}
s\\
\V p
\end{array}\right]+O\left(\D t^{3}\right)\\
= & \V{\theta}+\M{\Psi}\V{\omega}\D t-\frac{\left(\V{\omega}\cdot\V{\omega}\right)\D t^{2}}{8}\V{\theta}+O\left(\D t^{3}\right).
\end{align*}

\subsection{\label{Add:Torques}Torques}

In this section, we consider the case when a torque is generated by
a conservative potential $U_{\V{\varphi}}\left(\V{\varphi}\right)$,
so that $\V{\tau}=-\partial U_{\V{\varphi}}/\partial\V{\varphi}$.
Here $\V{\varphi}$ represents the oriented angle associated with
orientation. For the purposes of this discussion, we neglect the dependence
of potential on location, since this will have no bearing on the torque.
Consider extending the energy to depend on a quaternion $U\left(\V{\theta}\right)$
such that when $\norm{\V{\theta}}=1$, we have $U_{\V{\varphi}}\left(\V{\varphi}\right)=U(\V{\theta}_{\V{\varphi}})$.
We want to be able to write the torque, $\V{\tau}$ in terms of $U(\V{\theta})$
without needing to convert first to $\V{\varphi}$. 

Only quaternions with unit norm represent a viable orientation, and
therefore the value of the potential off of this constraint has no
physical meaning and should not affect the torque in any way. The
projected gradient of $U\left(\V{\theta}\right)$ on the unit 4-sphere
is
\[
\frac{\tilde{\partial}U}{\partial\V{\theta}}=\frac{\partial U}{\partial\V{\theta}}-\left(\V{\theta}\cdot\frac{\partial U}{\partial\V{\theta}}\right)\V{\theta}=\M P_{\theta}\frac{\partial U}{\partial\V{\theta}},
\]
where $\M P_{\theta}=\M I-\V{\theta}\V{\theta}^{T}$, and it can easily
be checked that $\M{\Psi}^{T}\M P_{\theta}=\M{\Psi}^{T}$. In section
\ref{sub:Quaternions}, we saw that 
\[
d\V{\theta}=\frac{1}{2}\left[\begin{array}{c}
-\V p^{T}\\
s\M I+\M P
\end{array}\right]d\V{\varphi}=\frac{1}{2}\left[\begin{array}{c}
-\V p\cdot d\V{\varphi}\\
s\, d\V{\varphi}+\V p\times d\V{\varphi}
\end{array}\right],
\]
so that the change in potential energy due to a small rotation $d\V{\varphi}$
is
\[
dU=-\V{\tau}\cdot d\V{\varphi}=\frac{\tilde{\partial}U}{\partial\V{\theta}}\cdot d\V{\theta}=-\frac{1}{2}\left[\frac{\tilde{\partial}U}{\partial s}\V p\cdot d\V{\varphi}-\frac{\tilde{\partial}U}{\partial\V p}\cdot\left(s\, d\V{\varphi}+\V p\times d\V{\varphi}\right)\right].
\]
Using the vector identity $\V a\cdot\left(\V b\times\V c\right)=\V c\cdot\left(\V a\times\V b\right),$
we can rewrite this as
\[
\V{\tau}\cdot d\V{\varphi}=\frac{1}{2}\left[\frac{\tilde{\partial}U}{\partial s}\V p-\left(s\M I-\M P\right)\frac{\tilde{\partial}U}{\partial\V p}\right]\cdot d\V{\varphi},
\]
leading to the identification of torque as
\begin{equation}
\V{\tau}=\frac{1}{2}\left[\frac{\tilde{\partial}U}{\partial s}\V p-\left(s\M I-\M P\right)\frac{\tilde{\partial}U}{\partial\V p}\right]=-\M{\Psi}^{T}\frac{\tilde{\partial}U}{\partial\V{\theta}}=-\M{\Psi}^{T}\M P_{\theta}\frac{\partial U}{\partial\V{\theta}}=-\M{\Psi}^{T}\frac{\partial U}{\partial\V{\theta}}.\label{eq:tau_dU_dq_Appendix}
\end{equation}

\section{\label{Add:ThermalDriftRigid}Stochastic Drift Terms}

Here we show that the temporal integrators introduced in Section \ref{sec:TemporalIntegrators}
generate the correct stochastic drift terms, more precisely, they
are first-order weakly accurate integrators. We will find it convenient
in the following calculations to consider the drift term separated
into multiple pieces as done in (\ref{eq:translational_drift_pieces_indices}).
We first derive (\ref{eq:rotationDriftExpanded}), which we use in
the following subsections. We start with the form of the drift written
in (\ref{eq:overdamped_rot}), denoting for simplicity $\M M\equiv\M M_{\V{\omega}\V{\tau}}$
and using indicial notation with Einstein's implied summation convention
for clarity, 
\begin{align*}
\left[\partial_{\V{\theta}}\cdot\widetilde{\M M}\right]_{i}=\partial_{j}\left(\widetilde{M}_{ij}\right)= & \partial_{j}\left(\Psi_{ik}M_{kl}\Psi_{jl}\right)\\
= & \left(\partial_{j}\Psi_{ik}\right)M_{kl}\Psi_{jl}+\Psi_{ik}\left(\partial_{j}M_{kl}\right)\Psi_{jl}+\Psi_{ik}M_{kl}\left(\partial_{j}\Psi_{jl}\right)\\
= & \left(\partial_{j}\Psi_{ik}\right)M_{kl}\Psi_{jl}+\Psi_{ik}\left(\partial_{j}M_{kl}\right)\Psi_{jl}\\
= & -\frac{1}{4}M_{kk}\theta_{i}+\Psi_{ik}\left(\partial_{j}M_{kl}\right)\Psi_{jl}
\end{align*}
where we used (\ref{eq:div_Psi}) to go from the second to the third
line. To go from the third to the fourth line we used the relationship
$\left(\partial_{j}\Psi_{ik}\right)\Psi_{jl}=-\delta_{kl}\theta_{i}/4$,
which can be shown by a straightforward calculation. In (somewhat
ambiguous) matrix notation, we can write
\begin{equation}
\partial_{\V{\theta}}\cdot\widetilde{\M M}=\M{\Psi}\left(\partial_{\V{\theta}}\M M\right):\M{\Psi}^{T}-\frac{1}{4}\mbox{Tr}\left(\M M\right)\V{\theta},\label{eq:div_M_tilde}
\end{equation}
which we use in proving first-order weak accuracy of our numerical
schemes next.

We also derive a similar relation for the drift including translational
degrees of freedom, as given in (\ref{eq:translational_drift_pieces_indices}).
We use Einstein's implicit summation notation, where Greek indices
range over components of location $\V q$, $s,t$, and $u$ range
over components of orientation $\V{\theta}$, and $i$ represents
any component of $\V x$. We now expand the $i$-th component of the
stochastic drift using the chain rule and (\ref{eq:div_Psi}),

\begin{align}
\left\{ \partial_{\V x}\cdot\left(\M{\Xi}\M N\M{\Xi}^{T}\right)\right\} _{i}= & \left[\begin{array}{c}
\left(\partial_{\beta}M_{i\beta}^{\V v\V F}\right)+\partial_{s}\left(M_{it}^{\V v\V{\tau}}\Psi_{st}\right)\\
\partial_{\alpha}\left(\Psi_{is}M_{s\alpha}^{\V{\omega}\V F}\right)+\partial_{s}\left(\Psi_{it}M_{tu}^{\V{\omega}\V{\tau}}\Psi_{su}\right)
\end{array}\right]\nonumber \\
= & \left[\begin{array}{c}
\left(\partial_{\beta}M_{i\beta}^{\V v\V F}\right)+\left(\partial_{s}M_{it}^{\V v\V{\tau}}\right)\Psi_{st}\\
\Psi_{is}\left(\partial_{\alpha}M_{s\alpha}^{\V{\omega}\V F}\right)+\Psi_{it}\left(\partial_{s}M_{tu}^{\V{\omega}\V{\tau}}\right)\Psi_{su}
\end{array}\right]+\left[\begin{array}{c}
0\\
\left(\partial_{s}\Psi_{it}\right)\left(M_{tu}^{\V{\omega}\V{\tau}}\right)\Psi_{su}
\end{array}\right]\nonumber \\
= & \Xi_{im}\left(\partial_{n}N_{mp}\right)\Xi_{np}+\left[\begin{array}{c}
0\\
-\frac{1}{4}M_{ss}^{\V{\omega}\V{\tau}}\theta_{i}
\end{array}\right].\label{eq:translational_drift_pieces_derivation}
\end{align}

\subsection{\label{sub:FixmanRotation}Fixman's Method}

To show that (\ref{eq:overdamped_rot_strato}) is equivalent to (\ref{eq:overdamped_rot}),
we can use the general identity that given two matrices $\M A\left(\V x\right)$
and $\M B\left(\V x\right)$,
\begin{align}
\M A\circ\M B\,\M{\mathcal{W}}\equiv & \frac{1}{2}\left(\partial_{\V x}\M A\right):\left(\M B\M B^{T}\M A^{T}\right)+\M A\M B\,\M{\mathcal{W}}\label{eq:ito_strato_kinetic}\\
= & \frac{1}{2}\left(\partial_{\V x}\cdot\left(\M A\M B\M B^{T}\M A^{T}\right)-\M A\,\partial_{\V x}\cdot\left(\M B\M B^{T}\M A^{T}\right)\right)+\M A\M B\,\M{\mathcal{W}},\nonumber 
\end{align}
in law, where $\left\{ \left(\partial_{\V x}\M A\right):\left(\M B\M B^{T}\M A^{T}\right)\right\} _{i}=\left(\partial_{l}A_{ij}\right)B_{jk}B_{mk}A_{lm}$,
and we used the product rule to obtain the second line of (\ref{eq:ito_strato_kinetic}).
Applying this identity we obtain 
\begin{align}
 & \sqrt{2k_{B}T}\,\M{\Xi}\M N\circ\M N^{-\frac{1}{2}}\V{\mathcal{W}}\label{eq:drift_N}\\
= & \left(k_{B}T\right)\partial_{\V x}\cdot\left(\M{\Xi}\M N\M{\Xi}^{T}\right)-\left(k_{B}T\right)\M{\Xi}\M N\left(\partial_{\V x}\cdot\M{\Xi}^{T}\right)+\sqrt{2k_{B}T}\,\M{\Xi}\M N^{\frac{1}{2}}\V{\mathcal{W}}\nonumber \\
= & \left(k_{B}T\right)\partial_{\V x}\cdot\left(\M{\Xi}\M N\M{\Xi}^{T}\right)+\sqrt{2k_{B}T}\,\M{\Xi}\M N^{\frac{1}{2}}\V{\mathcal{W}},\nonumber 
\end{align}
where we used (\ref{eq:div_Psi}). 

To show that scheme (\ref{eq:Fixman}) produces the correct drift
terms, we consider the drift for $\V q$ and $\V{\theta}$ separately.
In the following expression, Greek indices range only over components
corresponding to $\V q$ and not those corresponding to $\V{\theta}$,
and indices $s,\, t,\, u,$ and $v$ correspond to only components
of $\V{\theta}$. All other indices range over every variable. Letting
$\D x_{k}^{p,n+\frac{1}{2}}=\sqrt{k_{B}T\D t}\;\Xi_{kl}N_{lm}^{\frac{1}{2}}W_{m}^{n,1}$
be the stochastic term from the increment $x_{k}^{p,n+\frac{1}{2}}-x_{k}^{n}$,
the stochastic drift generated for $\V q$ by the corrector stage
is equal to 
\begin{align*}
\D{_{\text{th}}q_{\alpha}^{n}}= & \sqrt{k_{B}T\D t}\left(\partial_{k}N_{\alpha j}\right)\D x_{k}^{p,n+\frac{1}{2}}N_{jl}^{-\frac{1}{2}}\left(W_{l}^{n,1}+W_{l}^{n,2}\right)\\
= & k_{B}T\D t\left(\left(\partial_{\beta}N_{\alpha j}\right)N_{\beta m}^{\frac{1}{2}}W_{m}^{n,1}+\left(\partial_{s}N_{\alpha j}\right)\Psi_{st}N_{tp}^{\frac{1}{2}}W_{p}^{n,1}\right)\times\\
 & \, N_{jl}^{-\frac{1}{2}}\left(W_{l}^{n,1}+W_{l}^{n,2}\right),
\end{align*}
where all matrices are evaluated at $x^{n}$ and the term involving
$\Psi_{st}$ comes from expanding the Rotate procedure in the predictor
stage. After taking expectation and noting that, for example, $N_{\alpha\beta}=M_{\alpha\beta}^{\V v\V F}$,
we obtain the stochastic drift
\begin{equation}
\av{\D{_{\text{th}}q_{\alpha}^{n}}}=k_{B}T\D t\left(\left(\partial_{\beta}M_{\alpha\beta}^{\V v\V F}\right)+\left(\partial_{s}M_{\alpha t}^{\V v\V{\tau}}\right)\Psi_{st}\right),\label{eq:q_drift_Fixman}
\end{equation}
which matches the first row of the second line in (\ref{eq:translational_drift_pieces_derivation})
as required.

For the drift in the $\V{\theta}$ direction, we expand the Rotate
procedure in the corrector stage to first order in $\D t$ to obtain
\begin{align}
\D{_{\text{th}}\theta_{s}^{n}}= & \Psi_{st}\left(\sqrt{k_{B}T\D t}\left(\partial_{k}N_{ti}\right)\D{x_{k}}^{p,n+\frac{1}{2}}N_{ij}^{-\frac{1}{2}}\left(W_{j}^{n,1}+W_{j}^{n,2}\right)\right)\nonumber \\
 & -\frac{k_{B}T\D t}{8}\left(\left(W_{i}^{n,1}+W_{i}^{n,2}\right)N_{ti}^{\frac{1}{2}}N_{tj}^{\frac{1}{2}}\left(W_{j}^{n,1}+W_{j}^{n,2}\right)\right)\theta_{s}\nonumber \\
= & \left(k_{B}T\D t\right)\Psi_{st}\left(\left(\partial_{\alpha}N_{ti}\right)N_{\alpha k}^{\frac{1}{2}}W_{k}^{n,1}\right.\nonumber \\
 & \left.+\left(\partial_{u}N_{ti}\right)\Psi_{uv}N_{vk}^{\frac{1}{2}}W_{k}^{n,1}\right)N_{ij}^{-\frac{1}{2}}\left(W_{j}^{n,1}+W_{j}^{n,2}\right)\nonumber \\
- & \frac{k_{B}T\D t}{8}\left(\left(W_{i}^{n,1}+W_{i}^{n,2}\right)N_{ti}^{\frac{1}{2}}N_{tj}^{\frac{1}{2}}\left(W_{j}^{n,1}+W_{j}^{n,2}\right)\right)\theta_{s}.\label{eq:fixman_theta_drift}
\end{align}
After taking expectation we obtain the deterministic drift
\begin{equation}
\av{\D{_{\text{th}}\theta_{s}^{n}}}=k_{B}T\D t\left[\Psi_{st}\left(\partial_{\alpha}M_{t\alpha}^{\V{\omega}\V F}\right)+\Psi_{st}\left(\partial_{u}M_{tv}^{\V{\omega}\V{\tau}}\right)\Psi_{uv}-\frac{1}{4}M_{tt}^{\V{\omega}\V{\tau}}\theta_{s}^{n}\right],\label{eq:theta_drift_Fixman}
\end{equation}
which matches the second row of the second line in (\ref{eq:translational_drift_pieces_derivation})
as required. Note that a direct application of the Euler-Heun scheme
\cite{EulerHeun} to the (\ref{eq:overdamped_rot_strato}) would require
the final update of orientation to be
\[
\V{\theta}^{n+1}=\V{\theta}^{n}+\M{\Psi}^{p,n+\frac{1}{2}}\V{\omega}^{p,n+\frac{1}{2}}\D t.
\]
Using $\M{\Psi}^{p,n+\frac{1}{2}}$ instead of $\M{\Psi}^{n}$ here
generates the drift term $-\mbox{Tr}\left(\M M\right)\V{\theta}/4$;
here we obtain that part of the stochastic drift by using the Rotate
procedure instead of a simple additive update of the quaternions.

\subsection{Random Finite Difference Scheme}

To show that the random finite difference term generates the correct
drift, we need to show that $\delta^{-1}\left(\widetilde{\M N}-\M N^{n}\right)\widetilde{\V W}$
is a good approximation to $\partial_{\V x}\left(\M N\right):\M{\Xi}^{T}$
in expectation. We use the convention that Greek indices correspond
to translational degrees of freedom, $s$ and $t$ correspond to angular
degrees of freedom, and the remaining indices are summed over all
variables. Expanding the RFD term gives 
\begin{align*}
\D{_{\text{th}}x_{i}^{n}}=\frac{k_{B}T}{\delta}\left(\widetilde{N}_{ij}-N_{ij}^{n}\right)\widetilde{W_{j}}= & \frac{k_{B}T}{\delta}\partial_{k}\left(N_{ij}\right)\D{\tilde{x}}_{k}\widetilde{W}_{j}+O(\delta),
\end{align*}
where $\D{\widetilde{x}}_{k}=\tilde{x}_{k}-x_{k}^{n}$. Expanding
the increment $\D{\tilde{x}}_{k}$ gives
\begin{align*}
\frac{k_{B}T}{\delta}\partial_{k}\left(N_{ij}\right)\D{\tilde{x}}_{k}\widetilde{W}_{j}= & k_{B}T\left(\partial_{\alpha}\left(N_{ij}\right)\widetilde{W}_{\alpha}\widetilde{W}_{j}+\partial_{s}\left(N_{ij}\right)\Psi_{st}\widetilde{W}_{t}\widetilde{W}_{j}\right).
\end{align*}
Taking expectation gives the desired result
\begin{align*}
\av{\D{_{\text{th}}x_{i}^{n}}}= & k_{B}T\,\partial_{k}\left(N_{ij}\right)\Xi_{kj}.
\end{align*}

\section{\label{Add:RotationalDiff}Planar Rotational Diffusion Coefficient}

In Section \ref{sub:Boomerang}, we computed the two dimensional rotational
diffusion coefficient by measuring the change in angle $\theta$ of
the bisector of the boomerang projected onto the $x-y$ plane. This
is a convenient notion of rotational diffusion when the boomerang
lies flat, in which case it can be used to measure the $z-z$ component
of $\M M_{\V{\omega}\V{\tau}}$. However, in the case of general three
dimensional motion, the relationship between $D_{\theta}(\tau)$ and
$\M M_{\V{\omega}\V{\tau}}$ is not as simple. In this appendix, we
derive the relationship between the short time diffusion coefficient
in $\theta$ and the rotational mobility.

We consider the boomerang at an initial location and orientation,
$\V x=\left(\V q,s,\V p\right)$ and let $\V v=(v_{1},v_{2},v_{3})$
be the unit vector pointing in the direction of the bisector. We let
$\M M=\M M_{\V{\omega}\V{\tau}}(\V x)$ be the rotational mobility
evaluated at the initial configuration. Finally, we let $\M Q$ be
the projection operator that projects vectors onto the $xy$ plane.
We then consider the change in the angle $\theta$ between the projected
bisector and the $x$ axis after a rotation over a small time increment
$\D t$\@. Let $\V{\phi}$ be the angle of rotation over this time
increment, and $\M R$ be the rotation matrix that applies this small
rotation. For small $\V{\phi}$, the change in the scalar angle $\theta$
is 
\begin{align*}
\D{\theta}= & \frac{\norm{\left(\M Q\V v\right)\times\left(\M Q\M R\V v\right)}}{\norm{\M Q\V v}^{2}}+\M O\left(\norm{\V{\phi}}^{2}\right)
\end{align*}
For small $\V{\phi}$, we approximate the rotation matrix as 
\begin{align*}
\M R= & 2\left[\V p\V p^{T}+s\M P+\left(s^{2}-\frac{1}{2}\right)\M I\right]=\M I+\M{\Phi}+O\left(\norm{\V{\phi}}^{2}\right),
\end{align*}
where $\M{\Phi}$ is the cross product matrix for $\V{\phi}$, i.e.,$\M{\Phi}\V x=\V{\phi}\times\V x$.
Using this approximation to $\M R$, we get an expression for the
instantaneous planar diffusion coefficient $\chi_{\theta}\left(\V x\right)$,
\begin{align}
\lim_{\D t\to0}\,\frac{\avv{\D{\theta}^{2}}}{2k_{B}T\D t}= & M_{33}-\alpha\left(2v_{1}v_{3}M_{13}+2v_{2}v_{3}M_{23}\right)+\alpha^{2}v_{3}^{2}\left(v_{1}^{2}M_{11}+v_{2}^{2}M_{22}+2M_{12}v_{1}v_{2}\right),\label{eq:chi_theta_theory}
\end{align}
where $\alpha=\left(v_{1}^{2}+v_{2}^{2}\right)^{-1}$. The average
short time projected rotational diffusion coefficient is then $\chi_{\theta}=\av{\chi_{\theta}\left(\V x\right)}$,
where the average is taken with $\V x$ distributed according to the
equilibrium Gibbs-Boltzmann distribution (\ref{eq:GibbsBoltzmannWithLocation}).

\section{\label{add:SphereMob}Hydrodynamic Mobility of a Sphere Near a Wall}

A low-order approximation of the perpendicular and parallel translational
mobilities of a sphere next to a no-slip boundary is derived by Swan
and Brady \cite{StokesianDynamics_Wall} as a generalization of the
Rotne-Prager tensor using Blake's image construction \cite{blake1971note}.
Neglecting stresslet contributions, this approximation gives the self-mobility
\begin{eqnarray}
\frac{\mu_{\perp}(h)}{\mu_{0}} & = & 1-\frac{9a}{8h}+\frac{a^{3}}{2h^{3}}-\frac{a^{5}}{8h^{5}}\label{eq:SingleWallBlob}\\
\frac{\mu_{\parallel}(h)}{\mu_{0}} & = & 1-\frac{9a}{16h}+\frac{2a^{3}}{16h^{3}}-\frac{a^{5}}{16h^{5}},\nonumber 
\end{eqnarray}
where $\mu_{0}=\left(6\pi\eta a\right)^{-1}$ is the mobility in an
unbounded domain. We do not reproduce the lengthier formula for the
other components of the mobility.

More accurate formulas for the self-mobility of a sphere near a wall
are available. A very good approximation to the perpendicular mobility
is given by a semi-empirical rational relation approximation to an
exact series of Brenner \cite{SingleWallPerpMobility},
\begin{align}
\frac{\mu_{\perp}(h)}{\mu_{0}}= & \frac{6\left(\frac{h}{a}\right)^{2}+2\left(\frac{h}{a}\right)}{6\left(\frac{h}{a}\right)^{2}+9\left(\frac{h}{a}\right)+2}.\label{eq:SingleWallHardSpherePerp}
\end{align}
The hard sphere approximation to the parallel mobility is given by
a combination of a near-wall expression derived using lubrication
theory and a truncated expansion in powers of $a/h$ which is more
accurate further from the wall. The near-wall calculation given by
Goldman and Brenner \cite{NearWallSphereMobility} gives 
\begin{equation}
\frac{\mu_{\parallel}(h)}{\mu_{0}}=\frac{2\left(\ln\left(\frac{h}{a}\right)-0.9543\right)}{\left(\ln\left(\frac{h}{a}\right)\right)^{2}-3.188\ln\left(\frac{h}{a}\right)+1.591}\label{eq:SingleWallHardSphereLub}
\end{equation}
and it is used when $h-a\leq0.03a$. When the sphere is further from
the wall, we calculate the parallel mobility from the exact power
series expansion truncated to fifth order \cite{ConfinedSphere_Sedimented},
\begin{align}
\frac{\mu_{\parallel}(h)}{\mu_{0}}= & 1-\frac{9a}{16h}+\frac{a^{3}}{8h^{3}}-\frac{45a^{4}}{256h^{4}}-\frac{a^{5}}{16h^{5}}.\label{eq:SingleWallHardSphereParallel}
\end{align}

We were unable to find more accurate expansions for the rotation-rotation
and rotation-translation components of the mobility of a sphere near
a wall. Therefore, we compute them from a cubic spline fit to the
numerical mobility obtained from a sphere discretized with 162 blobs.
We observe that this well-resolved multiblob model provides a rather
accurate approximation, as confirmed by comparing the numerical translational
mobilities to the above theoretical expansions.


\end{document}